\documentclass[prd,aps,showpacs,preprintnumbers,amssymb]{revtex4}
\usepackage{graphicx}% Include figure files
\usepackage{epsf}
\usepackage{amsmath}
\usepackage{epstopdf}
\usepackage{multirow}
\usepackage{bm}
\usepackage{dcolumn}

\def\e3p{$\eta \rightarrow 3 \pi$}

\newcolumntype{M}[1]{>{\centering\arraybackslash}m{#1}}
\newcolumntype{N}{@{}m{0pt}@{}}

\begin{document}

\title{%
\hfill{\normalsize\vbox{%
\hbox{}
 }}\\

\vskip 1cm

{Spinless  mesons and glueballs mixing patterns in SU(3) flavor limit}}

\author{Amir H. Fariborz
$^{\it \bf a}$~\footnote[1]{Email:
 fariboa@sunypoly.edu}}

\author{Mars Lyukova
	$^{\it \bf b}$~\footnote[2]{Email:
		Mars.Lyukova@stonybrook.edu}}

\affiliation{$^{\bf \it a}$ Department of Mathematics and Physics, SUNY Polytechnic Institute, Utica, NY 13502, USA}

\affiliation{$^{\bf \it b}$ Department of Physics and Astronomy, Stony Brook University, Stony Brook, NY, 11794-3800, USA}

\date{\today}

\begin{abstract}
	
We present a detailed study of the interactions among scalar and pseudoscalar mesons and glueballs within the framework of the generalized linear sigma model  in the SU(3) flavor limit. The basis of our approach is to develop a global understanding of light scalar and pseudoscalar mesons (up to around 2 GeV) and thereby explore the underlying mixings among composite quark matter fields and glueballs.  The chiral  sector of the   Lagrangian is formulated in terms of two chiral nonets  representing quark-antiquarks and tetraquarks (in the leading order, this sector contains terms with eight or fewer number of quak or antiquark lines).    The Lagrangian also contains a sector that represents   scalar and pseudoscalar glueballs and their interactions with  the matter chiral fields,  in a manner that the axial and trace anomalies of QCD are exactly realized.     In this construct,  the model has two scalar octets and two pdeudoscalar octets (each a linear combination of two- and four-quark components), as well as three  scalar SU(3) singlets and three pseudoscalar SU(3) singlets (each a linear combination of two- and four-quark components as well as a glueball component).    With the inputs of the experimental masses of $a_0(980)$,  $a_0(1450)$, and the masses of  $\pi(137)$, $\pi(1300)$ and their decay constants, we perform an extensive numerical simulation to determine the boundaries of the  parameter space of the model.  We further  incorporate  experimental data on the mass spectrum of eta states as well as on several decay widths and decay ratios of $f_0$ states to  zoom in on the parameter space and make predictions for the substructure of pseudoscalar and scalar SU(3) octets and singlets as well as for the pseudoscalar and scalar  glueball masses.  We find the scalar and pseudoscalar glueball masses around 1.6 and 2.0 GeV,  respectively.

\end{abstract}
\maketitle

\section{Introduction}

One of the most fascinating features of QCD is the possibility that the gluons form colorless bound states with distinct spins, called glueballs \cite{Fritsch,Scharre}.  While the experimental detection of glueballs remains an open question, numerous theoretical works have been conducted and various mechanisms for understanding glueballs have been proposed \cite{PDG}.   Lattice QCD has given estimates of glueball mass spectrum  \cite{Morningstar} and continues to make progress in better understanding the complex nature of these states from a fundamental point of view.   However, the monumental complexity of mixing among glueballs and quark composites  should not  be underestimated and highlights the need for a broad-based investigation of these states from different angles.   Understanding  glueballs is not an standalone problem and naturally requires employing low-energy QCD frameworks  that have confronted various experimental data and can be used to infer information about glueballs. 
Our current understanding of low-energy QCD (which directly or indirectly involves glueballs) is based on progress made in 
chiral perturbation theory \cite{ChPT1} and its extensions such as chiral unitary approach \cite{ChUA1}-\cite{ChUA9} (for a recent review see \cite{ChUA9}) and inverse amplitude method \cite{IAM1}-\cite{IAM4};   lattice QCD  \cite{LQCD1}-\cite{McNeile:2000xx}; QCD sum-rules \cite{SVZ}-\cite{Huang}; linear sigma models \cite{Schechter1}-\cite{LMGR_19}; as well as many other inspiring  non-perturbative techniques \cite{jaffe}-\cite{Amsler:1995tu} (see \cite{PDG} and \cite{07_KZ,Pelaez:2015qba} for comprehensive reviews).

Glueballs with distinct quantum numbers can mix with  mesons of the same quantum numbers made of quarks,     and as a result,  their experimental detection becomes a nontrivial (if not impossible) adventure (see reviews \cite{07_KZ,Pelaez:2015qba,21_FL}).  For example, the scalar and pseudoscalar glueballs,  which are the focus of the present work,  respectively mix with isosinglet scalar and pseudoscalar mesons and their physical features get concealed in the experimental data on these spinless mesons.   
Particularly,  the case of scalar glueballs becomes even more complex due to the general puzzle surrounding the light scalar meson spectroscopy which has triggered an avalanche of  investigations over the past several decades. 
The lightest scalars below 1 GeV show deviations from conventional quark-antiquark states and that raises the  possibility  of these states having  more complex substructures such as four-quark composites \cite{jaffe} or molecular structures  \cite{Weinstein:1990gu}
or a mixed combination of the two.   Moreover, some of the scalars above 1 GeV [such as $f_0(1500)$ and $f_0(1710)$] seem  likely to  contain significant glue components and that adds to the complexity of their study.   The isosinglet pseudoscalar meson sector is also complex and should be understood on the basis of pseudoscalar glueball mixing with quark-made mesons.    Even though  the light pseudoscalar mesons below 1 GeV are  well understood, the situation above 1 GeV remains rather unclear with  an overpopulation of isosinglet pseudoscalars including $\eta(1295)$, 
$\eta(1405)$, $\eta(1475)$, $\eta(1760)$ and $\eta(2225)$ which are all listed in PDG \cite{PDG}.   These states have posed challenges of their own,  both experimentally and theoretically,  and it is likely that some of them are more complex than simple quark-antiquark  mesons.   There are various points of view about some of these states such as, for example, the discussions  in the literature \cite{07_KZ} that $\eta(1405)$ and $\eta(1475)$ are in fact two apparent states of a single underlying state called  $\eta(1440)$ which can be the radial excitation of $\eta$.   Some other eta states are considered to be
possible non quark-antiquark candidates, such as,  for example,  the interpretation of $\eta(1295)$ as an exotic particle (multiquark, glueball or hybrid) in \cite{07_KZ}, or the possibility of $\eta(1405)$ and $\eta(1475)$ being dynamically generated in $\eta f_0(980)$ and $\pi a_0(980)$ channels  investigated in \cite{ChUA8}.

In this work we aim to gain insight into the glueball-meson mixing by probing the substructure of the  scalar and pseudoscalar mesons below 2 GeV within the generalized linear sigma model with glueballs in the SU(3) flavor limit.    The basis of our investigation is to exploit the generalized linear sigma model  predictions for the global picture of scalar and pseudoscalar mesons and their underlying mixing patterns to probe spinless glueballs which appear as part of the substructure of the SU(3) singlet states.  
In the absence of glueballs, the properties of scalar and pseudoscalar mesons up to around  2 GeV,  particularly their quark-antiquark and four-quark admixtures,  have been previously studied \cite{000_BFMNS,Jora1,Jora2,Jora3,Jora4,Jora41,Jora42,Jora43,Jora5}. The  first step in constructing generalized linear sigma model   was taken in \cite{000_BFMNS} and then formalized in \cite{Jora1} (and subsequent works) in which the mass spectrum was studied,   almost entirely,  based on the underlying chiral symmetry and the resulting generating equations and Ward identity-type relations.  The model is formulated in terms of two chiral nonets (a quark-antiquark and a four-quark) and exhibits chiral symmetry and its breakdown,  both spontaneously and explicitly through quark mass terms.    The model also incorporates an effective instanton term that exactly mocks up the U(1)$_{\rm A}$ anomaly.    In the work of \cite{Jora1} the  potential of the model is not specified (i.e. the least model-independent approach), and while in this approach the pseudoscalar states can be studied,   in the scalar sector, only  the strange scalar kappa can be investigated in this method.    Therefore,  relying solely on the underlying symmetry provides limited results and we need to make specific potentials to be able to make more refined predictions. 
To make complete predictions for the scalar and pseudoscalar masses and mixings, the potential had to be modeled  which led to works in  \cite{Jora2}-\cite{Jora7}.  Also for modeling the potential, we cannot only rely on chiral symmetry, its breakdown,  and anomalies because there are infinite number of terms that contribute to the potential and for practical purposes we need to find a guiding  scheme to limit these number of terms.  In \cite{Jora5},  a guiding procedure was defined in which the contributing terms with fewer quark-antiquark lines were considered to be favored.    In this scheme, the leading potential contains terms with eight or fewer quark lines.  Various low-energy processes were studied in this framework in its leading order and it was found that while light pseudoscalar mesons below 1 GeV are dominantly quark-antiquark states, the  light scalars have reversed substructure, typically with  dominant four-quark components.   
% ---------------------------------
The generalized linear sigma model is applied to low-energy processes such as $\pi\pi$ scattering in \cite{11_GLSM_pipi} and $\pi K$ scattering in \cite{15_GLSM_piK}.  Motivated by large $N_c$ approximation,  the scattering amplitudes are approximated by tree diagrams, and the effect of quantum loops are approximated by K-matrix unitarization method. Such unitarization processes are particularly  necessary in order to approximate the rescattering effects when broad resonances $f_0(500)$ and $K_0^*(800)$ are probed in $\pi\pi$ and $\pi K$ scatterings, respectively. In this  work we  consider only the tree-level  amplitudes to calculate the decay width  of  scalar mesons to two pseudoscalar meson.
% ----------------------------------

Recently \cite{Jora25},  the generalized linear sigma model was extended to include two glueballs (a scalar and a pseudoscalar glueball). In \cite{Jora25},  the most general Lagrangian was formally developed that includes, in addition to  chiral symmetry (and its spontaneous and explicit breakdown), terms that represent interaction of glueballs with chiral nonets in a manner that exactly realizes the axial and trace anomalies of QCD.    This most general Lagrangian contains (even in the leading order) a large number of free parameters that need to be determined by fitting the theoretical predictions of the model to available experimental data. Since various theoretical predictions are  nonlinear functions of the model parameters, a fit of these predictions to available data by a brute force numerical method is neither an efficient nor a  reliable starting point due to the fact that the minimization of fitting functions (such as $\chi^2$ function or $\chi$ function used in \cite{Jora1}) that depend on many parameters are often a jagged function with many local minima.   Therefore, in such minimizations it is possible to end up at a non-physical minimum in the multidimensional parameter space of the model.   Our general strategy is to first push the model to  exactly solvable limits by imposing physically meaningful conditions, and once the parameters are determined in these solvable limits, gradually relax the conditions and study the evolution of the parameter space.      In \cite{Jora25}, we considered a decoupling limit, in which the scalar glueball  does not mix with quark composite fields.    In this exactly solvable limit, scalar glueball still plays an important role in stabilizing the QCD vacuum but refrains from mixing with quark composite fields,  and hence,  becomes a pure glueball.  Moreover, we showed in \cite{Jora26} that the model predicts  the pure scalar glueball mass $m_h = 2 h_0$ where $h_0$ is the glueball condensate.   In order to determine $h_0$ the  connection between $h_0$ and dimension four gluon condensate was worked out in  \cite{Jora26} with the value of the gluon condensate taken from the  QCD sum-rules.  This resulted in $h_0$ in the range of  0.887 to 0.974 GeV which in turn results in an estimate of the pure scalar glueball mass in the range of 1.77-1.95 GeV.   However, there is a significant discrepancy between the estimates of gluon condensate in QCD Sum-Rules and in Lattice QCD,  and therefore,  estimation of $h_0$ which is of central importance in our investigation requires further scrutiny.  We also determined,  in the decoupling limit,  the initial model predictions for masses as well as for quark and glue contents.    In the decoupling limit, the model is pushed to SU(3)$_{\rm V}$ subgroup with additional non-mixing restrictions imposed on the glueballs.

To take the first step away from the decoupling limit (while remaining in the flavor SU(3) limit) in \cite{19_FJL}, we relaxed the additional conditions that prevented the glueballs from mixing with quark-antiquark and four-quark nonets.   
%-------------------------------
Since glueballs are flavor singlets, it seems useful to first try to understand their interactions with distinct quark composites (quark-antiquarks and four-quarks) in SU(3) flavor limit which treats the light flavors on the same footing and significantly simplifies the studies of these interactions.   Within this limit,  the SU(3) singlets are of a special importance because they mix with glueballs and play a significant role in such analysis.  Therefore,  SU(3) flavor limit can be considered a reasonable starting point for understanding  the interactions of  glueballs with quark composites. Of course,  for a more robust understanding of glueball interactions with mesons the SU(3) breaking effects should be included in order to be able to construct a more realistic picture for the substructure of scalar and pseudoscalar mesons and thereby study their interactions with glueballs.  In \cite{19_FJL}, our optimizatopn favored $h_0 = 0.8-0.81$ GeV.  In this work, we probe $h_0$ more broadly and within the generalized linear sigma model framework without taking any inputs from either QCD Sum-Rules or Lattice QCD.       We resort to extensive numerical simulation and investigate the mass spectrum of scalar and pseudoscalar mesons over a broad range of $h_0$ from 0.20 to  0.87 GeV.
%--------------------------------------

In Sec. II, we give a brief overview of the generalized linear sigma model with glueballs as well as our notation and 
setup in the SU(3) flavor limit.   We then give  our strategy for determination of the boundaries of the parameter space  in Sec. III.   Section IV is devoted to our simulation results and  examination of the  glueball condensate, followed  by a summary, conclusion and directions for future works in Sec. V.   Several detailed formulas are given in the appendix \ref{A_formulas} and several tables of numerical details are given in appendix \ref{A_num_res}.

%%%%%%%%%%%%%%%%%%%%%%%%%%%%%%%%%%%%%%%%%%%%%%%%%%%%%%%%%%%%%%%%%%%%%%%%%%%

\section{Brief Review of the Theoretical framework}

%%%%%%%%%%%%%%%%%%%%%%%%%%%%%%%%%%%%%%%%%%%%%%%%%%%%%%%%%%%%%%%%%%%%%%%%%%%

In this section we give a brief review of the theoretical model developed in \cite{Jora1}-\cite{Jora7} and  extended to include scalar and pseudoscalar glueballs in \cite{Jora25,Jora26,19_FJL}.   The model is constructed in terms of 3$\times$3 matrix
chiral nonet fields:
\begin{equation}
M = S +i\phi, \hskip 2cm
M^\prime = S^\prime +i\phi^\prime,
\label{sandphi1}
\end{equation}
which are in turn defined in terms of ``bare'' scalar meson nonets $S$ (a quark-antiquark scalar nonet) and $S'$ (a four-quark scalar nonet), as well as ``bare'' pseudoscalar meson nonets $\phi$ (a quark-antiquark pseudoscalar nonet) and $\phi'$ (a four-quark pseudoscalar nonet).
Chiral fields $M$ and $M'$ transform in the same way under
chiral SU(3) transformations
\begin{eqnarray}
M &\rightarrow& U_L\, M \, U_R^\dagger,\nonumber\\
M' &\rightarrow& U_L\, M' \, U_R^\dagger.
\label{E_CS}
\end{eqnarray}
% ======================================
To see how $M$ and $M'$  transform under chiral symmetry according to  (\ref{E_CS}),  we consider 
their schematic quark substructure. Chiral nonet $M$ is a $q \bar q$ composite:
\begin{equation}
M_a^b \propto {\left( q_{bA} \right)}^\dagger \gamma_4 \frac{1 + \gamma_5}{2} q_
{aA}=\left[{\bar q}_R\right]^{bA}q_{LaA},
\label{M}
\end{equation}
where $a$ and $b$ are flavor indices and $A$ is color index and  the left and the right handed
projections are respectively $q_L = \frac{1}{2}\left( 1 + \gamma_5 \right) q$ and $q_R = \frac{1}{2}\left( 1 - \gamma_5 \right) q$.   This explicitly illustrates the transformation
(\ref{E_CS}) for $M$.        The U(1)$_A$ transformation acts as $q_{aL}
\rightarrow e^{i\nu} q_{aL}$, $q_{aR} \rightarrow e^{-i\nu} q_{aR}$ and
results in:
\begin{equation}
M \rightarrow e^{2i\nu} M. 
\label{MU1A}
\end{equation}

% ======================================
The quark substructure of the $M'$ is understandably more complicated and there are several four-quark constructs that all transform  according to  (\ref{E_CS}).    One of these constructs has a ``molecular'' substructure can be defined in terms of $M$:
\begin{equation}
M_a^{(2)b} \propto \epsilon_{acd} \epsilon^{bef}
{\left( M^{\dagger} \right)}_e^c {\left( M^{\dagger} \right)}_f^d.
\label{M2}
\end{equation}
Under U(1)$_A$ it transforms as 
\begin{equation}
M^{(2)} \rightarrow e^{-4i\nu} M^{(2)},
\end{equation}
which differs from U(1)$_A$ transformation of $M$ given in (\ref{MU1A}).  
Two other substructures can be constructed in terms of diquarks-antidiquarks with two distinct 
spin and color underpinnings  that still transform    according to  (\ref{E_CS}).   
In the first case the two composite quarks belong to a $\bar 3$ of color and form a spin singlet.   Defining 
\begin{eqnarray}
L^{gE} = \epsilon^{gab} \epsilon^{EAB}q_{aA}^T C^{-1} \frac{1 +
	\gamma_5}{2} q_{bB}, \nonumber \\
R^{gE} = \epsilon^{gab} \epsilon^{EAB}q_{aA}^T C^{-1} \frac{1 - \gamma_5}{2}
q_{bB}, 
\end{eqnarray}
where $C$ is the charge conjugation matrix of the Dirac theory, we can form chiral nonet:
\begin{equation}
M_g^{(3)f} \propto {\left( L^{gA}\right)}^\dagger R^{fA}.
\label{M3}
\end{equation}
which transforms in the same way as $M^{(2)}$ under
SU(3)$_L \times$ SU(3)$_R$ and U(1)$_A$.  Another possibility is when the two composite quarks form a $6$ representation of color and have spin 1.  Using
\begin{eqnarray}
L_{\mu \nu,AB}^g = L_{\mu \nu,BA}^g = \epsilon^{gab} q^T_{aA} C^{-1}
\sigma_{\mu \nu} \frac{1 + \gamma_5}{2} q_{bB}, \nonumber \\
R_{\mu \nu,AB}^g = R_{\mu \nu,BA}^g = \epsilon^{gab} q^T_{aA} C^{-1}
\sigma_{\mu \nu} \frac{1 - \gamma_5}{2} q_{bB}, 
\end{eqnarray}
where $\sigma_{\mu \nu} = \frac{1}{2i} \left[ \gamma_\mu, \gamma_\nu \right]
$, we can define another four-quark chiral nonet:
\begin{equation}
M_g^{(4) f} \propto {\left( L^{g}_{\mu \nu,AB}\right)}^\dagger R^{f}_{\mu \nu,AB},
\label{M4}
\end{equation}
which also transforms like $M^{(2)}$ and $M^{(3)}$
under all of SU(3)$_L \times$ SU(3)$_R$ and 
U(1)$_A$.   It can be shown (using Fierz transformation) that there is a linear relationship among 
$M^{(2)}$, $M^{(3)}$ and $M^{(4)}$.   Since in formulation of the present framework (which is formulated at the mesonic level) there is no direct contact with the specific underlying color and spin configurations, the model  cannot distinguish among the three substructures (\ref{M2}),     (\ref{M3}) and  (\ref{M4}), and therefore, our $M'$ is in principle some unknown linear combination of these four-quark configurations. In other words,  what the model can distinguish is quark-antiquark versus four-quark (due to different U(1)$_A$ transformations), but is not able to favor one type of four-quark configuration versus the other.   Under chiral transformation both $M$ and $M'$ transform according to (\ref{E_CS}), but they transform differently under  U(1)$_A$ transformation: 
\begin{eqnarray}
M &\rightarrow& e^{2i\nu}\, M,  \nonumber\\
M' &\rightarrow& e^{-4i\nu}\, M'.
\label{U1A}
\end{eqnarray}

% ======================================

The effective Lagrangian is constructed in terms $M$ and $M'$ and should display chiral symmetry (and its beakdown), explicit symmetry breaking due to quark masses as well as the axial and trace anomalies of QCD according to:
\begin{eqnarray}
&&\partial^{\mu}J^5_{\mu}=\frac{g^2}{16\pi^2}N_F\tilde{F}F=G,
\nonumber\\
&&\theta^{\mu}_{\mu}=\partial^{\mu}D_{\mu} = -\frac{\beta(g^2)}{2g}FF=H,
\label{intr64553}
\end{eqnarray}
where $F$ is the SU(3)$_{\rm C}$ field tensor, $\tilde{F}$ is its dual, $N_F$ is the number of flavors, $\beta(g^2)$ is the beta function for the coupling constant, $J^5_{\mu}$ is the axial current and $D_{\mu}$ is the dilatation current.

With identifications $H=h^4$, $G=g h^3$, where $h$ and $g$ respectively represent the scalar and pseudoscalar glueball fields, the generalized linear sigma model Lagrangian augmented to include scalar and pseudoscalar glueballs has the general structure \cite{Jora25}
\begin{eqnarray}
{\cal L}&=&-\frac{1}{2}{\rm Tr}(\partial^{\mu}M\partial_{\mu}M^{\dagger})-\frac{1}{2}{\rm Tr}(\partial^{\mu}M'\partial_{\mu}{M'}^{\dagger})
-{1\over 2} (\partial_{\mu} h)(\partial_{\mu} h) - {1\over 2} (\partial_{\mu} g) (\partial_{\mu} g)
-V, \nonumber \\
- V&=& f + f_{\rm A} + f_{\rm S} + f_{\rm SB}.
\label{inlgr567}
\end{eqnarray}
where $f(M, M', g, h)$ is invariant under chiral, axial and scale transformations, and in leading order is
\begin{eqnarray}
f &=&
- \left(
u_1 h^2 {\rm Tr}[MM^{\dagger}]
+ u_2{\rm Tr}[MM^{\dagger}MM^{\dagger}]+
u_3 h^2 {\rm Tr}[M^{\prime}M^{\prime \dagger}]+ u_4 h (\epsilon_{abc}\epsilon^{def}M^a_dM^b_eM^{\prime c}_f+h.c.)+
\right.
\nonumber \\
&&
\left.\hskip .5cm  u_5 h^4 + u_6  h^2 g^2  + \cdots\right),
\label{pot201867} 
\end{eqnarray}
where $u_1\cdots u_6$ are the unknown constants that need to be determined from experiment.    The instanton term $f_{\rm A}$ breaks axial symmetry and is given by
\begin{eqnarray}
f_{\rm A} &=&
i{G\over 12} \left[ \gamma_1\ln \left(\frac{\det M}{\det M^{\dagger}}\right)+\gamma_2\ln\left(\frac{{\rm Tr}(MM^{\prime\dagger})}{{\rm Tr}(M^{\prime}M^{\dagger})}\right)\right],
\label{axial2018967}
\end{eqnarray}
where $\gamma_1$ and $\gamma_2$  are arbitrary parameters that must satisfy the constraint: $\gamma_1+\gamma_2=1$ \cite{Jora25}.   The next term $f_{\rm S}$ breaks scale symmetry
\begin{eqnarray}
f_{\rm S} &=&
-H \left\{
\lambda_1 \ln\left(\frac{H}{\Lambda^4}\right)
+\lambda_2 \left[\ln\left(\frac{\det M}{\Lambda^3}\right)+\ln\left(\frac{\det M^{\dagger}}{\Lambda^3}\right)\right]
\right.
\nonumber \\
&&
\left.
\hskip .8cm
+ \lambda_3\left[ \ln\left(\frac{{\rm Tr} MM^{\prime\dagger}}{\Lambda^2}\right)+\ln\left(\frac{{\rm Tr}M' M^{\dagger}}{\Lambda^2}\right)
\right]\right\}.
\label{scale7756}
\end{eqnarray}
where $\lambda_1$, $\lambda_2$ and $\lambda_3$ are also arbitrary parameters that must fulfill the condition: $4\lambda_1+6\lambda_2+4\lambda_3=1$ \cite{Jora25}.
The potential is invariant under ${\rm U(3)}_{\rm L} \times {\rm  U(3)}_{\rm R}$ with the exception of the $f_A$ term which breaks ${\rm U(1)}_{\rm A}$.
The leading explicit symmetry breaking term has the form of quark mass term:
\begin{eqnarray}
f_{SB}=2 {\rm Tr}[{\cal A} S]
\label{sym3528637}
\end{eqnarray}
where ${\cal A}={\rm diag}(A_1,A_2,A_3)$ is a  matrix proportional to the three light quark masses.  We are interested in the SU(3) limit where
\begin{eqnarray}
&&\alpha_1 = \alpha_2 = \alpha_3 = \alpha
\nonumber\\
&&\beta_1 = \beta_2 = \beta_3 = \beta
\nonumber \\
&&A_1 = A_2 = A_3 = A 
\label{u3_sol}
\end{eqnarray}
In this limit the minimum equations are:
\begin{eqnarray}
\left\langle {{\partial V} \over {\partial S_1^1}} \right\rangle_0  &=&
\left\langle {{\partial V} \over {\partial S_3^3}} \right\rangle_0 =
8\, u_4\, h_0\,\alpha\,\beta + 2\, u_1 \, { h_0}^{2}\alpha
+ 4\, u_2 \, {\alpha}^{3} +
{2\over 3}\,{\frac {{h_0}^{4} \left( 3\,\lambda_2+\lambda_3 \right) }{
		\alpha}}
-2\,A = 0,
\label{E_Vmin_qq}
\\
% ------------------------------------------------------------------------------
\left\langle {{\partial V} \over {\partial {S'}_1^1}} \right\rangle_0  &=&
\left\langle {{\partial V} \over {\partial {S'}_3^3}} \right\rangle_0  =
{{2h_0}\over{3\beta}}
\left(
6\,\alpha^{2}\beta\, u_4 + 3\, u_3\,h_0\, {\beta}^{2}+{
		 h_0}^{3}\lambda_3
\right) = 0,
\label{E_Vmin_4q}
\\
% ------------------------------------------------------------------------------
\left\langle {{\partial V} \over {\partial h}} \right\rangle_0  &=&
8\,\ln  \left( {\frac {3\alpha\,\beta}{{\Lambda}^{2}}} \right) 
{ h_0}^{3}\lambda_3 + 8\,\ln  \left( {\frac {{\alpha}^{3}}{{\Lambda}^{3
		}}} \right) { h_0}^{3}\lambda_2 + 4\,\ln  \left( {\frac {{ h_0}^{
		4}}{{\Lambda}^{4}}} \right) { h_0}^{3}\lambda_1\nonumber \\
&&
+ 4\, \left( \lambda_1 +  u_5 \right) { h_0}^{3}
+ 6\, \left(  u_1 \,{\alpha}^{2} + u_3\,{\beta}^{2} \right)  h_0
+12\,{\alpha}^{2}\beta\, u_4  = 0.
\label{E_Vmin_h}
\end{eqnarray}
The first two equations represent the stability of vacuum with respect to variation of quark-antiquark  field $S_1^1$ (or $S_3^3$), and four-quark field ${S'}_1^1$ (or ${S'}_3^3$), respectively.  The last equation determines  the stability of vacuum with respect to the scalar glueball field $h$.

The octet ``8'' and singlet ``0'' mass matrices in the SU(3) limit are (where $Y$ refers to the scalars and $N$ to the pseudoscalars):
\begin{eqnarray}
\left(Y^2_8\right)_{11} &=&
{1\over \alpha^2}
\left(12\,u_2\,{\alpha}^{4}-4\, u_4 \,  h_0\,\beta\,{
	\alpha}^{2}+2\, u_1\,{ h_0}^{2}{\alpha}^{2}-2\,{ h_0}^{4}
\lambda_2\right)
\nonumber\\
\left(Y^2_8\right)_{12} &=& 		
{{2h_0}\over {3\alpha\beta}}
\left( -6\,{\alpha}^{2}\beta\,  u_4 +{h_0}^{3}\lambda_3 \right)
\nonumber\\
\left(Y^2_8\right)_{22} &=& 2 u_3 h_0^2		
\end{eqnarray}
\begin{eqnarray}
\left(Y^2_0\right)_{11} &=&
{1 \over {3\alpha^2}}\left[
\left( -6\,\lambda_2-2\,\lambda_3 \right) h_0^{4}+6
\, u_1\, h_0^2{\alpha}^{2}+24\, u_4 \, h_0\,\beta\,
{\alpha}^{2}+36\, u_2\,{\alpha}^{4}
\right]
\nonumber\\
\left(Y^2_0\right)_{12} &=&
8\,u_4\, h0\,\alpha
\nonumber\\
\left(Y^2_0\right)_{13} &=&
{4\over {\sqrt{3}\alpha}}\,
\left( 6\,{\alpha}^{2}\beta\,{\it u_4}+3\,{\it
	u_1}\,{\it h_0}\,{\alpha}^{2}+6\,{h_0}^{3}\lambda_2+2\,{h_0}^{3
}\lambda_3 \right)
\nonumber\\
\left(Y^2_0\right)_{22} &=&
-{{2h_0^2}\over {3 \beta^2}} \,
\left( -3\, u_3\, \beta^{2}+ h_0^2\lambda_3 \right)
\nonumber\\		
\left(Y^2_0\right)_{23}		&=&	
{{4\sqrt{3}}\over {3\beta}}
\left( 3\,{\alpha}^{2}\beta\,{\it u_4}+3\, u_3\, h_0\,{\beta}^{2}+2\, h_0^{3}\lambda_3 \right)
\nonumber\\
\left(Y^2_0\right)_{33}		&=&	
24\,\ln  \left( {\frac {{\alpha}^{3}}{{\Lambda}^{3}}} \right) h_0^2 \lambda_2
+12\,\ln  \left( {\frac {h_0^4}{{\Lambda}^{4}}}
\right) {h_0}^{2}\lambda_1+24\,\ln  \left( {\frac {\alpha\,\beta
	}{{\Lambda}^{2}}} \right) {h_0}^{2}\lambda_3+24\,\ln  \left( 3
\right) { h_0}^{2}\lambda_3 + \left( 28\,\lambda_1 + 12\,{\it u_5}
\right) {h_0}^{2}
\nonumber\\
&&
+6\, u_1 \,{\alpha}^{2}+6\, u_3 \,{\beta}
^{2}
\label{Y20}
\end{eqnarray}

\begin{eqnarray}
\left(N^2_8\right)_{11} &=&
{1\over {\alpha^2}}
\left(
4\, u_2 \, \alpha^4 + 4\,  u_4 \, h_0 \, \beta \, \alpha^2 + 2\,  u_1 \, h_0^2 \alpha^2 + 2\, h_0^4
\lambda_2
\right)
\nonumber\\
\left(N^2_8\right)_{12} &=&
{{2 h_0}\over {3 \alpha \beta}}
\left( 6\, \alpha^2 \beta\,  u_4 +  h_0^3 \lambda_3 \right)
\nonumber\\
\left(N^2_8\right)_{22} &=&
2\, u_3\,  h_0^2
\end{eqnarray}
\begin{eqnarray}
\left(N^2_0\right)_{11} &=&
{1\over {3\alpha^2}}
\left[
\left(
6\,\lambda_2 + 2\,\lambda_3 \right) h_0^4 + 6 \,  u_1 \, h_0^2 \alpha^2 - 24\, u_4 \,  h_0 \, \beta\,
\alpha^2 + 12\, u_2 \, \alpha^4
\right]
\nonumber\\
\left(N^2_0\right)_{12} &=& -8\, u_4 \, h_0\, \alpha
\nonumber\\
\left(N^2_0\right)_{13} &=&
{{\sqrt{3}h_0^3}\over {18\alpha}}
\left( 2\,\gamma_1 + 1 \right)
\nonumber\\
\left(N^2_0\right)_{22} &=&
{{2h_0^2}\over {3\beta^2}}\,
\left( 3\,u_3\,\beta^2 + h_0^2\lambda_3 \right)
\nonumber\\
\left(N^2_0\right)_{23} &=&
{{\sqrt {3} h_0^3}\over {18\beta}}\,  \left( -1+\gamma_1 \right)
\nonumber\\
\left(N^2_0\right)_{33} &=& 2\, u_6\, h_0^2.
\nonumber\\
\end{eqnarray}

The octet physical states
\begin{equation}
\Psi_{8^+}  =
\left[
\begin{array}{cc}
\psi_{8^+}^{(1)}\\
\psi_{8^+}^{(2)}
\end{array}
\right],
% --------------------------------------
\hskip .75cm
% --------------------------------------
\Psi_{8^-}  =
\left[
\begin{array}{cc}
\psi_{8^-}^{(1)}\\
\psi_{8^-}^{(2)}
\end{array}
\right],
\end{equation}
diagonalize $\left[Y_8^2\right]$ and $\left[N_8^2\right]$ respectively and are related to the octet ``bare'' states
\begin{equation}
B_{8^+}=
\left[
\begin{array}{c}
f_8\\
f'_8
\end{array}
\right],
%---------------------------------------------------
\hskip .75cm
%---------------------------------------------------
B_{8^-}=
\left[
\begin{array}{c}
\eta_8\\
\eta'_8
\end{array}
\right],
\label{F_eta8_OS}
\end{equation}
by
\begin{eqnarray}
\Psi_{8^+} &=&
\left[{K_{8^+}}\right]^{-1}
B_{8^+},
\nonumber\\
\Psi_{8^-} &=&
\left[{K_{8^-}}\right]^{-1}
B_{8^-},
\end{eqnarray}
therefore
\begin{eqnarray}
{\widetilde \Psi}_{8^+}
 \left[ Y_8^2\right]_{\rm diag}  \Psi_{8^+} & = &
{\widetilde  B_{8^+}}  \left[Y_8^2\right] B_{8^+}, \nonumber \\
{\widetilde \Psi}_{8^-}
 \left[ N_8^2\right]_{\rm diag}  \Psi_{8^-} & = &
 {\widetilde  B_{8^-}}  \left[N_8^2\right] B_{8^-},
\end{eqnarray}
where $\left[ Y_8^2\right]_{\rm diag} = {\rm diag}\left(m_{8^+}, {m'}_{8^+}\right)$ and
$\left[ N_8^2\right]_{\rm diag} = {\rm diag}\left(m_{8^-}, {m'}_{8^-}\right)$ are diagonalized mass matrices that contain the physical octet masses.

Similarly, the singlet physical states
\begin{equation}
\Psi_{0^+}  =
\left[
\begin{array}{cc}
\psi_{0^+}^{(1)}\\
\psi_{0^+}^{(2)}\\
\psi_{0^+}^{(3)}
\end{array}
\right],
% --------------------------------------
\hskip .75cm
% --------------------------------------
\Psi_{0^-}  =
\left[
\begin{array}{cc}
\psi_{0^-}^{(1)}\\
\psi_{0^-}^{(2)}\\
\psi_{0^-}^{(3)}
\end{array}
\right],
\end{equation}
diagonalize $\left[Y_0^2\right]$ and $\left[N_0^2\right]$ respectively and are related to the singlet ``bare'' states
\begin{equation}
B_{0^+}=
\left[
\begin{array}{c}
f_0\\
f'_0\\
h
\end{array}
\right],
%---------------------------------------------------
\hskip .75cm
%---------------------------------------------------
B_{0^-}=
\left[
\begin{array}{c}
\eta_0\\
\eta'_0\\
g
\end{array}
\right],
\label{F_eta_OS}
\end{equation}
by
\begin{eqnarray}
\Psi_{0^+} &=&
\left[{K_{0^+}}\right]^{-1}
B_{0^+},
\nonumber\\
\Psi_{0^-} &=&
\left[{K_{0^-}}\right]^{-1}
B_{0^-},
\end{eqnarray}
therefore
\begin{eqnarray}
{\widetilde \Psi}_{0^+}
 \left[ Y_0^2\right]_{\rm diag}  \Psi_{0^+} & = &
{\widetilde  B_{0^+}}  \left[Y_0^2\right] B_{0^+}, \nonumber \\
{\widetilde \Psi}_{0^-}
 \left[ N_0^2\right]_{\rm diag}  \Psi_{0^-} & = &
 {\widetilde  B_{0^-}}  \left[N_0^2\right] B_{0^-}.
\label{m0+_def}
\end{eqnarray}
where $\left[ Y_0^2\right]_{\rm diag} = {\rm diag}\left(m_{0^+}, {m'}_{0^+}, {m''}_{0^+}\right)$ and
$\left[ N_0^2\right]_{\rm diag} = {\rm diag}\left(m_{0^-}, {m'}_{0^-}, {m''}_{0^-}\right)$ are diagonalized mass matrices that contain the physical singlet masses.

In summary,  the fields in this model are tabulated in Table \ref{T_MMpgh}.  In the absence of SU(3) flavor symmetry breaking (the framework of the present work) it can be seen that there are two SU(3) octet mixings (each described by a $2\times2$ rotation matrix) as well as two SU(3) singlet mixings (each described by a $3\times 3$ rotation matrix).    When SU(3) flavor symmetry is broken into the isospin symmetry,  each isosinglet state within each octet also mixes with the SU(3) singlets of appropriate parity.  This leads to $5\times 5$ rotation matrices among the singlets (or higher dimensional rotation matrices if additional glueballs are introduced).    Note that the conventional singlet-octet mixing (or the $\eta$-$\eta'$ mixing at the physical level) is studied within the quark-antiquark nonet in the absense of the four-quark nonet (or additional glueballs) and os of a different type than the mixing that we study in this work.

%   New Table

\begin{table}[ht]
	\caption{Chiral nonets and glueballs in this model (first row); breakdown of each chiral nonet in terms of its scalar and pseudoscalar nonets (second row); and SU(3) singlets and octets (third row).}
	\begin{center}
		\begin{tabular}{c|c|c|c|c|c|c|c|c|c}
			\hline
				\multicolumn{4}{c|}{$M$}&\multicolumn{4}{|c|}{$M'$}& $g$ & $h$\\
			\hline
			\multicolumn{2}{c}{$S$}&\multicolumn{2}{|c}{$\phi$}& 	\multicolumn{2}{|c|}{$S'$}&\multicolumn{2}{|c|}{$\phi'$} & --& --\\			\hline
		$8^+$ & $0^+$ & $8^-$& $0^-$ & $8^+$	& $0^+$ & $8^-$& $0^-$ & $0^-$ & $0^+$\\
	%	\hline\\
		%	\multicolumn{2}{c}{}&X\\
			%\hline
		%	X&X&X\\
			\hline
		\end{tabular}
		\end{center}
	\label{T_MMpgh}
\end{table}

%%%%%%%%%%%%%%%%%%%%%%%%%%%%%%%%%%%%%%%%%%%%%%%%%%%%%%%%%%%%%%%%%%%%%%%%%%%%%%

\section{Boundaries of the parameter space of the model}

%%%%%%%%%%%%%%%%%%%%%%%%%%%%%%%%%%%%%%%%%%%%%%%%%%%%%%%%%%%%%%%%%%%%%%%%%%%
Taking into account the relationship among $\gamma_i$ terms in axial anomaly (\ref{axial2018967}) as well as the relationship among $\lambda_i$ terms in  trace  anomaly (\ref{scale7756}),  there are 14 remaining  free parameters that need to be determined in this model:  $u_1\cdots u_6$, $\gamma_1$,   $\lambda_2$, $\lambda_3$, $\Lambda$, $A$, $\alpha$, $\beta$, $h_0$.    It is not practical to  tackle the determination of these 14 parameters by attempting to form and solve exact mathematical equations which take inputs from experiment when the experimental situation is still evolving.  
The model includes five scalar mesons and five pseudoscalar mesons below and above 1 GeV.   Some of the pseudoscalars above 1 GeV have not been fully understood, and similarly for scalars the situation is  complicated and the experimental data is even more limited.  Even if there were sufficient number of experimental inputs, the basic physical quantities computed in this model (such as masses and decay properties) are computed through systems of coupled  nonlinear equations which are not completely solvable analytically. 
Moreover, regardless of availability and reliability of experimental data as well as mathematical complications in solving these systems,  both for pseudoscalars and scalars, there is an over population of experimental states that requires searching their properties to find out if any of them are compatible with the current model.  Therefore, overall, an exact analytical method is not feasible and we are forced to proceed with a mixed analytical-numerical approach in which we incorporate experimental inputs to solve for some of the parameters, but ultimately we are left with imposing constraints and probing the parameter space of the model through an optimization process.

%=======================================
Our initial strategy for determination of these parameters is given in detail in  Refs. \cite{Jora26} and  \cite{19_FJL} and will be extended in the present work.   For the convenience of the readers, here in this section  we give a summary of this strategy and  parameter determination.      When dealing with a large number of parameters that are determined from a set of nonlinear mathematical relationships, the minimization function  is often a rapidly changing function over the parameter space and can contain many local minia.  Therefore,  the optimization can potentially become quite nontrivial because  the numerical algorithm can, in principle, converge to any of the local minima,  and that makes the minimization process quite challenging.  Our general strategy  is to probe the free parameters of the model by first studying a special (and physically meaningful) limit at which the model is exactly solvable.   This special limit was studied in \cite{Jora26} where a decoupling limit was considered in which the scalar glueball does not mix with quark components of the scalar mesons and becomes a pure scalar glueball.   In this limit, which reduces the symmetry to SU(3)$_{\rm V}$,  the scalar glueball still plays an important role in stabilizing the QCD vacuum and the scalar glueball mass becomes exactly $m_h = 2h_0$.  The glueball condensate was not determined within the framework of \cite{Jora26} and a value of $h_0$ in the range of  0.887-0.98 GeV was  used by connecting to QCD sum-rules. Since the model is exactly solvable in this limit,  we  use the values of the model parameters as a starting point when we relax the decoupling condition.   In \cite{19_FJL}, the effects of removing the decoupling condition while still maintaining SU(3)$_{\rm V}$ symmetry, was studied and the values of the model parameters were determined.      In the present work,  we continue the parameter determination by allowing $h_0$ to vary over a large range and use additional experimental inputs  to determine the best value of $h_0$ exclusively within the framework of the model. For the numerical determination of the parameters we scan the parameter space and form sets of acceptable parameter values that satisfy the overall range of experimental inputs.  The next step in our strategy is that once these simulation sets are determined, we then define our minimization function $\chi$ to search through  these sets for point(s) that can give the closest agreement with the target experimental values.    
 
%=======================================

  We solve for $A$ from the minimum equation 
(\ref{E_Vmin_qq}) and use the minimum equation  (\ref{E_Vmin_4q}) to establish the relationship:
\begin{eqnarray}
	2\alpha^2u_4+\beta h_0u_3+\frac{\alpha h_0^3\lambda_3}{3\alpha\beta}=0,
	\label{firdtmin76775}
\end{eqnarray}
which gives
\begin{eqnarray}
	[N_8^2]_{12}=-2h_0^2\frac{\beta }{\alpha}u_3.
	\label{E_N8_12_initial}
\end{eqnarray}
Using the parameterization of rotation matrices for the pseudoscalar octet we can write:
\begin{eqnarray}
	\left[N_8^2\right]_{12}&=& \sin \theta_{8^-} \cos \theta_{8^-} \left( {m'}_{8^-}^2 - m_{8^-}^2\right)
	=-2h_0^2\frac{\beta }{\alpha}u_3,
	\label{E_N8_12}\\
	\left[N_8^2\right]_{22}	&=&  \cos^2 \theta_{8^-} \, {m'}_{8^-}^2 +
	                                              \sin^2 \theta_{8^-} \, m_{8-}^2 = 2\, h_0^2\, u_3.
\label{E_N8_22}
\end{eqnarray}
The decay constants of the pseudoscalar octet are obtained from equations:
\begin{eqnarray}
	&&f_{8^-}=2\alpha\cos \theta_{8^-}-2\beta\sin\theta_{8^-},
	\nonumber\\
	&&{f'}_{8^-}=2\alpha\sin \theta_{8^-} + 2\beta\cos \theta_{8^-}.
	\label{E_fpi_fpip}
\end{eqnarray}
which gives
\begin{eqnarray}
&&\frac{\beta}{\alpha}=\frac{-\sin\theta_{8^-}\,
	f_{8^-} + \cos\theta_{8^-}\, f'_{8^-}}{\cos\theta_{8^-}\, f_{8^-} + \sin\theta_{8^-}f'_{8^-}}.
\label{E_beta_over_alpha}
\end{eqnarray}
Then, this result together with Eqs. (\ref{E_N8_12}) and (\ref{E_N8_22}) completely determine $\cos \theta_{8^-}$ in terms of pseudoscalar octet masses and decay constants:
\begin{eqnarray}
	\cos\theta_{8^-}=f_{8^-}m^2_{8^-}\Bigg[f_{8^-}^2\, m_{8^-}^4 + {f'}_{8^-}^2 \, {m'}_{8^-}^4\Bigg]^{-1/2}.
	\label{E_cos_theta_8m}
\end{eqnarray}
Substitution of this result into (\ref{E_fpi_fpip}) determines $\alpha$ and $\beta$ and subsequently $u_3$ from (\ref{E_N8_22}), all  in terms of pseudoscalar octet masses and decay constants as well as the glueball condensate $h_0$.  It is natural to identify  the two pseudoscalar octet states with $\pi(137)$ that has well-known mass and decay constant and with $\pi(1300)$  with known experimental bounds on its mass and an expected range for its decay constant.  This allows numerical computation of  $\alpha$, $\beta$ and the pseudoscalar octet mixing angle $\theta_{8^-}$.   The glueball condensate $h_0$  is not known experimentally, but it has  been studied within our model in Refs. \cite{Jora25,Jora26,19_FJL}.  For our numerical analysis we use inputs
\begin{eqnarray}
m_{8^-} &=& 137 \,\,{\rm MeV}
\nonumber\\
{m'}_{8^-} &=& 1300 \pm 100 \,\,{\rm MeV}
\nonumber \\
f_{8^-} &=& 131 \,\,{\rm MeV}
\nonumber\\
{f'}_{8^-} &=&  -0.6939 \pm 0.06939 \,\,{\rm MeV}
\nonumber\\
m_{8^+} &=& 980 \,\,{\rm MeV}
\nonumber\\
{m'}_{8^+} &=& 1474\,\,{\rm MeV}
\nonumber \\
h_0 &=& 0.20 \rightarrow 0.87 \,\,{\rm GeV} 
\label{inputs}
\end{eqnarray}
Note that in this work we have significantly extended the range of glueball condensate $h_0$ compared to the work of \cite{19_FJL} in which a much tighter range of 0.80-0.81 GeV was used. The overall relationships among parameters lead to a complicated numerical system that requires significant parameter search and   optimization. The experimental  uncertainties of the inputs in (\ref{inputs})  result in a set of predicted  points in  the three dimensional ($\alpha$, $\beta$, $u_3$) parameter subspace:
\begin{equation}
S_\mathrm{I} = \left\{
\alpha, \beta, u_3 \,\Bigg| \, {\rm Generated \,\, from \,\, inputs \,\, of \,\,} (\ref{inputs})
\right\}
\label{S_I}
\end{equation}

Using  $\alpha$, $\beta$, $u_3$ from set $S_\mathrm{I}$, together with the trace and determinant of the octets we can determine $u_1$, $u_2$, $u_4$ and $\lambda_2$.   Then, setting $\lambda_1=\frac{11}{36}$  from a first order trace anomaly result, together with  the condition $4\lambda_1+6\lambda_2+4\lambda_3=1$,  we can calculate $\lambda_3$.   The determinant equations
\begin{eqnarray}
\det \left(N_8^2\right) &=&  m^2_{8^-} {m'}^2_{8^-},
\label{E_det_N8}\\
\det\left( Y_8^2\right)	&=&  m^2_{8^+} {m'}^2_{8^+},
\label{E_det_Y8}
\end{eqnarray}
can be used to solve for $u_1$ and $u_2$:
\begin{eqnarray}
u_1 &=& 
\left( 
144\,\alpha^4 \beta^2  h_0^2 u_4^2 
+288 \, \alpha^2 \beta^3 h_0^3 u_3 \, u_4 
-48 \, \alpha^2 \beta\,h_0^5 \lambda_3\, u_4 
-108 \,\beta^4 h_0^4 u_3^2 
+144 \, \beta^2 h_0^6 \lambda_2 \, u_3 
+4 \, h_0^8 \lambda_3^2 \right. \nonumber \\
&& \left. 
-27 \, m_{8^-}^2  m_{8^-}'^2 \alpha^2 \beta^2 
+9 \, m_{8^+}^2 m_{8^+}'^2 \alpha^2 \beta^2 \right) / 
\left( 72\,\alpha^2 \beta^2 h_0^4 u_3 \right) ,
\nonumber \\
u_2	&=&  \left( 
144 \, \alpha^4 \beta^2 h_0^2 u_4^2
+144 \, \alpha^2 \beta^3 h_0^3 u_3\, u_4
-48 \, \alpha^2 \beta \, h_0^5 \lambda_3 \, u_4
-36 \, \beta^4 h_0^4 u_3^2 
+72 \, \beta^2 h_0^6 \lambda_2 \, u_3 \right. \nonumber \\
&& \left. 
+4 \, h_0^8 \lambda_3^2 
-9\, m_{8^-}^2 m_{8^-}'^2\alpha^2 \beta^2
+9\, m_{8^+}^2 m_{8^+}'^2\alpha^2 \beta^2 \right) /
\left( 144\,\alpha^4\beta^2 h_0^2 u_3 \right),
\label{E_u1_u2}
\end{eqnarray}
which upon substitution into trace of scalar octet leads to a quadratic equation in $u_4$ and $\lambda_2$, therefore, for every input of $\alpha$, $\beta$ and $u_3$ from set $S_\mathrm{I}$,  we cannot uniquely solve for $u_4$ and $\lambda_2$.   This creates a large set of points for $u_4$ and $\lambda_2$ [as well as $u_1$ and $u_2$ using (\ref{E_u1_u2})] which  requires imposing physical constrains.
We select values of $u_4 $ and $\lambda_2$ that give scalar singlet masses below 2 GeV.  This generates a set of points for $u_4$ and $\lambda_2$ :
\begin{equation}
	S_\mathrm{II} = \left\{
	u_1, u_2, u_4, \lambda_2 \,\Bigg| \, m^2_{8^+} > 0, {m'}^2_{8^+}> 0,
 m^2_{0^+} > 0, {m'}^2_{0^+}> 0, {m''}^2_{0^+}> 0,	
	{\rm Tr}\left(Y_0^2\right)\le 12 \,\,\mathrm{GeV}^2 \right\}.
	\label{S_II}
\end{equation}
We find that  $\lambda_2$ is limited in the range of $-0.04$ to $-0.03$.  

Parameters $\gamma_1$ and $u_6$ can be probed when constrains from the pseudoscalar singlet system are considered.
We examine whether there are values of $\gamma_1$ and $u_6$ (together with choices of $\alpha$, $\beta$ and $u_3$ from set $S_\mathrm{I}$ and choices of  $u_1$, $u_2$, $u_4$, $\lambda_2$ from set $S_\mathrm{II}$) such that the three pseudoscalar singlets of the model approach three of the experimentally known $\eta$ states. There are several $\eta$ states listed in PDG \cite{PDG}:
\begin{eqnarray}
m^{\rm exp.}[\eta (547)] &=& 547.862 \pm
0.017\, {\rm
	MeV},\nonumber \\
m^{\rm exp.}[\eta' (958)] &=& 957.78 \pm 0.06
\, {\rm
	MeV}.
\nonumber\\
m^{\rm exp.}[\eta (1295)] &=& 1294 \pm 4\, {\rm
	MeV},\nonumber \\
m^{\rm exp.}[\eta (1405)] &=& 1408.8 \pm 1.8 \,
{\rm
	MeV},
\nonumber \\
m^{\rm exp.}[\eta (1475)] &=& 1476 \pm 4\, {\rm
	MeV},\nonumber \\
m^{\rm exp.}[\eta (1760)] &=& 1751 \pm 15 \,
{\rm
	MeV},
\nonumber \\
m^{\rm exp.}[\eta (2225)] &=& 2221^{+13}_{-10} \,
{\rm 	MeV}.
\label{E_eta_exp}
\end{eqnarray}
We will refer to these $\eta$ states by $\eta_i$ with $i=1\cdots 7$. There are 35 scenarios for selecting three of the $\eta$ states from  this list.  
% -----------------------
Note that the mixing among singlets studied here is of a different type (which involves mixing among quark-antiquarks, four-quarks and glueballs) and  not the standard octet-singlet mixing in which $\eta'(958)$ is found to be closer to the singlet state \cite{99_scadron}.  
In this work,  we further examine the assignment of these pseudoscalar singlet states to physical $\eta$'s by imposing additional constrains and allowing a wider range of $h_0$.    
% -------------------------
To generate  range of $\eta$ masses in (\ref{E_eta_exp}), we numerically scan  $u_6$ and $\gamma_1$ and  impose basic constraints on predicted isosinglet masses.  This results in  the following set:  
\begin{equation}
	S_\mathrm{III} = \left\{
	u_6, \gamma_1 \,\Bigg| \, m^2_{0^-} > 0, {m'}^2_{0^-}> 0, {m''}^2_{0^-}> 0,
	{\rm Tr}\left(N_0^2\right)\le 12 \,\,\mathrm{GeV}^2 \right\}.
	\label{S_III}
\end{equation}
Note that the pseudoscalar singlet masses also depend on other parameters
which are selected from sets $S_\mathrm{I}$ and $S_\mathrm{II}$.   
The three sets $S_\mathrm{I}$, $S_\mathrm{II}$ and $S_\mathrm{III}$ define a rough boundaries for the parameter space of the model based on imposing basic constraints and expectations.   To narrow down this parameter space, we need to impose further conditions and experimental inputs.

Although the remaining two parameters $\Lambda$ and $u_5$ appear in the mass matrix (\ref{Y20}),  when minimum equation (\ref{E_Vmin_h}) is invoked these two parameters get eliminated from the mass matrix and 
their values are not needed in mass spectrum or decay analysis of this model.   We will discuss these parameters in the Sec. V.

%%%%%%%%%%%%%%%%%%%%%%%%%%%%%%%%%%%%%%%%%%%%%%%%%%%%%%%%%%%%%%%%%%%%%%%%%%%

\section{Simulation resutls}

%%%%%%%%%%%%%%%%%%%%%%%%%%%%%%%%%%%%%%%%%%%%%%%%%%%%%%%%%%%%%%%%%%%%%%%%%%%

The boundaries of the parameter space of the model is numerically defined by the three sets $S_\mathrm{I}$, $S_\mathrm{II}$ and $S_\mathrm{III}$ determined in previous section.  
We now search through this parameter space for a region  that is most consistent with specific experimental data that we use as constraints.    Note that the masses of pseudoscalar and scalar octets are already used in determination of the boundaries of the parameter space in previous section.  In this section we impose two sets of experimental data on the SU(3) singlet states: The mass spectrum of the etas (\ref{E_eta_exp}) and the decay properties of the insosinglet scalar mesons.

\subsection{Imposing constraints from $\eta$ system}

We investigate
which of the 35 scenarios for identifying our model predictions for the three pseudoscalar singlets with three of the $\eta$ states is favored.  We use function $\chi$ (introduced in \cite{Jora1}),  which measures the agreement of our model prediction for masses $m_{0^-}$, ${m'}_{0^-}$ and   ${m''}_{0^-}$ with the central values of the experimental masses, i.e.
\begin{equation}
\chi_{ijk} =
{
	{\left| {m}_{0^-} - {\widehat m}^{\rm exp}_{\eta_i} \right|}
	\over
	{{\widehat m}^{\rm exp}_{\eta_i} }
}
+
{
	{\left| {m'}_{0^-} - {\widehat m}^{\rm exp}_{\eta_j} \right|}
	\over
	{{\widehat m}^{\rm exp}_{\eta_j} }
}
+
{
	{\left| {m''}_{0^-} - {\widehat m}^{\rm exp}_{\eta_k} \right|}
	\over
	{{\widehat m}^{\rm exp}_{\eta_k} }
}
\end{equation}
where $i,j,k = 1 \cdots 7$ with $i<j<k$ and hatted parameters refer to the central values of the experimental masses, i.e. $m^{\rm exp}_{\eta_i} = {\widehat m}^{\rm exp}_{\eta_i} \pm \delta m_{\eta_i}^{\rm exp}$.    We measure the goodness of each $\chi_{ijk}$ by comparing it with the overall percentage of experimental  uncertainty that we define by
\begin{equation}
\chi^{\rm exp}_{ijk} =
{
	{\delta m_{\eta_i}^{\rm exp}}
	\over
	{{\widehat m}^{\rm exp}_{\eta_i} }
}
+
{
	{\delta m_{\eta_j}^{\rm exp}}
	\over
	{{\widehat m}^{\rm exp}_{\eta_j} }
}
+
{
	{\delta m_{\eta_k}^{\rm exp}}
	\over
	{{\widehat m}^{\rm exp}_{\eta_k} }
}.
\label{E_chi_exp_ijk}
\end{equation}
Clearly, a scenario $ijk$ with $\chi_{ijk} \le \chi^{\rm exp}_{ijk}$ is within the overall experimental uncertainty.

\begin{figure}[!htb]
 	\centering
 	\includegraphics[width=5in]{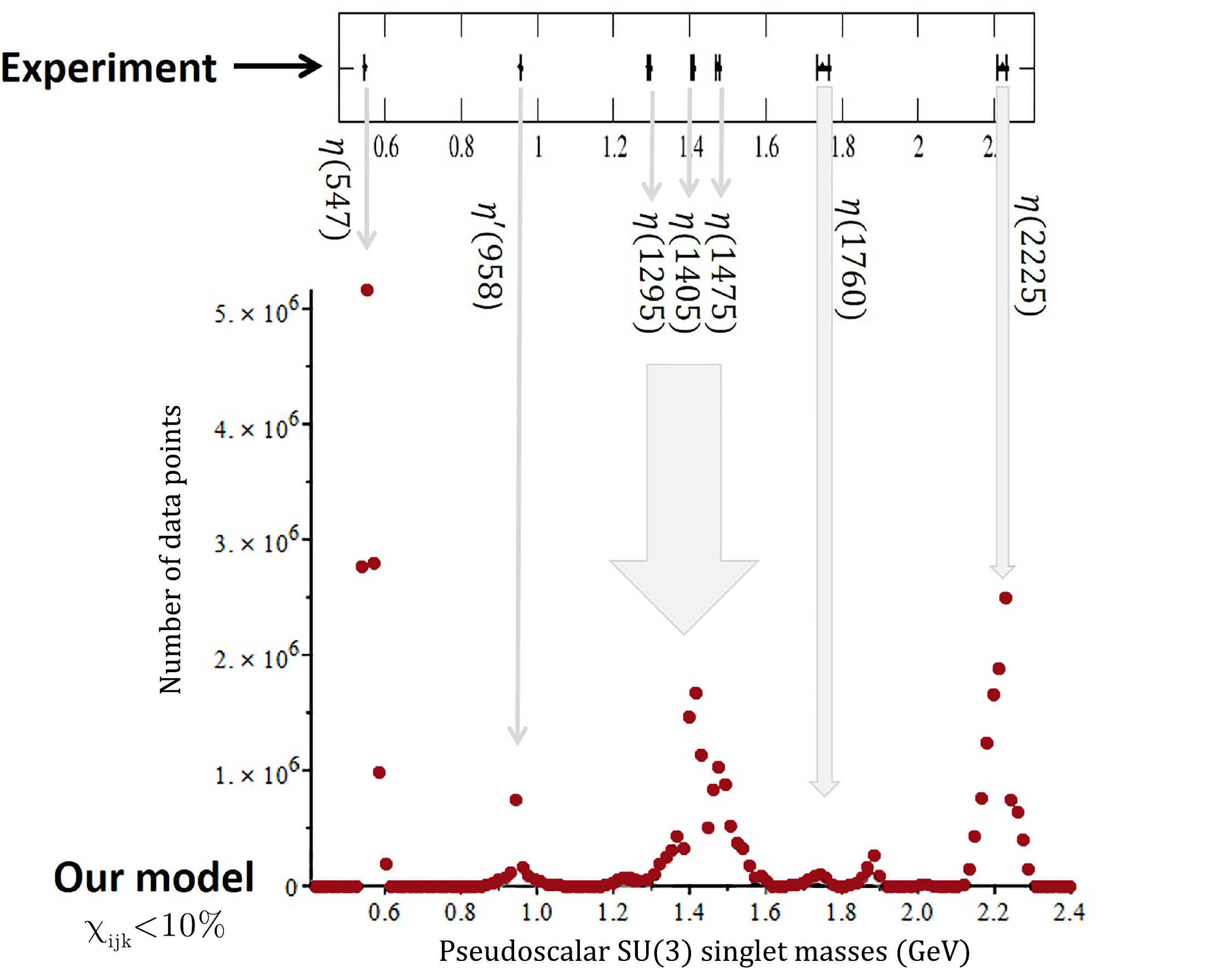} \\ [15pt]
 	\caption{Numerical simulation results with $\chi_{ijk} < 10\%$ for  the SU(3)
 		 singlet pseudoscalar masses (GeV).  The model clearly predicts $\eta(547)$ and $\eta(2225)$ and shows a  smearing of states around 1.4 GeV which may be identified with the interfering effects of $\eta(1295)$, $\eta(1405)$ and $\eta(1475)$.  The $\eta'(958)$ can be seen as well. An indication of $\eta(1760)$ together with a small enhancement around 1.9 GeV can also be seen albeit with insufficient statistics. }
 	\label{F_eta_masses_global}
 \end{figure}

For orientation, we start out by examining our simulation results obtained by only imposing that the $\chi_{ijk}$ function for all possible permutations is less than ten percent (note that $\chi_{ijk}^{\rm exp}$ is less than about 1\%).   This allows checking  how naturally compatible the model predictions are with experimental data for the $\eta$ system.    In Fig. \ref{F_eta_masses_global} (lower pannel) we have given an overview of our simulations for the pseudoscalar SU(3) singlet masses and compared them with the available experimental data on eta system reported in PDG \cite{PDG}  in the same figure (upper pannel).   The vertical axis in our simulation graph (lower pannel of Fig. \ref{F_eta_masses_global}) gives the number of simulations that resulted in a given mass for the  pseudoscalar SU(3) singlets which are given on the horizontal axis (in GeV).   A comparison with experiment shows that the most clear signal in our model is for 
 $\eta(547)$ (the first peak) with more than five million simulations predicted this state,  followed by 
$\eta(2225)$ (the last peak) with about 2.5 million simulations predicting this state,    and a less clear peak that can be compared with a 
group of states around 1.3-1.5 GeV region that hosts the experimental candidates $\eta(1295)$, $\eta(1440)$ and $\eta(1475)$. The 
 $\eta'(958)$ is also seen (the second peak).  The state that is not detected is $\eta(1760)$ even though there is a very small bump at that energy as well as another small bump around 1.9 GeV.  Both of these two small signals are statistically insignificant and we will see that they get eliminated when a tighter chi is imposed.

\begin{figure}[!htb]
	\centering
	\includegraphics[width=4in]{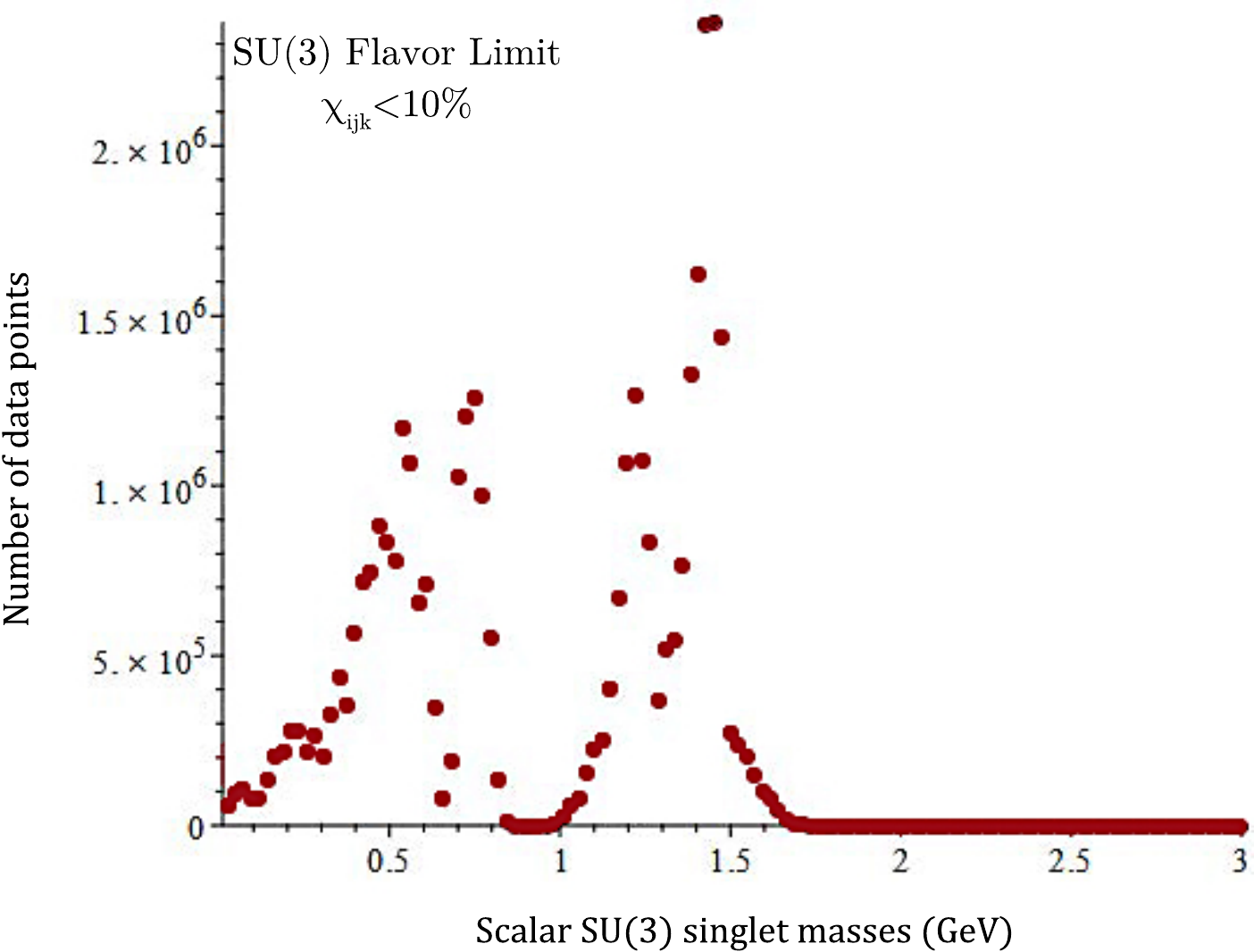} 
	\caption{Numerical simulation results with $\chi_{ijk} < 10\%$ for the three SU(3) singlet scalar masses (GeV).   The  peak in the range of 0.4-0.8 GeV is a clear indication of the sigma meson and the significant enhancements in the two regions of 1.2-1.4 GeV and 1.4-1.5 above 1 GeV may be identified with $f_0(1370)$ and $f_0(1500)$.
		 }
	\label{F_f0_masses_global}
\end{figure}

The case of scalar SU(3) singlets is considerably more complex as can be seen in Fig. \ref{F_f0_masses_global} 
 in which only one sharp peak  together with a smearing of other states appear.   
 The most clear signal (with about 2.5 million simulations) is around 1.4-1.5
 GeV together with a nearby enhancement in the range of 1.2-1.4 GeV  which can be indications of $f_0(1500)$ and 
 $f_0(1370)$, respectively.     Below 1 GeV the figure shows a light scalar meson around 0.4-0.8 GeV that should represent the broad and interfering sigma meson.

\begin{figure}[!htb]
	\centering
	\includegraphics[width=4in]{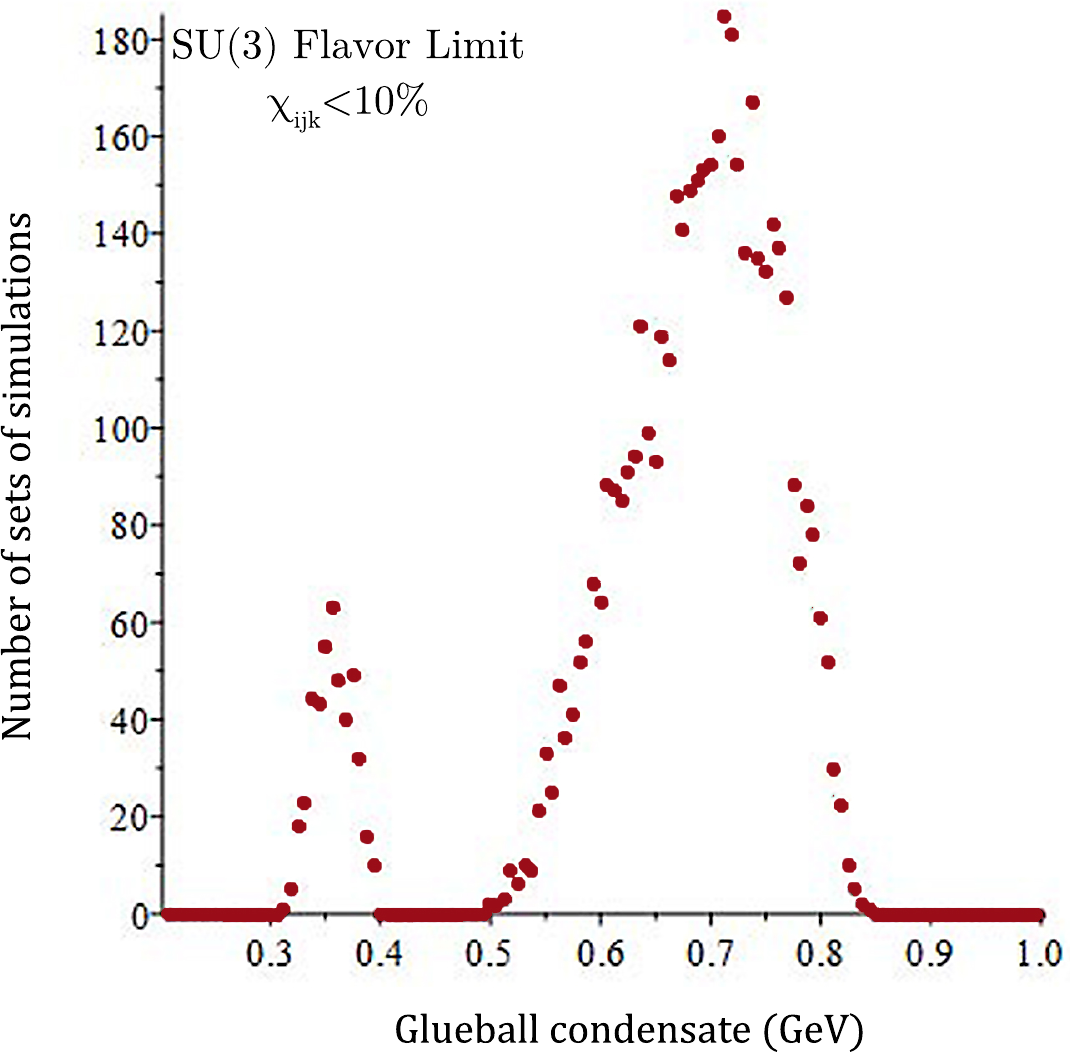}
	
	\caption{Numerical simulation results with $\chi_{ijk} < 10\%$ for glueball condensate $h_0=\langle h\rangle$ (horizontal axis, in GeV).  Two ranges between 0.3-0.4 GeV and 0.5-0.8 GeV can be seen. (Note that the vertical axis gives the number of \underline{\it sets} of simulations and that for each set there are numerous data points.)}
	\label{F_glueball_cond}
\end{figure}

 Probing the scalar glueball condensate is one of our important objectives and in Fig. \ref{F_glueball_cond} we have given the number of simulations versus this quantity.  
 We see that clearly there are two favored ranges one  $0.3-0.4$ GeV and another $0.5-0.8$ GeV. We found in the work of \cite{19_FJL} that $h_0\approx 0.8$ GeV which is near the larger peak.

\begin{figure}[!htb]
	\centering
	\includegraphics[width=3in]{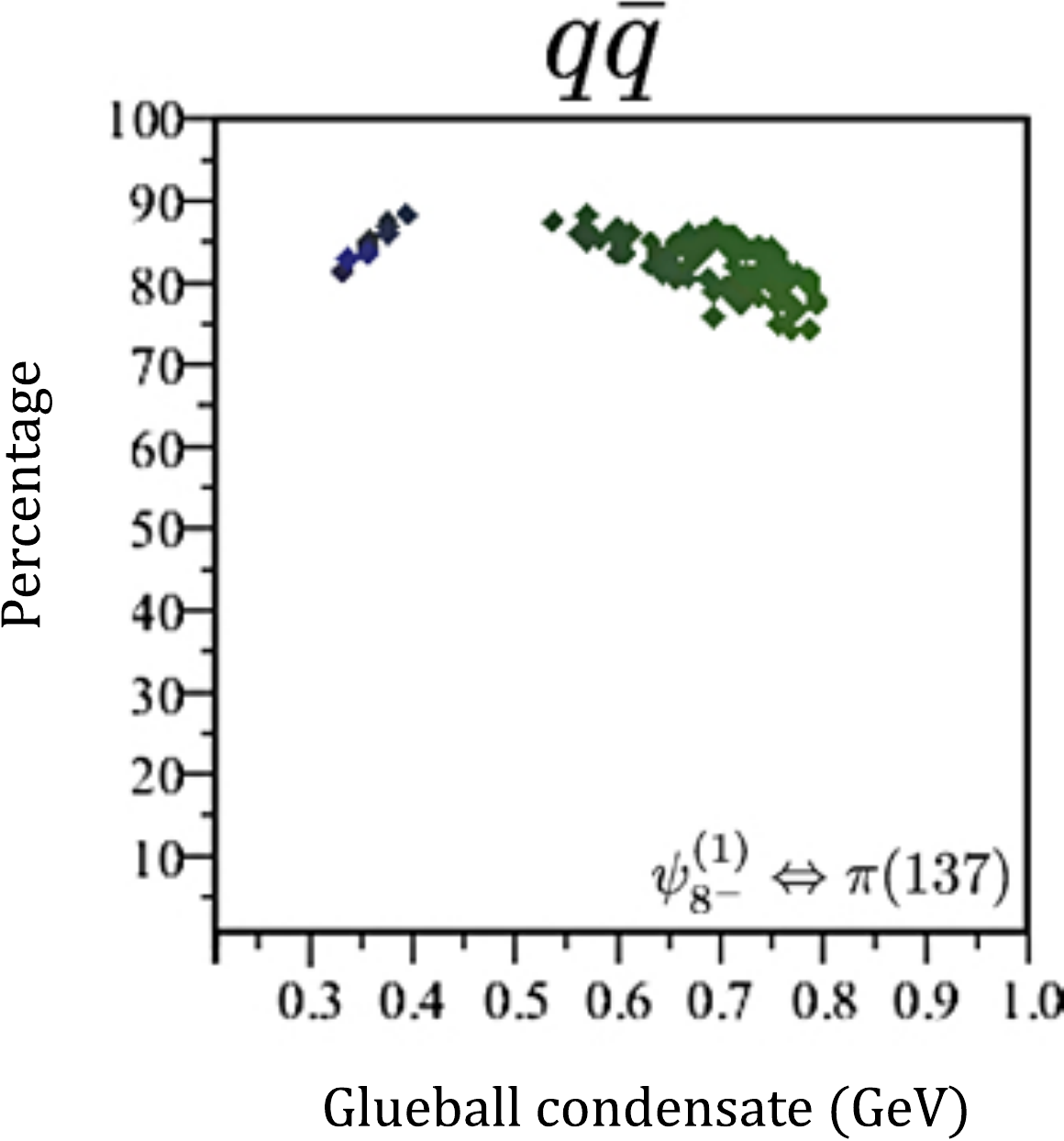}
	\includegraphics[width=3in]{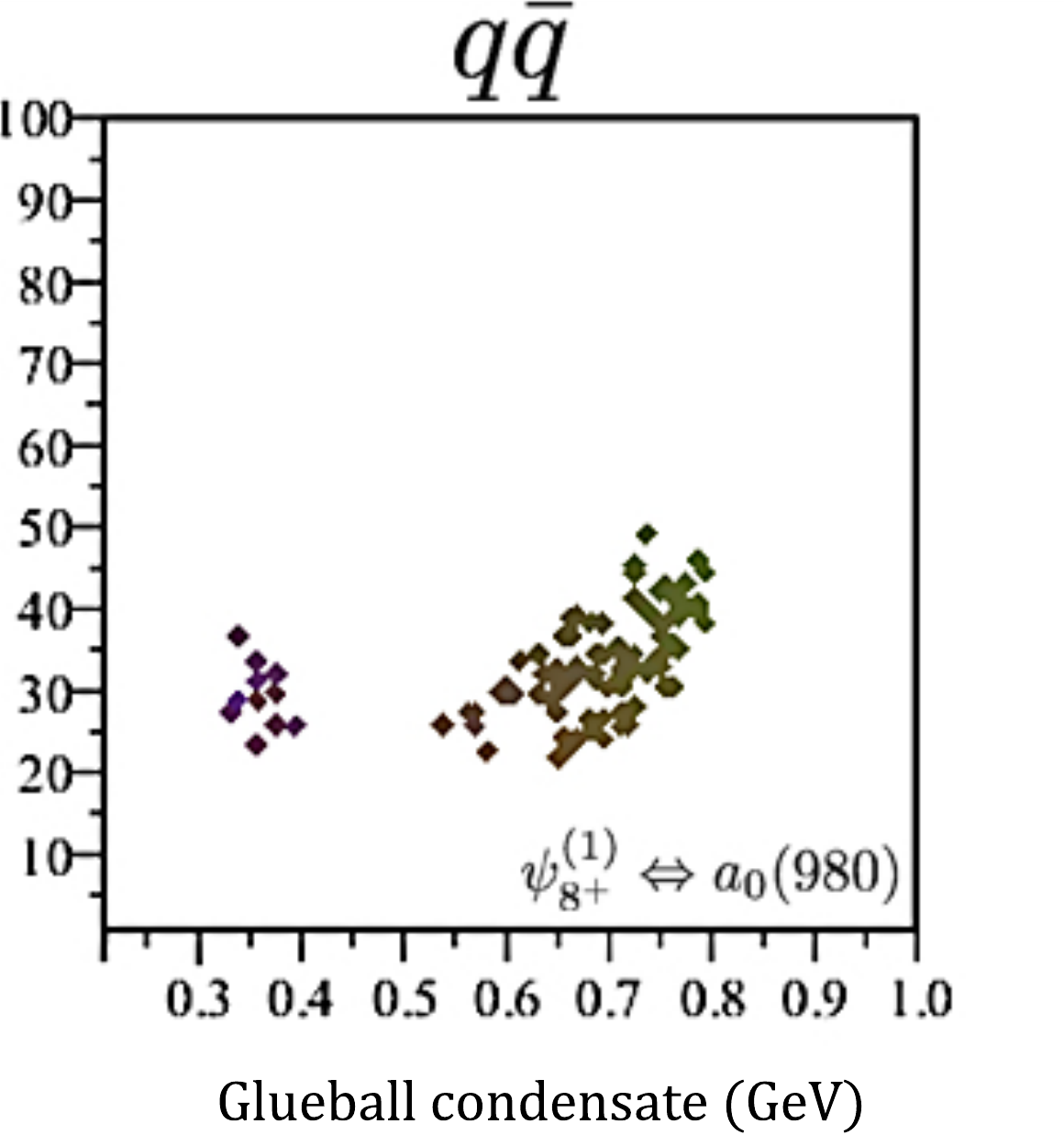}
	
	\caption{Numerical simulation results with $\chi< 10\%$ for the percentage of quark-antiquark component of the lighter pseudoscalar octet [left, identified with $\pi(137)$] and the lighter scalar octet [right, identified with $a_0(980)$],  versus the glueball condensate (horizontal axis,  in GeV).}  
	\label{F_global_8m_8p_qq}
\end{figure}

\begin{figure}[!htb]
	\centering
	\includegraphics[width=0.7\textwidth]{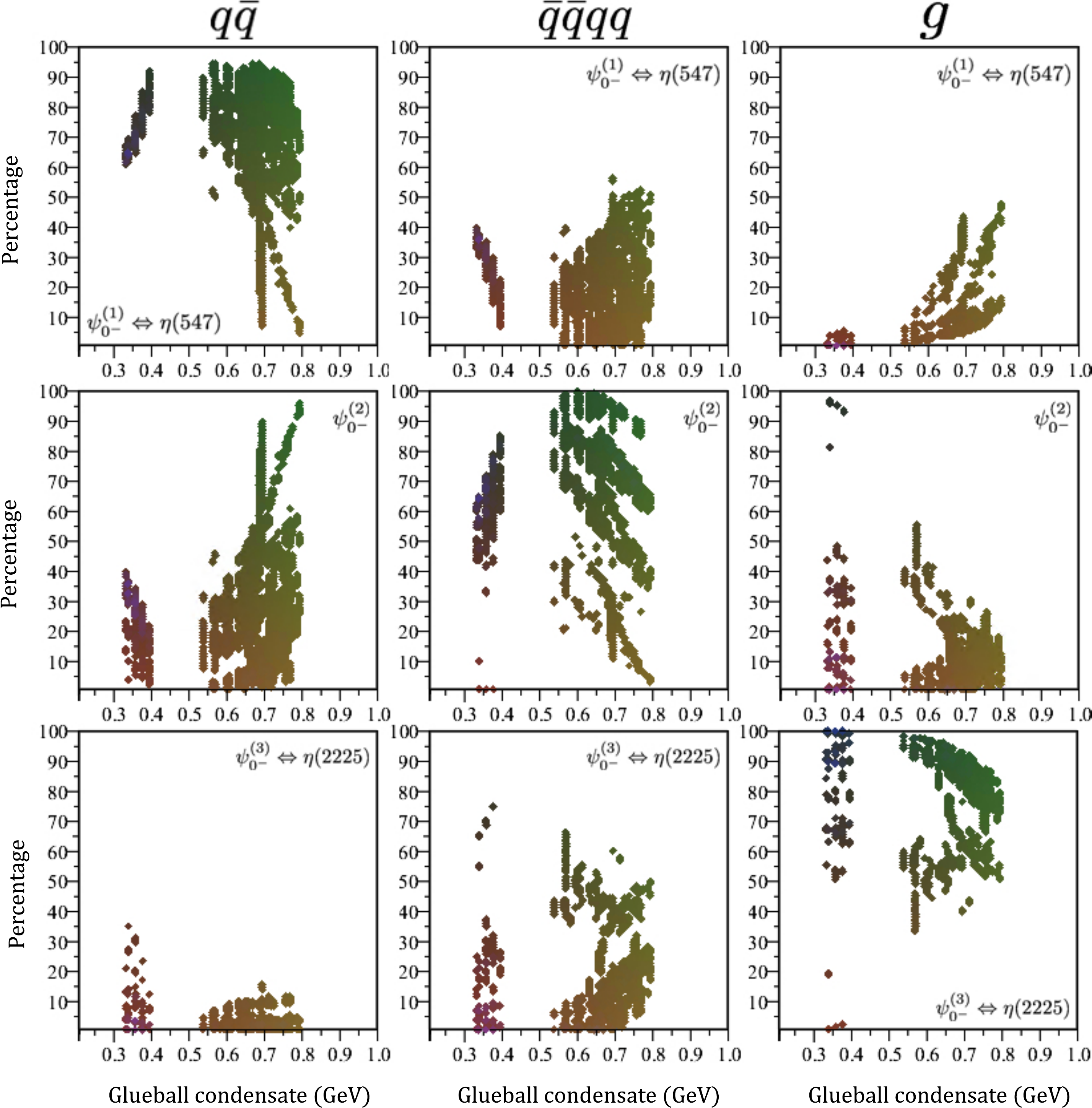}

	\caption{Numerical simulation results with $\chi< 10\%$ for the quark and glue components of the three pseudoscalar SU(3) singlets.  The three rows from top to bottom respectively represent the three SU(3) singlet states in ascending order of mass.  The three columns from left to right respectively represent quark-antiquark, four-quark and glue components.   The vertical axis in each graph represents the component percentage and the horizontal axis represents the  glueball condensate (in GeV).}
	\label{F_global_0m_comps}
\end{figure}

With the constraint of $\chi_{ijk} < 10\%$, we also examine the predictions for quark and glue components of pseudoscalars and scalars.  Fig. \ref{F_global_8m_8p_qq} shows the quark-antiquark component of the lighter pseudoscalar octet (left figure) as well as the same component for the lighter scalar octet (right figure).   These are both consistent with the established expectation that the light pseudoscalar octet (pion) is dominantly a quark-antiquark state while the lighest scalar octet $a_0(980)$ has a large four-quark component.   In this figure we see that the quark-antiquark component of pion is in 75-90\% range while that of $a_0(980)$ is in the range of 20-50\%.   The case of SU(3) singlets are more complex that can be understood within this numerical accuracy.  Figure \ref{F_global_0m_comps} gives the simulation results for the  components of the three pseudoscalar SU(3) singlets in ascending order of mass from the top row (the lightest state) to the bottom row (the heaviest state).  The three columns from left to right respectively represent quark-antiquark, four-quark and glue components.    The numerical accuracy is not sufficient for any quantitative estimates but a few qualitative conclusions can be made.    The glue component of the first two states (the first two rows) is relatively small contrary to the heaviest state that has a large glue component.   Also we can qualitatively infer that the four-quark component of the lightest state and the quark-antiquark component of the middle state are relatively smaller.  

\begin{figure}[!htb]
	\centering
	\includegraphics[width=0.7\textwidth]{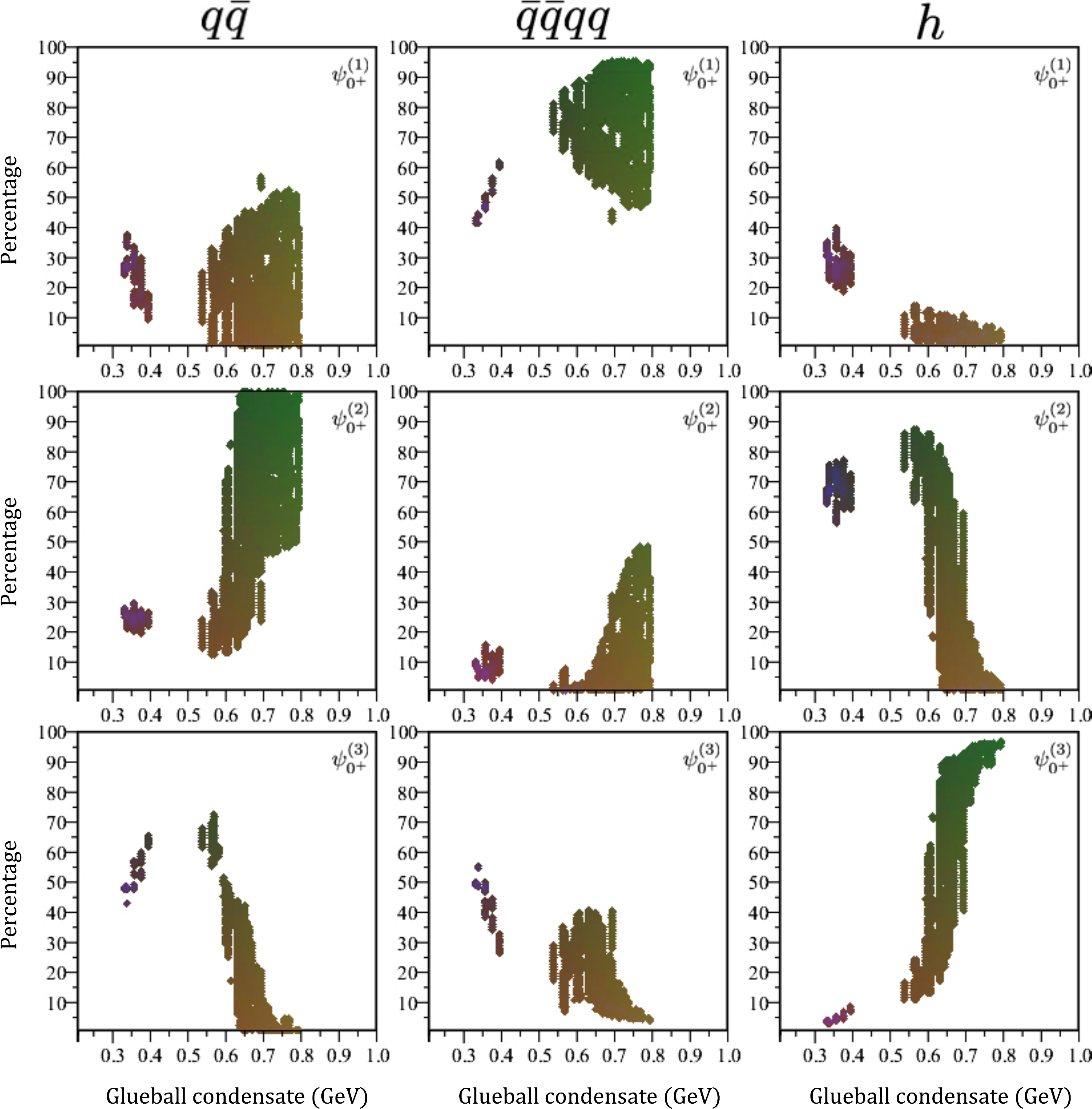}

	\caption{Numerical simulation results with $\chi< 10\%$ for the quark and glue components of the three scalar SU(3) singlets.  The three rows from top to bottom respectively represent the three SU(3) singlet states in ascending order of mass.  The three columns from left to right respectively represent quark-antiquark, four-quark and glue components.   The vertical axis in each graph represents the component percentage and the horizontal axis represents the  glueball condensate (in GeV).}
	\label{F_global_0p_comps}
\end{figure}

The simulations for the scalar SU(3) singlets given in Fig. \ref{F_global_0p_comps} provides only qualitative predictions.    The four-quark   component of the lightest scalar SU(3) singlet is large while its glue component is negligible.   The glue component can be large in the middle and heaviest scalar SU(3) singlet states but at this level of accuracy we cannot determine what the  glue share of each of these two states is.      We need to push the model closer to the experiment by  imposing tighter constraints. Imposing the constraints in stages allows us to correlate the results to various constraints.   For example we see that constraint $\chi_{ijk} < 10\%$ is not sufficient for understanding the components of SU(3) singlets, but the simulations for the components of SU(3) octets are in fact predictive.

We highlight the importance of imposing the constraints incrementally in order to be able to determine the compatibility of the model with experiment and its response to various inputs.
With $\chi_{ijk} \le 10\%$ we conclude that the pseudoscalar SU(3) singlet masses that the model predicts,  naturally approach experimental data even with imposing this minimum constraint.    However, since the masses of scalar SU(3) singlets are not accurately predicted, we need to impose stronger constraints.

\begin{figure}[!htb]
	\centering
	\includegraphics[width=6in]{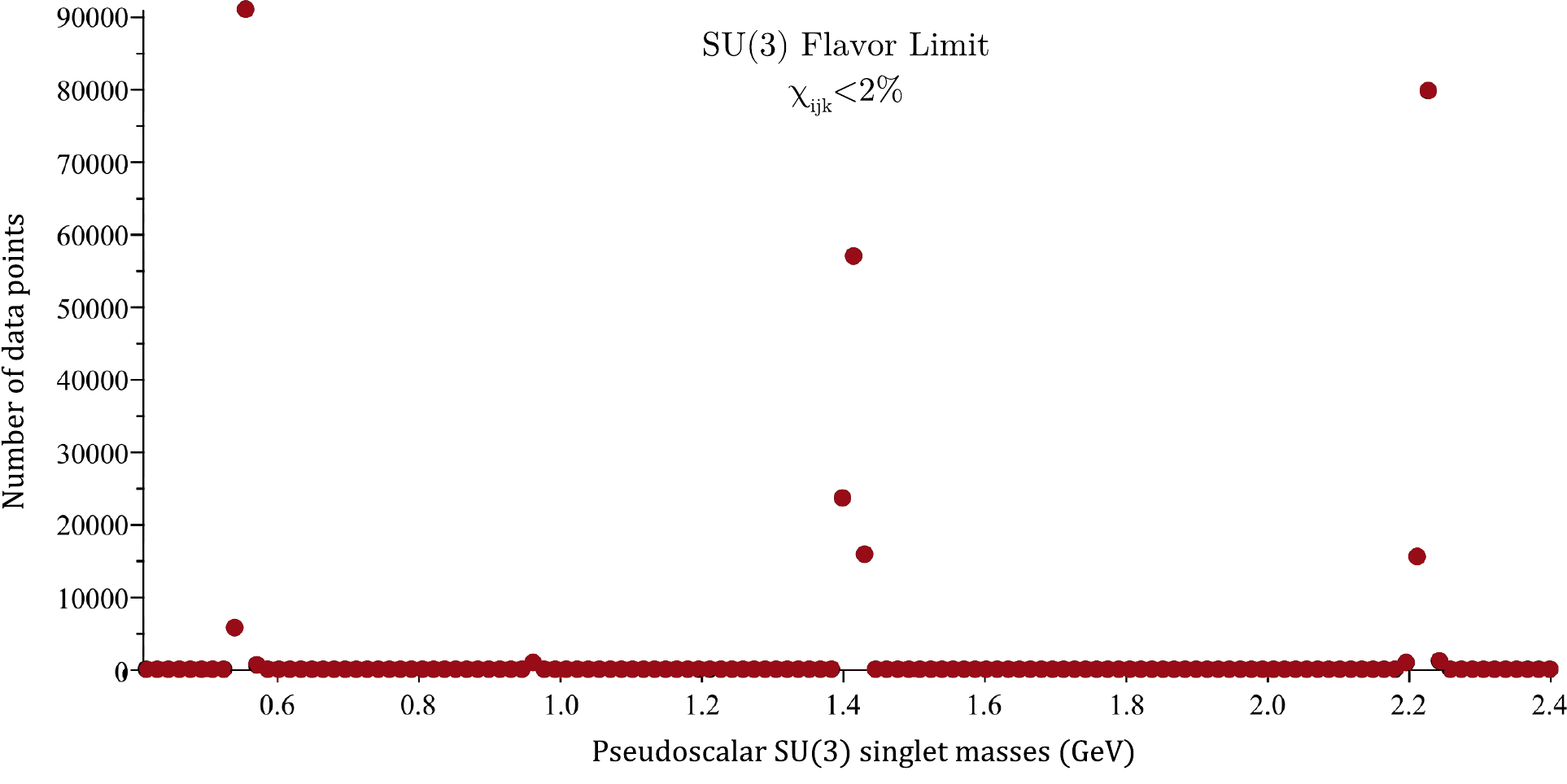}
	\caption{Numerical simulation results  for the pseudoscalar  SU(3) singlet  masses (horizontal axis,  in GeV) with $\chi< 2\%$.   Model predicts three distinct masses:  $\eta(547)$ (with about 90,000 simulation points), followed by one of the eta's around 1.4 GeV and the $\eta(2225)$. The vertical axis gives the number of simulations. }
	\label{F_eta_masses_2p}
\end{figure}

In order to see how the predictions evolve as we get closer to the experimental data for the $\eta$ system, as the next step, we tighten the constraint and impose  $\chi_{ijk} < 2\%$. This results in the SU(3) singlet pseudoscalar masses  in Fig. \ref{F_eta_masses_2p}.   It is evident that the states that survive this tighter constraint  are in fact the three peaks observed in Fig. \ref{F_eta_masses_global}.    Again we  see that $\eta(547)$ is predicted with a very sharp and clear signal, followed by the next mass near 1.5 GeV region which can be an indication of $\eta(1405)$  or $\eta(1475)$; and finally the heaviest state around 2.2 GeV which is a clear signal for $\eta(2225)$.    

\begin{figure}[!htb]
	\centering
	\includegraphics[width=6in]{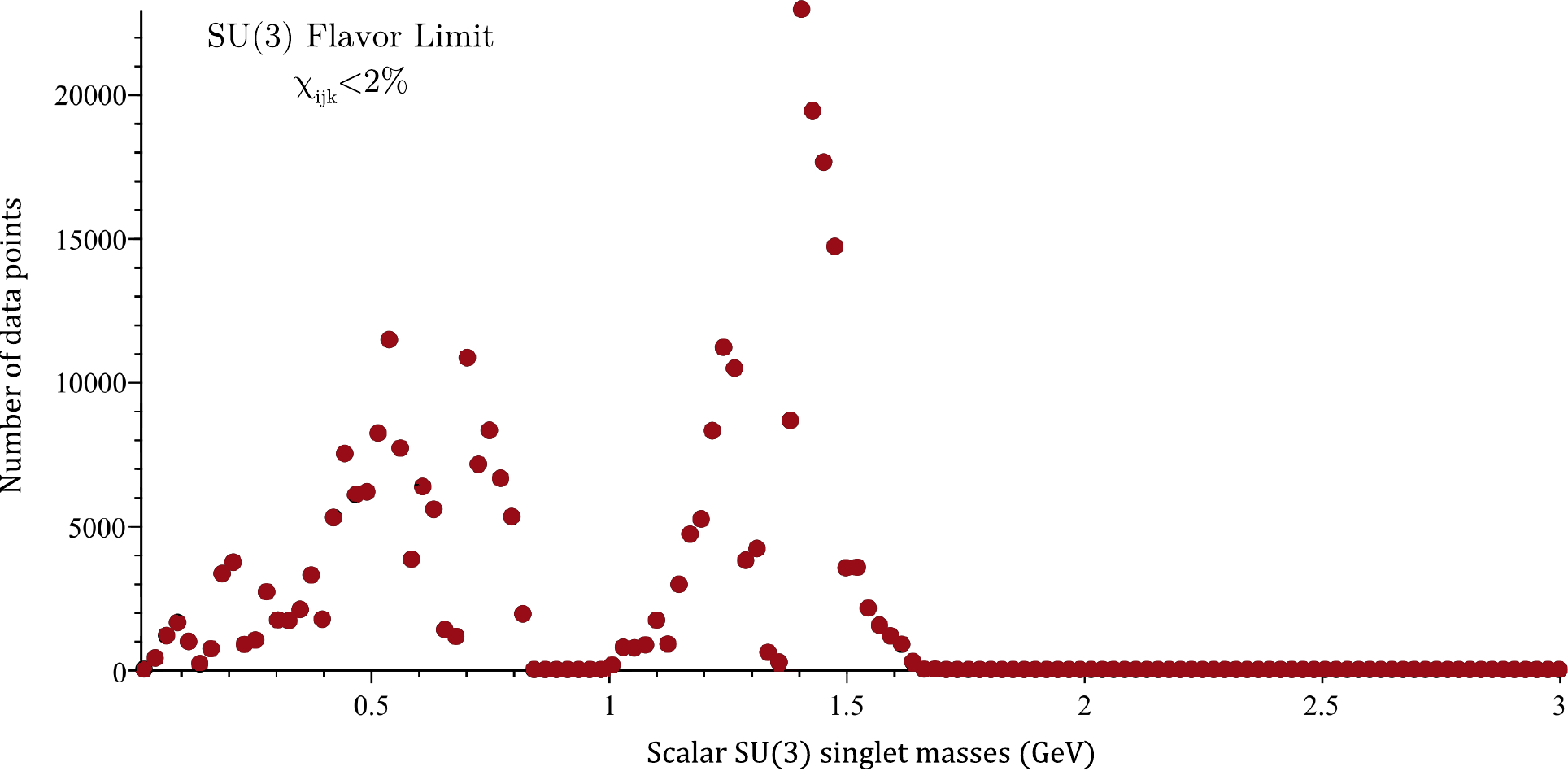}
	\caption{Numerical simulation results with $\chi< 2\%$ for the scalar SU(3) singlet  masses (horizontal axis,  in GeV).   Unlike the pseudoscalar singlets, the signals are not very clean, nevertheless, show two states in the range 1.2-1.5 GeV which can be identified with $f_0(1370)$ and $f_0(1500)$, and a light state below 1 GeV in the range of 0.4-0.8 GeV which is an indication of the sigma meson. The vertical axis gives the number of simulations.}
	
	\label{F_f0_masses_2p}
\end{figure}

The case of scalar singlets continues to remain less certain in this case as well,  as can be seen in Fig. \ref{F_f0_masses_2p} in which  simulations predict states with masses in the ranges of approximately less that 
0.8 GeV and 1.2 to 1.5 GeV, highlighting again the well-known complexity of understanding light scalar mesons.     The quark and glue components with $\chi_{ijk} < 2\%$ get more refined compared to  $\chi_{ijk} < 10\%$ and we skip them here and study these components when we impose our next constraint.


\begin{figure}[!htb]
	\centering
	\includegraphics[width=2in]{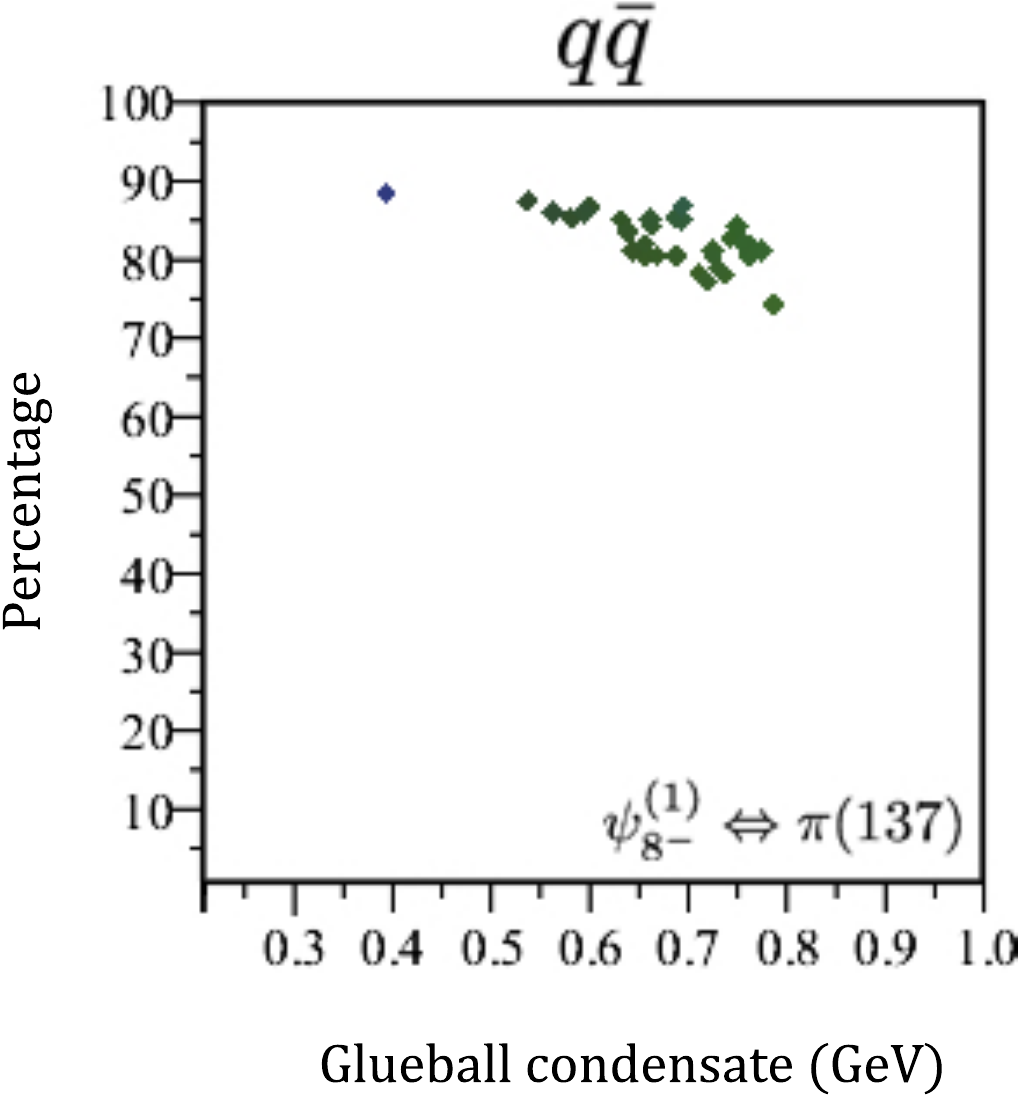}
	\includegraphics[width=2in]{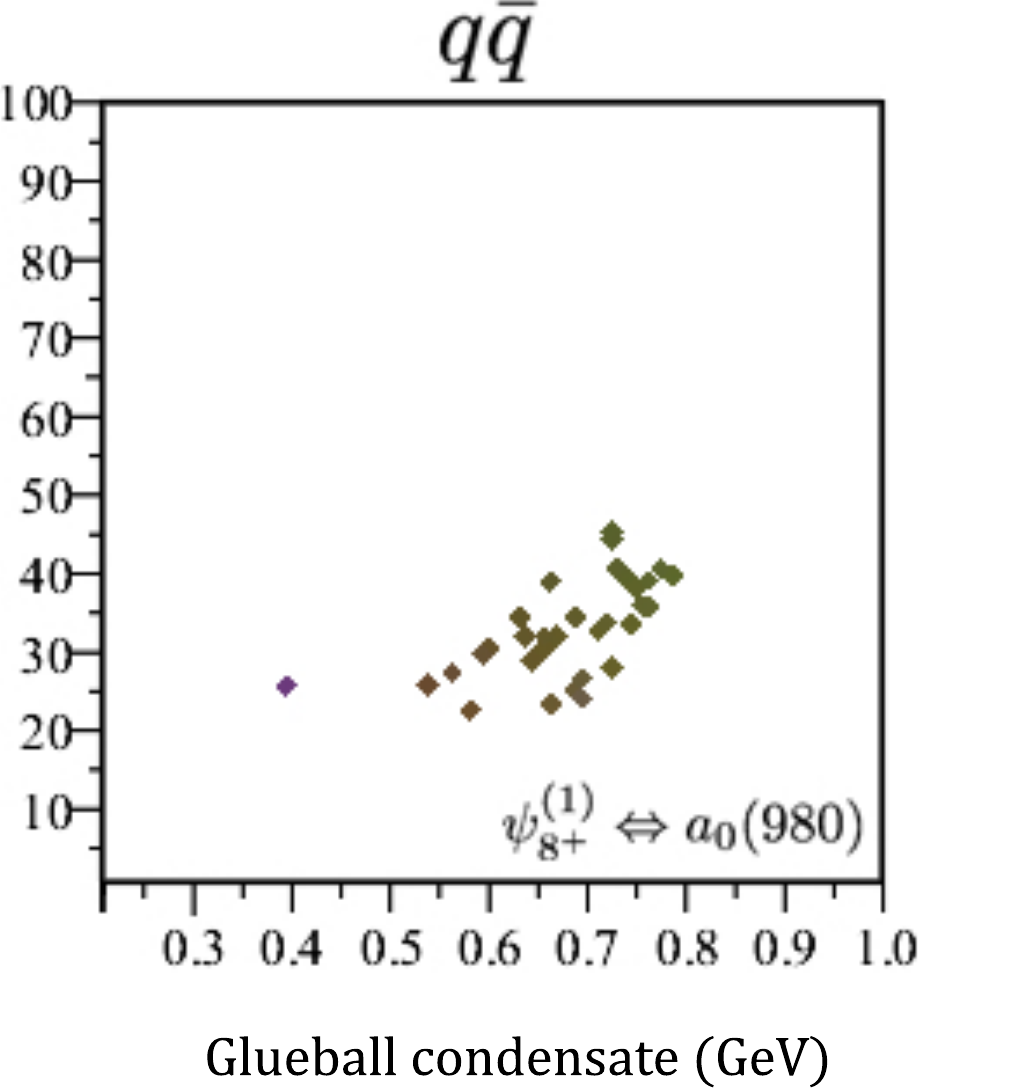}
	
	\caption{Numerical simulation results with $\chi< \chi^{\rm exp}$ for the percentage of quark-antiquark component of the lighter pseudoscalar octet [left, identified with $\pi(137)$] and the lighter scalar octet [right, identified with $a_0(980)$],  versus the glueball condensate (horizontal axis,  in GeV).
}
	\label{F_global_8m_8p_qq_chi_exp}
\end{figure}

\begin{figure}[!htb]
	\centering
	\includegraphics[width=0.7\textwidth]{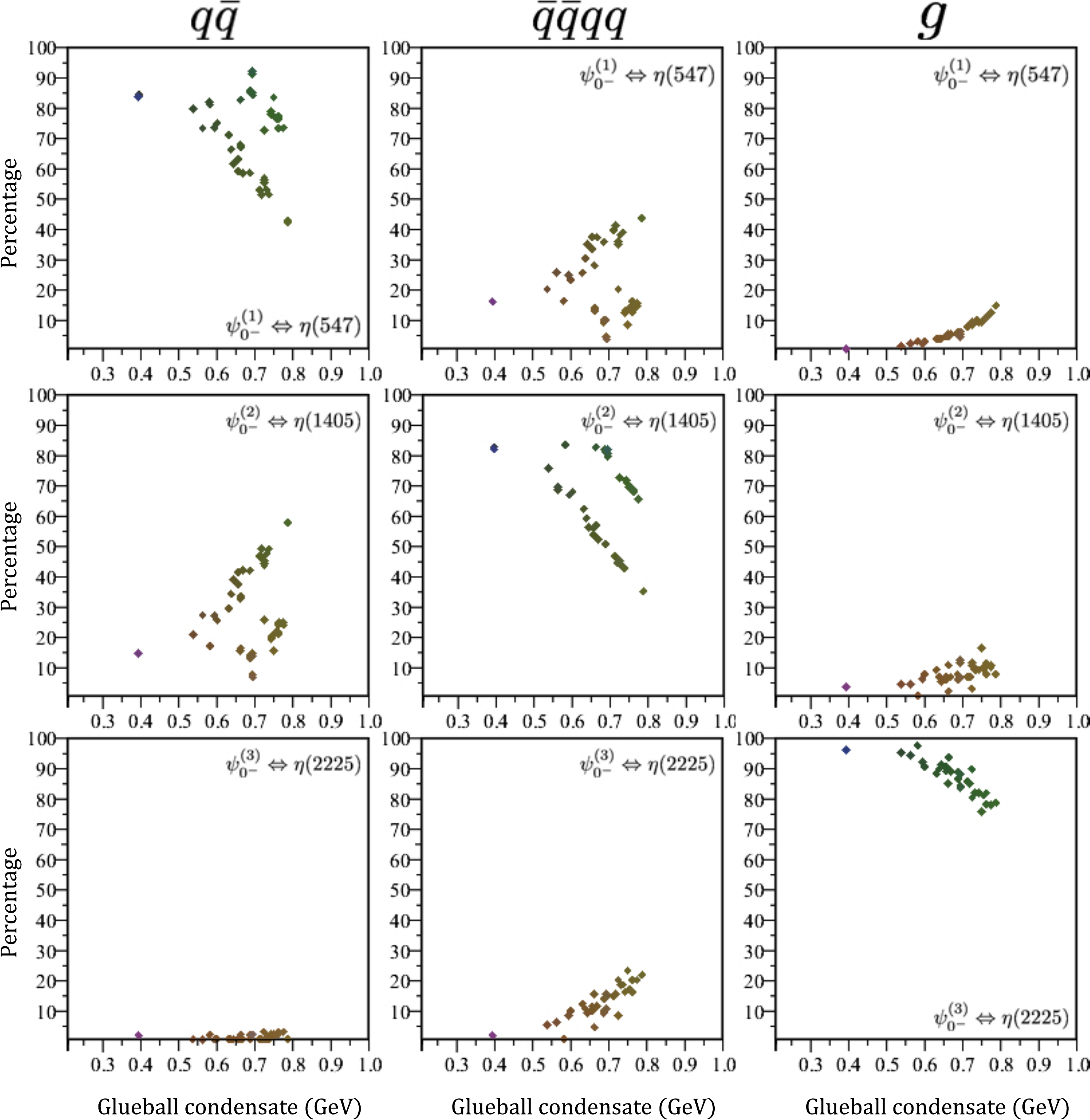}
	\caption{Numerical simulation results with $\chi< \chi^{\rm exp}$ for the quark and glue components of the three pseudoscalar SU(3) singlets.  The three rows from top to bottom respectively represent the three SU(3) singlet states in ascending order of mass.  The three columns from left to right respectively represent quark-antiquark, four-quark and glue components.   The vertical axis in each graph represents the component percentage and the horizontal axis represents the  glueball condensate (in GeV).}
	\label{F_global_0m_comps_chi_exp}
\end{figure}

Next, we impose the most stringent constraint  for identifying the three pseudoscalar SU(3) singlets predicted by the model with three of the seven experimental eta's from the list (\ref{E_eta_exp}).    This leads to 35 scenarios for which we examine whether $\chi_{ijk} \le \chi^{\rm exp}_{ijk}$.   We find that only scenario $ijk=147$ is compatible with our model.  Fig. \ref{F_global_8m_8p_qq_chi_exp} shows that the situation for  octet states does not change much when this tighter constraint is imposed.   This is of course understandable as this constraint is on the pseudoscalar SU(3) singlets.  On the other hand,  Fig. \ref{F_global_0m_comps_chi_exp} is more predictive of the quark and glue components of pseudoscalar SU(3) singlets compared to Fig. \ref{F_global_0m_comps}.    It can be seen that the lightest state (top row in Fig.  \ref{F_global_0m_comps_chi_exp} which is identified with $\eta(475)$) has a large quark-antiquark state, whereas the second state (middle row in  Fig.  \ref{F_global_0m_comps_chi_exp} which is identified with  $\eta(1405)$)  has a significant four-quark component.   The heaviest pseudoscalar SU(3) singlet state (the bottom row which is identified with $\eta(2225)$) is dominantly made of  glue.

\begin{figure}[!htb]
	\centering
	\includegraphics[width=0.7\textwidth]{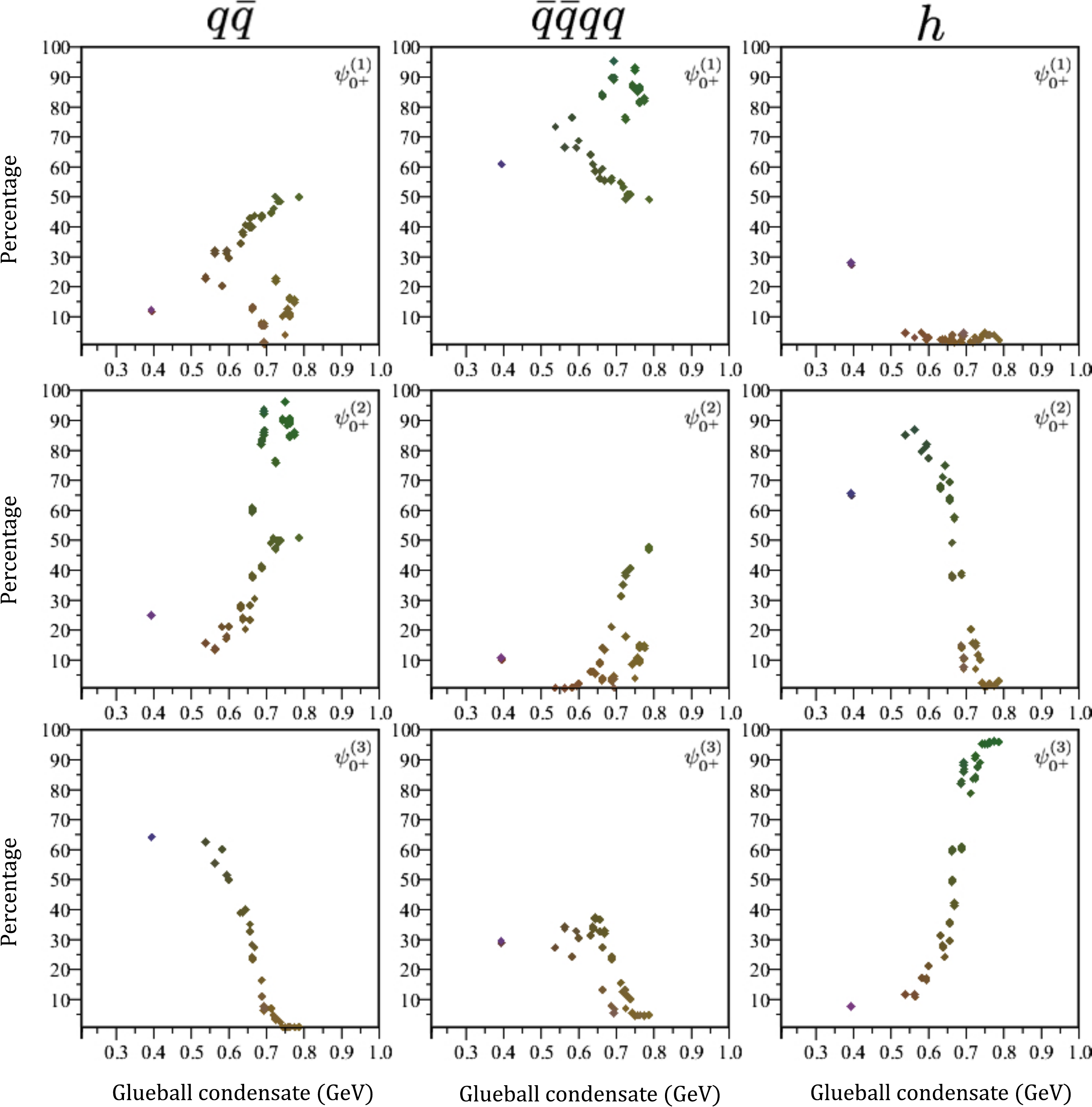}
	\caption{Numerical simulation results with $\chi< \chi^{\rm exp}$ for the quark and glue components of the three scalar SU(3) singlets.  The three rows from top to bottom respectively represent the three SU(3) singlet states in ascending order of mass.  The three columns from left to right respectively represent quark-antiquark, four-quark and glue components.   The vertical axis in each graph represents the component percentage and the horizontal axis represents the  glueball condensate (in GeV).}
	\label{F_global_0p_comps_chi_exp}
\end{figure}

The case of scalar SU(3) signlets shown in Fig. \ref{F_global_0p_comps_chi_exp} is practically  the same as
what we saw in Fig. \ref{F_global_0p_comps}.   To improve the predictions for scalar SU(3) singlets we need to incorporate specific experimental data on these states and that is what we will do in the next subsection.

\clearpage 

\subsection{Constraints from scalar meson decays}\label{ss_scalar_const}

Up to this point  we have incorporated the inputs (\ref{inputs}) that include the scalar and pseudoscalar octet masses, as well as constraints for the pseudoscalar SU(3) singlet masses according to the condition $\chi_{ijk} \le \chi_{ijk}^{\rm exp}$ discussed in previous subsection.   To further narrow down the  parameter space of the model, in this subsection we include experimental constraints for scalar SU(3) singlets.  Out of the five scalar mesons in this model (two SU(3) octets and three SU(3) singlets),  we have identified the two SU(3) octet scalars with the $a_0(980)$ and $a_0(1450)$ from the beginning [see Eq. (\ref{inputs})].   Then the   simulations of Fig. \ref{F_f0_masses_global} showed that the masses of the first and the second SU(3) singlets in this model are quite consistent with  the characteristics of $f_0(500)$ and $f_0(1370)$ and from this point on we make this identification.      The mass of the third SU(3) singlet is closer to $f_0(1500)$ but we cannot rule out its identification with  $f_0(1710)$  either, therefore,  we need more experimental data on scalars to constrain our simulations and be able to examine the mass spectrum as well as the quark and glue components.  With the identification
\begin{eqnarray}
\psi_{8^+}^{(1)} &\Leftrightarrow&  a_0(980) \nonumber \\
\psi_{8^+}^{(2)} &\Leftrightarrow&  a_0(1450) \nonumber \\
\psi_{0^+}^{(1)} &\Leftrightarrow&  f_0(500) \nonumber \\
\psi_{0^+}^{(2)} &\Leftrightarrow&  f_0(1370) \nonumber \\
\psi_{0^+}^{(3)} &\Leftrightarrow&  
f_0(1500) 
\hskip 0.25cm 
{\rm or}
\hskip 0.25cm  
f_0(1710)
\end{eqnarray}
we incorporate several experimental data on the partial decay widths of these scalar mesons and filter out simulations that do not support the above assignment.   However,  there are limited number of experimental data available on scalar systems and some of these have large uncertainties.   This means that we need to make additional assumptions about certain aspects of the data.  We incorporate the following experimental data from PDG \cite{PDG}:
\begin{eqnarray}
\Gamma^{\rm Total} [a_0(980)] &=& 50-100 \,\,\,  {\rm MeV}\nonumber \\
\Gamma^{\rm Total} [a_0(1450)] &=& 265 \pm 13 \,\,\,  {\rm MeV} \nonumber\\
m[f_0(500)] &=& 400-550 \,\,\, {\rm MeV} \nonumber \\
\Gamma^{\rm Total} [f_0(500)] &=& 400-700 \,\,\, {\rm  MeV}  \nonumber \\
m[f_0(1370)] &=& 1200-1500 \,\,\, {\rm MeV} \nonumber \\
\Gamma^{\rm Total} [f_0(1370)] &=& 200-500 \,\,\, {\rm MeV} \nonumber \\
m[f_0(1500)] &=& 1506 \pm 6 \,\,\, {\rm MeV} \nonumber \\
\Gamma^{\rm Total} [f_0(1500)] &=& 112\pm 9  \,\,\, {\rm MeV} \nonumber \\
m[f_0(1710)] &=& 1704 \pm 12 \,\,\, {\rm MeV} \nonumber \\
\Gamma^{\rm Total} [f_0(1710)] &=& 123 \pm 18 \,\,\, {\rm MeV} 
\label{PDG_scalars}
\end{eqnarray}
We define our target masses to be the central values of these experimental data.   To make use of the  total decay widths, we assume: (a) $a_0(980)$ and $a_0(1450)$ dominantly decay into $\pi\eta$ (b) $f_0(500)$ dominantly decays into $\pi\pi$; (c)The main decay channels of $f_0(1370)$, $f_0(1500)$ and $f_0(1710)$ are $\pi\pi$ and $\eta\eta$.   We consider two scenarios, one where the heaviest scalar SU(3) singlet state is identified with $f_0(1500)$, and another one where this heaviest SU(3) singlet is identified with $f_0(1710)$.  
Moreover, we also take into account several decay ratios provided  by WA102 experiment \cite{WA102_data}:
\begin{eqnarray}
{ {\Gamma [f_0(1370) \rightarrow \pi\pi]}\over {\Gamma^ [f_0(1370) \rightarrow KK]} } &=& 2.17 \pm 0.9 \nonumber \\ 
{ {\Gamma [f_0(1370) \rightarrow \eta\eta]}\over {\Gamma^ [f_0(1370) \rightarrow KK]} } &=& 0.35 \pm 0.30 \nonumber \\ 
{ {\Gamma [f_0(1500) \rightarrow \pi\pi]}\over {\Gamma^ [f_0(1500) \rightarrow \eta\eta]} } &=& 5.56 \pm 0.93 \nonumber \\ 
{ {\Gamma [f_0(1710) \rightarrow \pi\pi]}\over {\Gamma^ [f_0(1710) \rightarrow KK]} } &=&  0.20 \pm 0.03 \nonumber \\ 
{ {\Gamma [f_0(1710) \rightarrow \eta\eta]}\over {\Gamma^ [f_0(1710) \rightarrow KK]} } &=& 0.48 \pm 0.19 \nonumber \\
\label{WA102_data}
\end{eqnarray}
Note that for $f_0(1370)$ and $f_0(1710)$ the decay ratios to $\pi\pi$ over $\eta \eta$ are not directly given in (\ref{WA102_data}), therefore we  estimate these ratios using:
\begin{equation}
{\Gamma[f_0\rightarrow \pi\pi] \over {\Gamma[f_0\rightarrow \eta\eta]}} =
{{\Gamma[f_0\rightarrow \pi\pi] \over {\Gamma[f_0\rightarrow KK]}}\over 
	{{\Gamma[f_0\rightarrow \eta\eta] \over {\Gamma[f_0\rightarrow KK]}} } }
\end{equation}
which  gives:
\begin{eqnarray}
{\Gamma[f_0(1370)\rightarrow \pi\pi] \over {\Gamma[f_0(1370)\rightarrow \eta\eta]}} &=& 6.2^{+7.9}_{-6.2}\label{pp_ee_1370} \\
{\Gamma[f_0(1710)\rightarrow \pi\pi] \over {\Gamma[f_0(1710)\rightarrow \eta\eta]}} &=& 0.42 \pm 0.23
\label{pp_ee_1710}
\end{eqnarray}

To guide our numerical simulations toward the experimental data in (\ref{PDG_scalars}), (\ref{WA102_data}), (\ref{pp_ee_1370}) and (\ref{pp_ee_1710}) we define a chi function for the scalar states (for each of the two scenarios where the heaviest SU(3) singlet is identified with either $f_0(1500)$ or $f_0(1710)$):

\begin{eqnarray}
\chi_s(i) &=& 
{
	{\Big|m_{0^+} - {\widehat m}[f_0(500)] \Big|} 
	\over 
	{{\widehat m}[f_0(500)]}
} 
+
{
	{\Big|m'_{0^+} - {\widehat m}[f_0(1370)] \Big|} 
	\over 
	{{\widehat m}[f_0(1370)]}
}
+
{
	{\Big|m''_{0^+} - {\widehat m}[f_0(i)] \Big|} 
	\over 
	{{\widehat m}[f_0(i)]}
}\nonumber \\
&&+
{
	{\Big|\Gamma^1_{8\rightarrow 80} - {\widehat \Gamma[a_0(980)\rightarrow \pi \eta]}\Big|} 
	\over 
	{\widehat \Gamma[a_0(980)\rightarrow \pi\eta]}
}
+
{
	{\Big|\Gamma^2_{8\rightarrow 80} - {\widehat \Gamma[a_0(1450)\rightarrow \pi \eta]}\Big|} 
	\over 
	{\widehat \Gamma[a_0(1450)\rightarrow \pi\eta]}
}
+
{
	{\Big|\Gamma^1_{0\rightarrow 88} - {\widehat \Gamma[f_0(500)\rightarrow \pi \pi]}\Big|} 
	\over 
	{\widehat \Gamma[f_0(500)\rightarrow \pi\pi]}
} \nonumber \\
&&+
{
	{\Big|\Gamma^2_{0\rightarrow 88} + \Gamma^2_{0\rightarrow 00}- {\widehat \Gamma[f^{\rm Total}_0(1370)]}\Big|} 
	\over 
	{\widehat \Gamma[f_0^{\rm Total}(1370)]}
} +
{
	{\Big|\Gamma^3_{0\rightarrow 88}+\Gamma^3_{0\rightarrow 00} - {\widehat \Gamma^{\rm Total}[f_0(i)]}\Big|} 
	\over 
	{\widehat \Gamma[f_0^{\rm Total}(i)]}
} \nonumber \\
&& +
{
	{\Big|{\Gamma^2_{0\rightarrow 88}\over {\Gamma^2_{0\rightarrow 00}} } - 
	{ {{\Gamma[f_0(1370)\rightarrow \pi \pi]}\over 
			{\Gamma[f_0(1370)\rightarrow \eta \eta]}}}\Big|} 
	\over 
	{{\Gamma[f_0(1370)\rightarrow \pi \pi]}\over 
		{\Gamma[f_0(1370)\rightarrow \eta \eta]}}
}
 +
{
	{\Big|{\Gamma^3_{0\rightarrow 88}\over {\Gamma^3_{0\rightarrow 00} }} - 
		{ {{\Gamma[f_0(i)\rightarrow \pi \pi]}\over 
				{\Gamma[f_0(i)\rightarrow \eta \eta]}}}\Big|} 
	\over 
	{{\Gamma[f_0(i)\rightarrow \pi \pi]}\over 
		{\Gamma[f_0(i)\rightarrow \eta \eta]}}
}, \hskip 2cm i=1500 \hskip .15cm {\rm or} \hskip 0.15cm 1710
\label{chi_s}
\end{eqnarray}
where $m_{0^+}$, $m'_{0^+}$ and $m''_{0^+}$ are the computed scalar SU(3) singlet masses in this model defined in (\ref{m0+_def}),   and the hat represents central values of the experimental quantities.   The short notation used for the computed partial decay widths are defined in Table \ref{T_DW_short}.  Calculation of the coupling constant and the  partial decay widths are given in  appendix \ref{A_formulas}.  We compare the computed chi in (\ref{chi_s}) with the overall percent experimental uncertainty that we compute from:

\begin{table}[htbp]
			\caption{Short notations for partial decay widths of scalar states.}
	\begin{tabular}{ c | c}
		Short notation & Corresponding decay width\\
		& in SU(3) notation		   \\
		\hline \\
		$\Gamma^1_{8\rightarrow 80}$ & $\Gamma\left[\psi^{(1)}_{8^+}\rightarrow \psi^{(1)}_{8^-}  \psi^{(1)}_{0^-}\right] $  \\
		&  \\
		$\Gamma^2_{8\rightarrow 80}$ & $\Gamma\left[\psi^{(2)}_{8^+}\rightarrow \psi^{(1)}_{8^-}  \psi^{(1)}_{0^-}\right] $  \\
		&  \\
		$\Gamma^1_{0\rightarrow 88}$  & $\Gamma\left[\psi^{(1)}_{0^+}\rightarrow \psi^{(1)}_{8^-}  \psi^{(1)}_{8^-}\right] $   \\
		&  \\
		$\Gamma^2_{0\rightarrow 88}$ & $\Gamma\left[\psi^{(2)}_{0^+}\rightarrow \psi^{(1)}_{8^-}  \psi^{(1)}_{8^-}\right] $   \\
		&  \\
		$\Gamma^3_{0\rightarrow 88}$  & $\Gamma\left[\psi^{(3)}_{0^+}\rightarrow \psi^{(1)}_{8^-}  \psi^{(1)}_{8^-}\right] $  \\
		&  \\
		$\Gamma^2_{0\rightarrow 00}$  & $\Gamma\left[\psi^{(2)}_{0^+}\rightarrow \psi^{(1)}_{0^-}  \psi^{(1)}_{0^-}\right] $   \\
		&  \\
		$\Gamma^3_{0\rightarrow 00}$ & $\Gamma\left[\psi^{(3)}_{0^+}\rightarrow \psi^{(1)}_{0^-}  \psi^{(1)}_{0^-}\right] $  \\
	\end{tabular}
	\label{T_DW_short}
\end{table}

\begin{eqnarray}
\chi_s^{\rm exp.}(i) &=& 
{
	 {\Delta m[f_0(500)] } 
	\over 
	{{\widehat m}[f_0(500)]}
} 
+
{
	{\Delta m[f_0(1370)] } 
	\over 
	{{\widehat m}[f_0(1370)]}
}
+
{
	{\Delta m[f_0(i)] } 
	\over 
	{{\widehat m}[f_0(i)]}
}
+
{
	{\Delta \Gamma[a_0(980)\rightarrow \pi \eta]} 
	\over 
	{\widehat \Gamma[a_0(980)\rightarrow \pi\eta]}
}
\nonumber \\
&&
+
{
	{\Delta \Gamma[a_0(1450)\rightarrow \pi \eta]} 
	\over 
	{\widehat \Gamma[a_0(1450)\rightarrow \pi\eta]}
}
+
{
	{\Delta\Gamma[f_0(500)\rightarrow \pi \pi]} 
	\over 
	{\widehat \Gamma[f_0(500)\rightarrow \pi\pi]}
} 
+
{
	{\Delta\Gamma[f^{\rm Total}_0(1370)]} 
	\over 
	{\widehat \Gamma[f_0^{\rm Total}(1370)]}
} +
{
	{\Delta\Gamma^{\rm Total}[f_0(i)]} 
	\over 
	{\widehat \Gamma[f_0^{\rm Total}(i)]}
} \nonumber \\
&& +
{
	{ \Delta { {{\Gamma[f_0(1370)\rightarrow \pi \pi]}\over 
				{\Gamma[f_0(1370)\rightarrow \eta \eta]}}} } 
	\over 
	{{\Gamma[f_0(1370)\rightarrow \pi \pi]}\over 
		{\Gamma[f_0(1370)\rightarrow \eta \eta]}}
}
+
{
	{\Delta		{ {{\Gamma[f_0(i)\rightarrow \pi \pi]}\over 
				{\Gamma[f_0(i)\rightarrow \eta \eta]}}} } 
	\over 
	{{\Gamma[f_0(i)\rightarrow \pi \pi]}\over 
		{\Gamma[f_0(i)\rightarrow \eta \eta]}}
}, \hskip 2cm i=1500 \hskip .15cm {\rm or} \hskip 0.15cm 1710
\end{eqnarray}

\begin{figure}[!htb]
	\centering
	\includegraphics[width=6in]{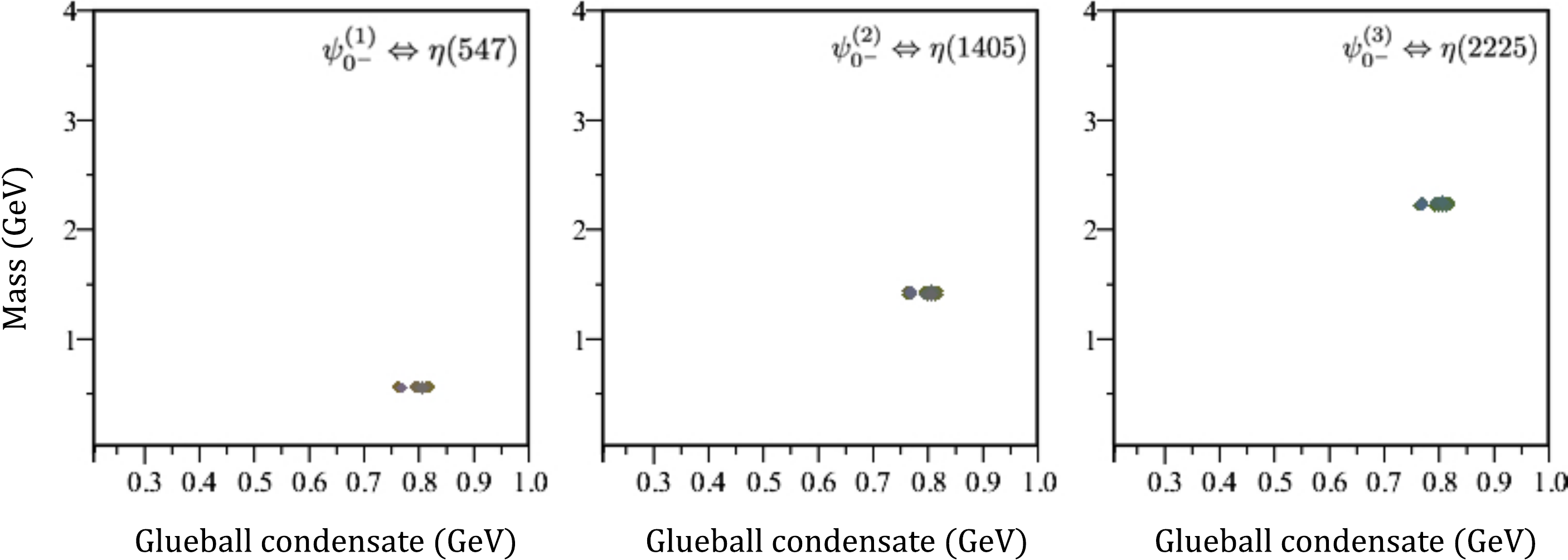}
	\includegraphics[width=6in]{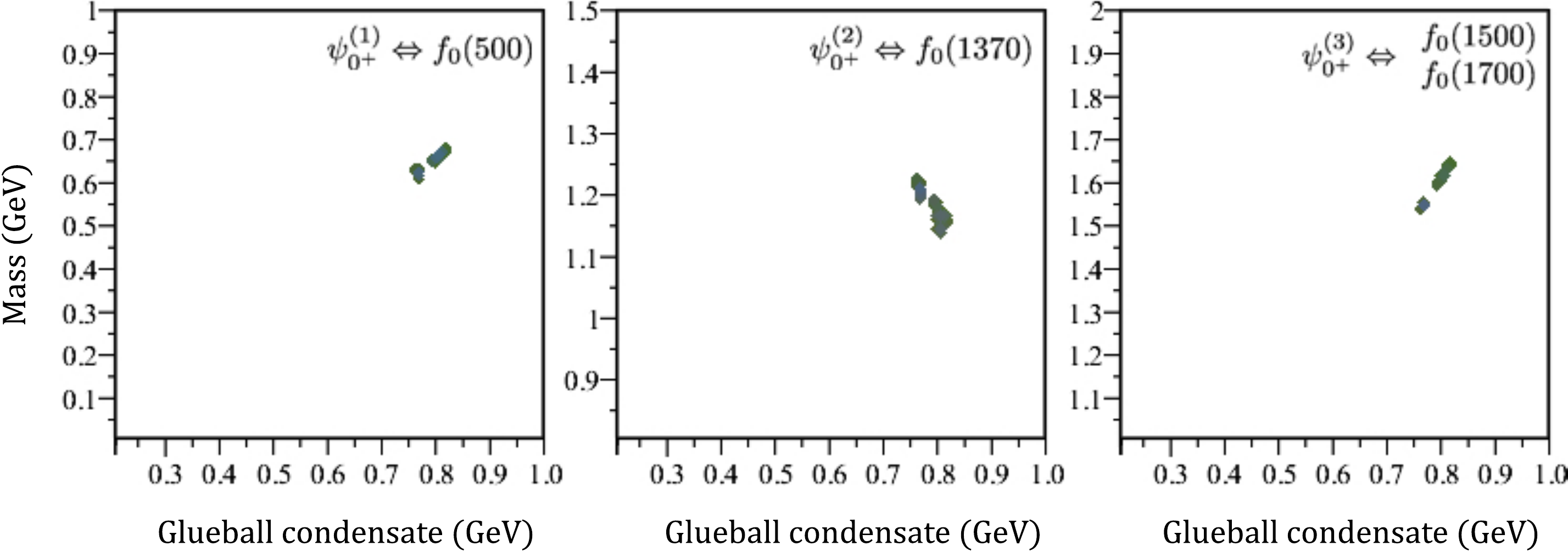}
	\caption{Numerical simulation results with $\chi_s(i) < 1.1 \chi_s^{\rm min}$ for the masses (in GeV) of the three pseudoscalar singlets (top) and the three scalar singlets (bottom),  versus the glueball condensate (in GeV).}
	\label{F_m0m_m0p_after_chi_s}
\end{figure}

In a unique situation where there is an exact agreement between the model predictions and the central values of \underline{\it all} experimental data that we have considered, we have $\chi_s(i) = 0$, but this is neither typically achievable nor physically that significant because of the existence of experimental uncertainties and the fact that the central values of the experimental data are often not any more important than any other values within the given experimental ranges.  Therefore, while we target the central values of the experimental data, we do not overemphasize their importance and allow numerical deviations to reflect the overall experimental uncertainties.  A more realistic approach is  to allow the model predictions to vary in such a way that the overall deviations from the central values are comparable to the overall relative experimental uncertainties.   This means that, $\chi_s(i) \le \chi_s^{\rm exp}(i)$,  instead of $\chi_s(i)=0$, gives a reasonable criterion for guiding our numerical search.  If the model is close to the exact limit, we expect to be able to impose $\chi_s(i) \le \chi_s^{\rm exp}(i)$ which, if upheld,  would show an overall agreement between model simulations and all experimental bounds.  However, in the present SU(3) approximation, this condition is  rather stringent and expectedly not satisfied when all experimental data are simultaneously targeted.   
We therefore study the minimum of $\chi_s(i)$ and only keep simulations that are within 10\% of the minimum, i.e. we filter out simulations that result in 
\begin{equation}
\chi_s(i) > 1.1 \, \chi_s^{\rm min}(i)
\end{equation}    
After this filter is imposed,  we find that the favored glueball condensate is around 0.8 GeV consistent with our previous investigation \cite{19_FJL}.   We also find that the predicted masses for the scalar and pseudoscalar  SU(3) singlets are quite close to some of the experimental candidates given in PDG \cite{PDG}.   The results are given in Fig. \ref{F_m0m_m0p_after_chi_s}, and show that, consistent with what we saw in previous subsection, the mass of the three pseudoscalar SU(3) singlet states are respectively very close to the mass of $\eta(547)$; one of the eta's in the 1.5 GeV range [$\eta(1405)$ or $\eta(1475)$];  and $\eta(2225)$.  In the same figure we also see that the mass of the three scalar SU(3) singlets are respectively consistent with  the mass of $f_0(500)$; $f_0(1370)$; and one of the two states $f_0(1500)$ or $f_0(1710)$.   

\begin{figure}[!htb]
	\centering
	\includegraphics[width=4in]{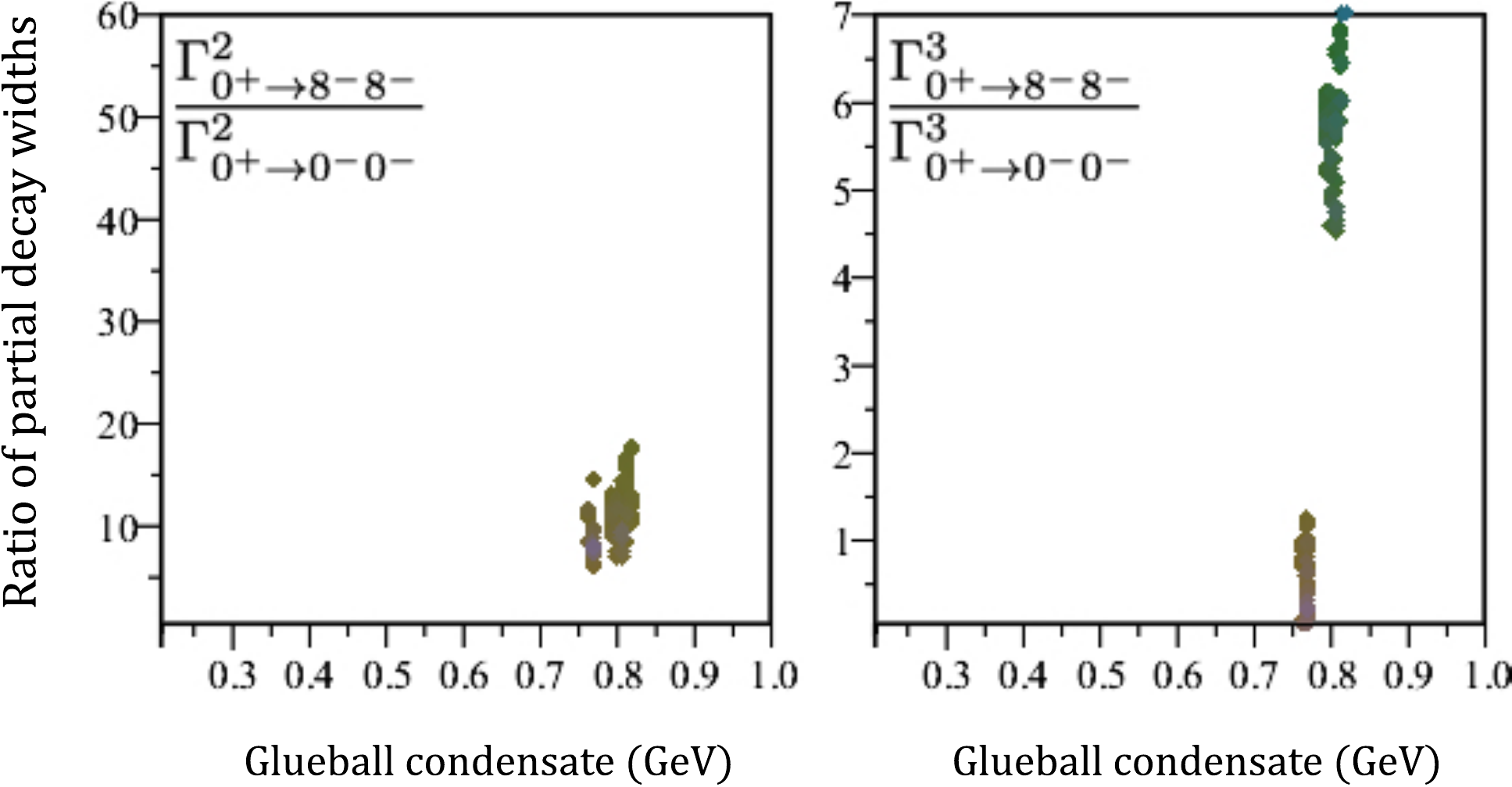}
	
	\caption{Numerical simulation results with $\chi_s(i) < 1.1 \chi_s^{\rm min}$ for the ratio of the partial decay widths to $\pi \pi$  over $\eta \eta$ of the scalar SU(3) singlets versus the glueball condensate (in GeV).  The left figure gives this ratio for the second SU(3) singlet scalar (i.e.,  ${\Gamma^2_{0^+\rightarrow 8^-8^-}\over \Gamma^2_{0^+\rightarrow 0^-0^-}}$) and the  right figure gives the same ratio for the heaviest  SU(3) singlet scalar (i.e.,  ${\Gamma^3_{0^+\rightarrow 8^-8^-}\over \Gamma^3_{0^+\rightarrow 0^-0^-}}$). (See Table \ref{T_DW_short} for the short notations.)   In this work, the second SU(3) singlet scalar is identified with $f_0(1370)$ and the heaviest SU(3) scalar can be identified with either $f_0(1500)$ or $f_0(1710)$.   The left plot is consistent with the experimental value for this decay ratio extracted for $f_0(1370)$ in (\ref{pp_ee_1370}).   The lower part of the right plot is consistent with experimental value for this ratio extracted for $f_0(1710)$ in (\ref{pp_ee_1710}), and the upper part of the same plot is consistent with the experimental value for this ratio given by WA102 collaboration for $f_0(1500)$ in (\ref{WA102_data}).   The right plot highlights the fact that our model clearly identifies $f_0(1500)$ and $f_0(1710)$ as possible candidates for the heaviest SU(3) singlet scalar with significant glue component,  but does not have the accuracy (at least not at its leading order) to unambiguously favor one of these states versus the other.} 
	\label{F_DecayRatios_after_chi_s}
\end{figure}

\begin{figure}[!htb]
	\centering
	\includegraphics[width=6in]{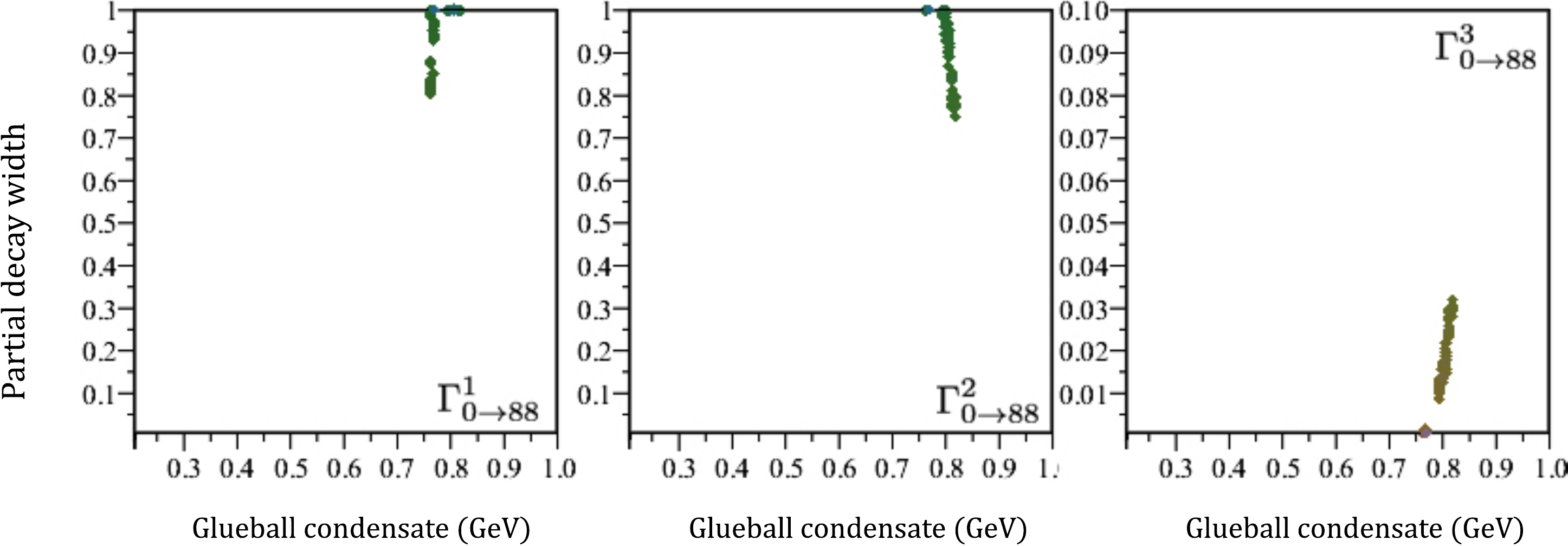} \\ [15pt]
	\includegraphics[width=4in]{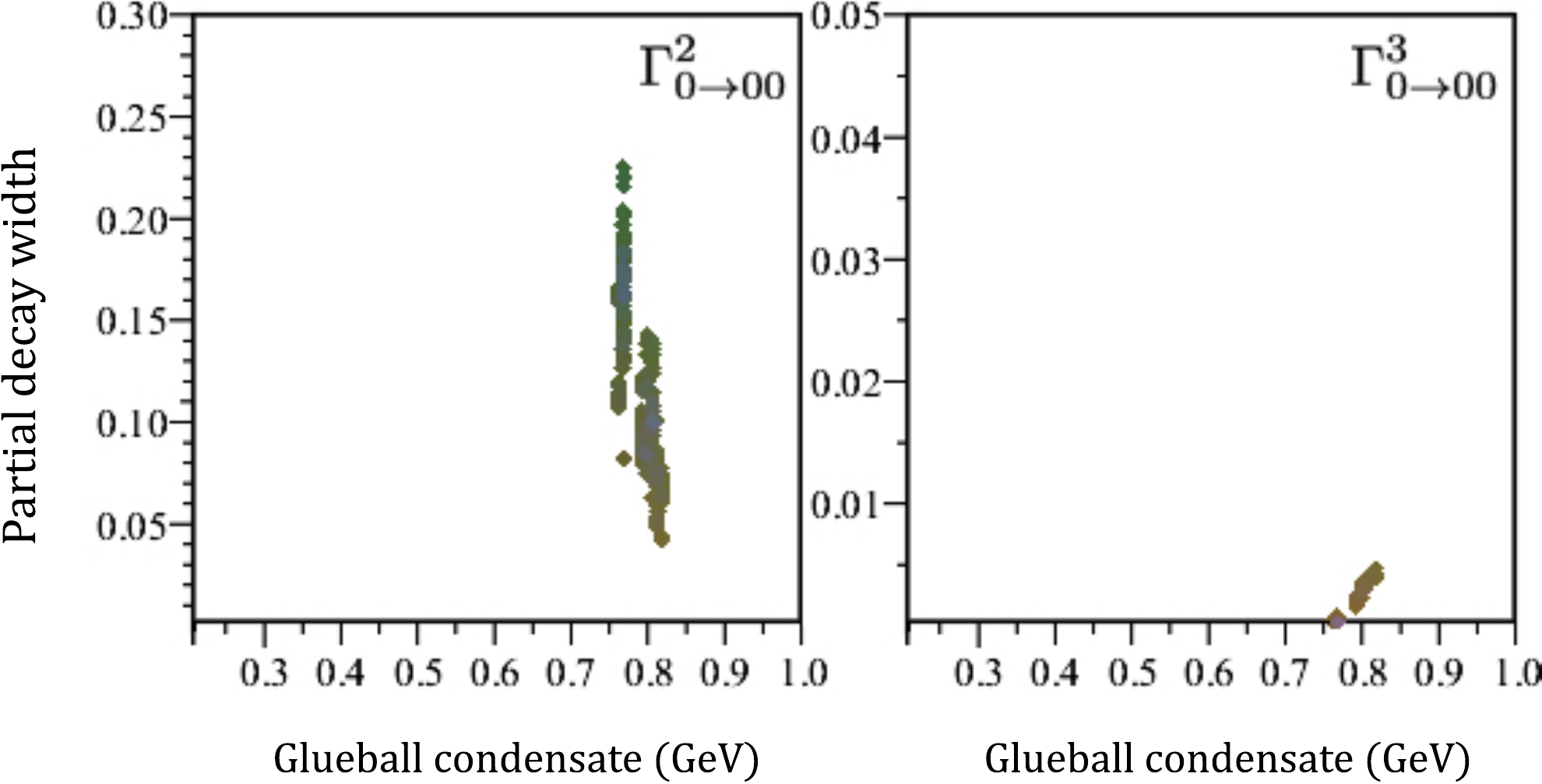} \\ [15pt]
	\includegraphics[width=4in]{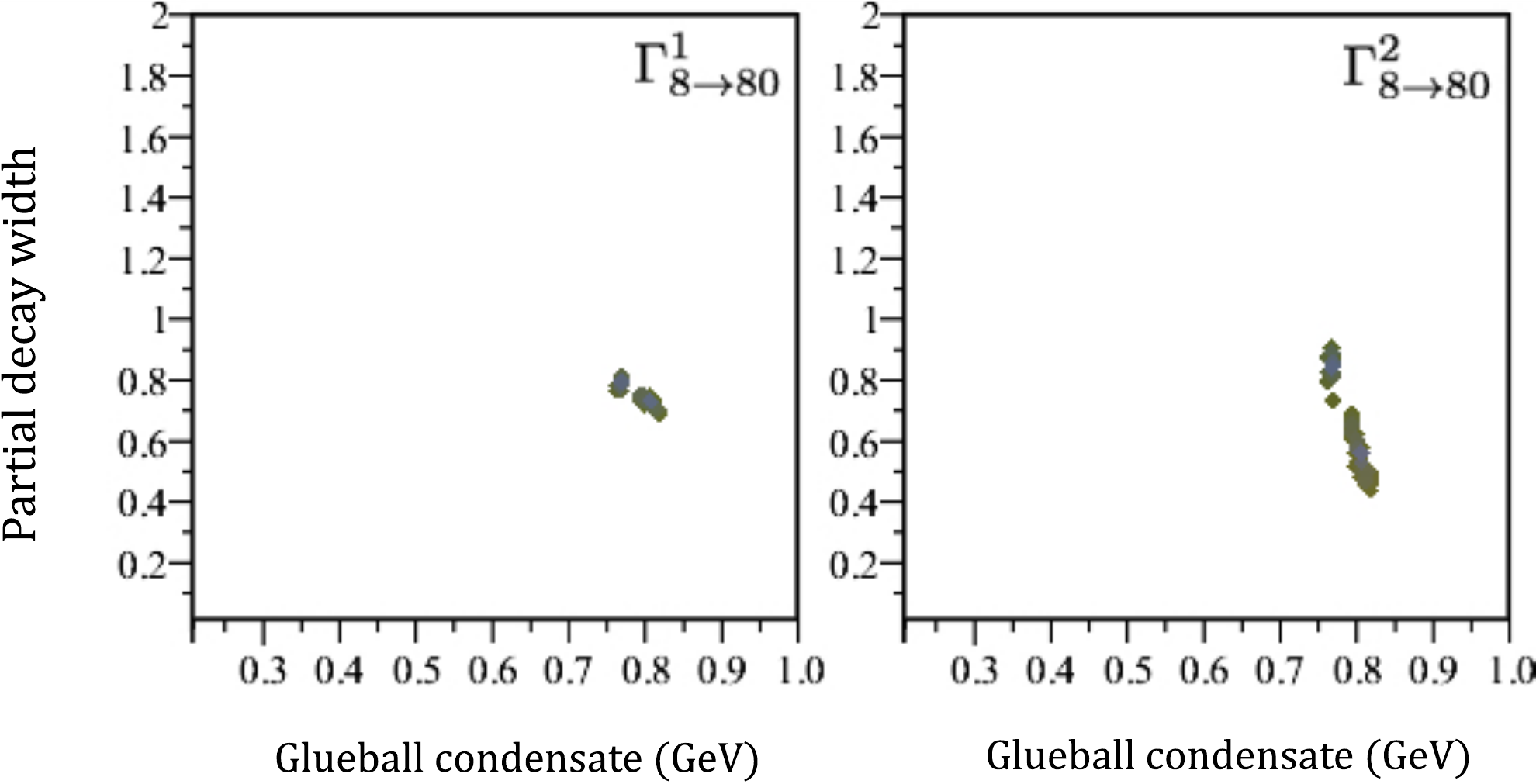}
	\caption{Numerical simulation results with $\chi_s(i) < 1.1 \chi_s^{\rm min}$ for the decay widths (in GeV) of the three SU(3) singlets into $\pi\pi$ (top) and the decay widths into $\eta\eta$ of the second and the heaviest SU(3) singlets (middle panel, second state left and the heaviest state right) and the decay widths into $\pi\eta$ of the SU(3) scalar octets (lowest panel, lighter state left, heavier state right) versus the glueball condensate (in GeV).}
	\label{F_DecayWidths_after_chi_s}
\end{figure}

While the decay ratios for the scalar SU(3) singlets, plotted in Fig. \ref{F_DecayRatios_after_chi_s},   agree with experimental data  (\ref{WA102_data}) as well as  (\ref{pp_ee_1370}) and (\ref{pp_ee_1710}), the decay widths are not accurately predicted as can be seen in Fig. \ref{F_DecayWidths_after_chi_s}.  Of course some of these decay widths are qualitatively consistent with the experiment such as the known  large decay widths for the $f_0(500)$ and $f_0(1370)$.   

\begin{figure}[!htb]
	\centering
	\includegraphics[width=5in]{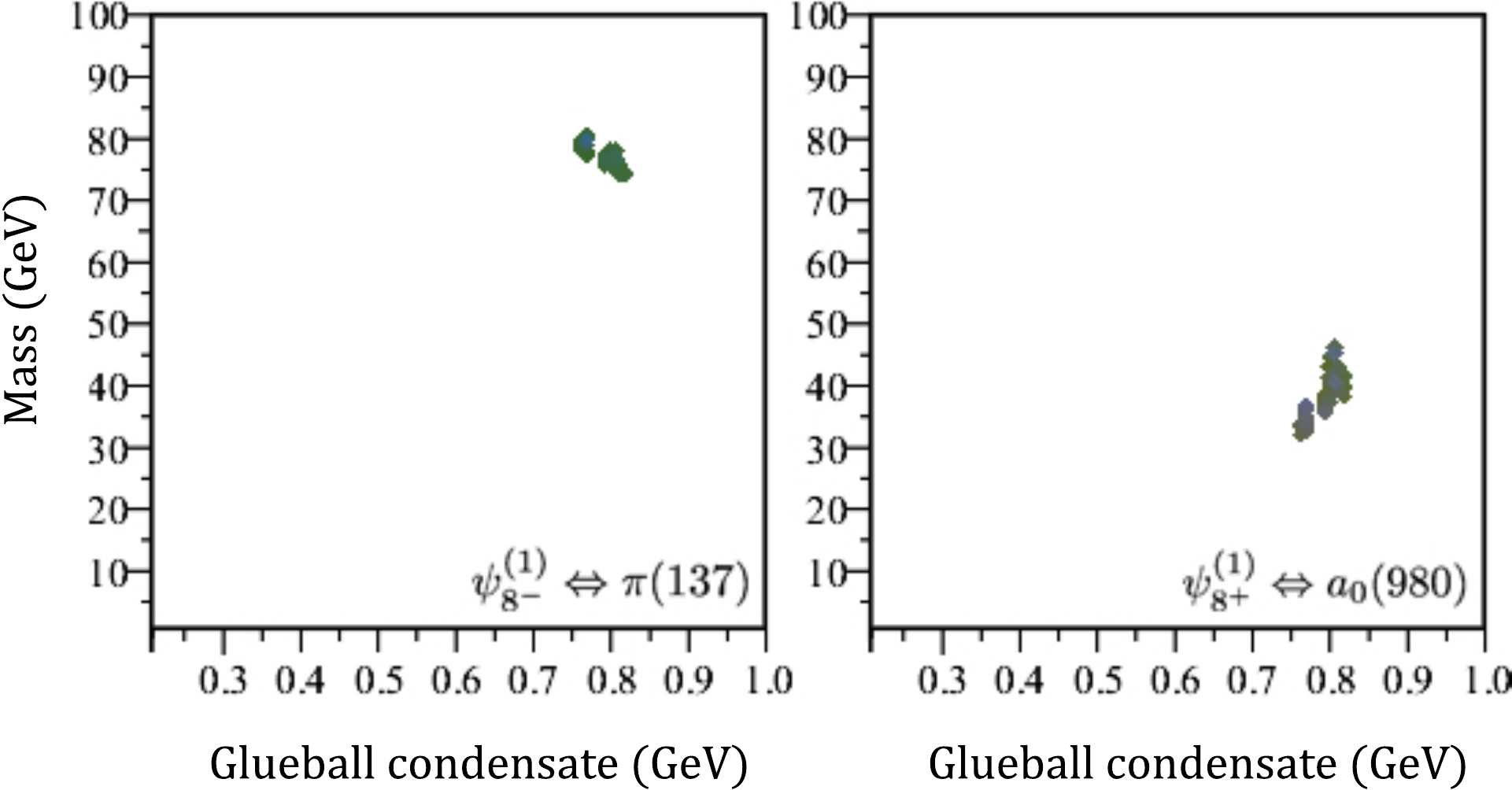}
	\caption{
				Numerical simulation results with $\chi_s(i) < 1.1 \chi_s^{\rm min}$ for the percentage of quark-antiquark component of the lighter pseudoscalar octet [left, identified with $\pi(137)$] and the lighter scalar octet [right, identified with $a_0(980)$],  versus the glueball condensate (horizontal axis,  in GeV)
				}
	\label{F_8m_comps_after_chi_s}
\end{figure}

\begin{figure}[!htb]
	\centering
	\includegraphics[width=0.7\textwidth]{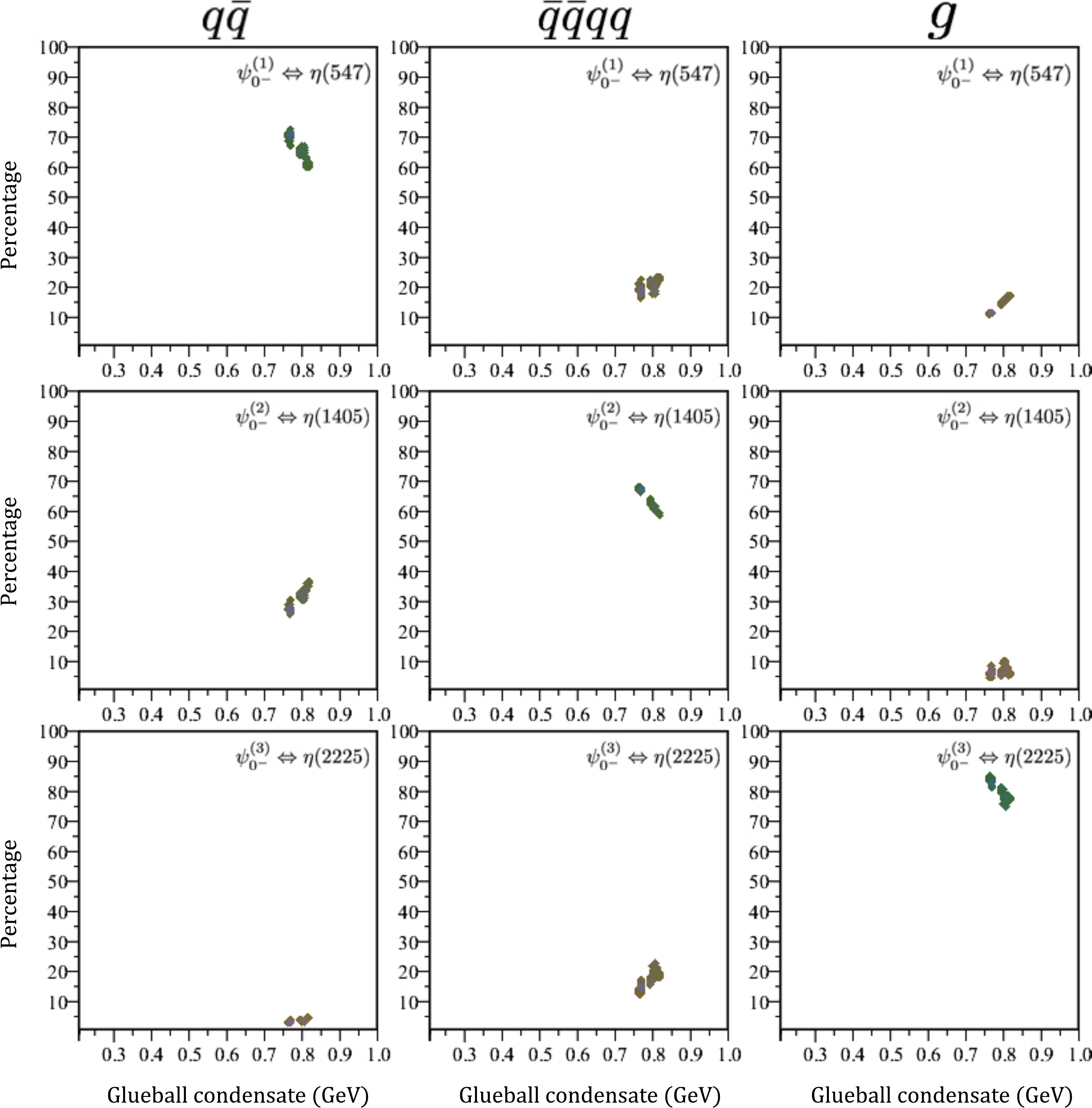}
	\caption{Numerical simulation results with $\chi_s(i) < 1.1 \chi_s^{\rm min}$ for the quark and glue components of the three pseudoscalar SU(3) singlets.  The three rows from top to bottom respectively represent the three  SU(3) singlet pseudoscalar states in ascending order of mass.  The three columns from left to right respectively represent quark-antiquark, four-quark and glue components.   The vertical axis in each graph represents the component percentage and the horizontal axis represents the  glueball condensate (in GeV).}
	\label{F_0m_comps_after_chi_s}
\end{figure}

\begin{figure}[!htb]
	\centering
	\includegraphics[width=0.7\textwidth]{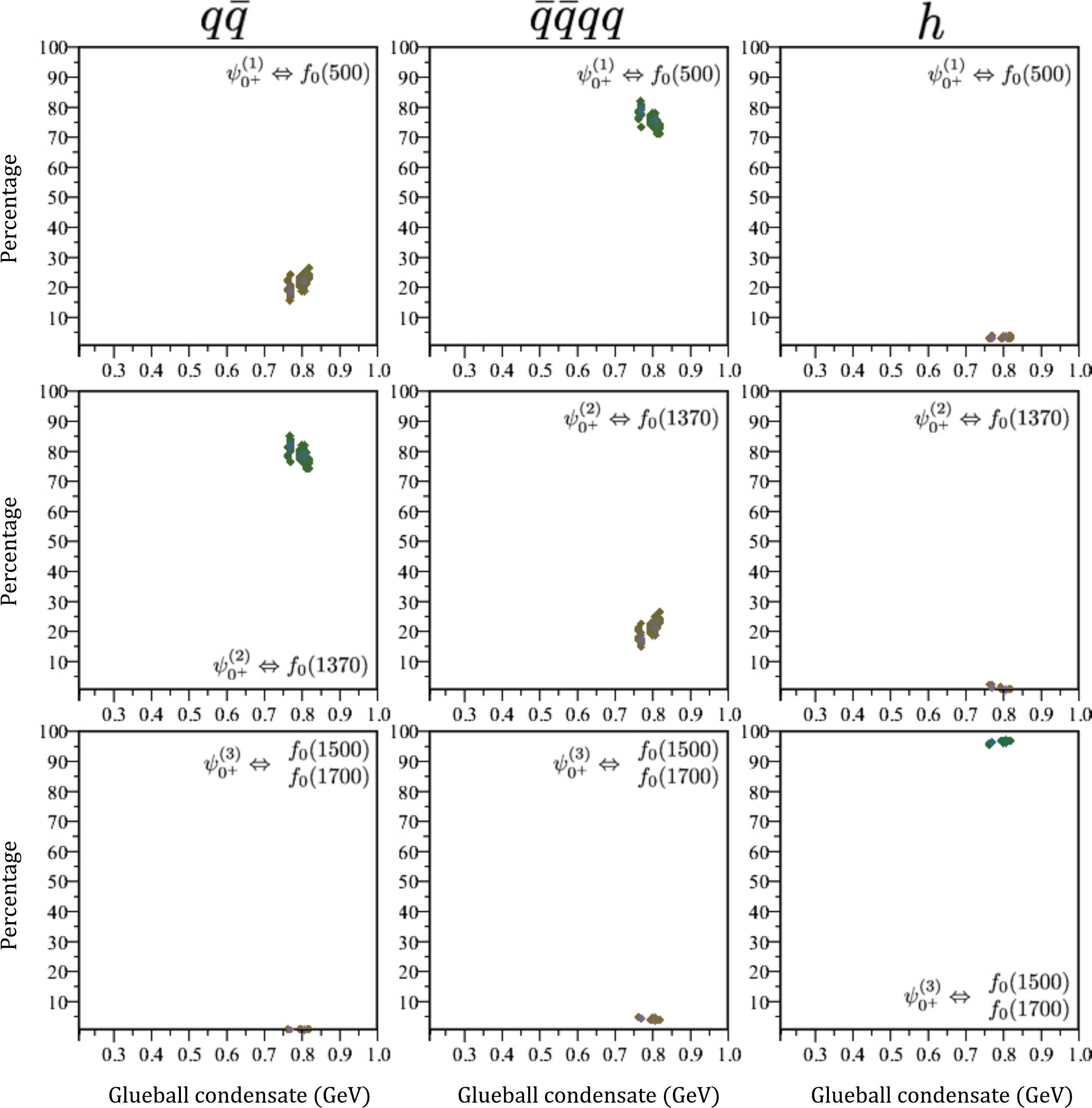}
	
	\caption{Numerical simulation results with $\chi_s(i) < 1.1 \chi_s^{\rm min}$ for the quark and glue components of the three scalar SU(3) singlets.  The three rows from top to bottom respectively represent the three  SU(3) singlet scalar states in ascending order of mass.  The three columns from left to right respectively represent quark-antiquark, four-quark and glue components.   The vertical axis in each graph represents the component percentage and the horizontal axis represents the  glueball condensate (in GeV).}
	\label{F_0p_comps_after_chi_s}
\end{figure}

Despite this lack of accuracy in predictions of the individual decay widths, the overall experimental inputs considered here result in a more refined prediction of the quark and glue components of scalar and pseudoscalar states.  These are presented in Figs. \ref{F_8m_comps_after_chi_s}, \ref{F_0m_comps_after_chi_s} and \ref{F_0p_comps_after_chi_s}.   The results show a clear pattern for the quark and glue components of the scalar and pseudoscalar SU(3) octets and singlets consistent with general expectations. The lighter pseudoscalar SU(3) octet (pion) is seen to consist of about 77\% quark-antiquark and 23\% four-quark and hence the reverse substructure is predicted for $\pi(1300)$ (with about 23\% quark-antiquark and 77\% four-quark).  On the other hand, the lighter scalar SU(3) octet, $a_0(980)$, is consist of around 38\% quark-antiquark and 62\% four-quark (and  the reverse of this for $a_0(1450)$ with about 38\% of four-quark and 62\% of quark-antiquark).

The lightest pseudoscalar SU(3) singlet, $\eta(547)$, is consist of about 66\% quark-antiquarks, 20\% four-quarks and  14\% glue, while the second state around 1.5 GeV [$\eta(1405)$] contains about 30\% quark-antiquarks, 63\% four-quarks and about 7\% glue; and the heaviest state, $\eta(2205)$,  is dominantly made of glue, about 79\%,  together with about  17\% four-quark and about 4\%  quark-antiquark component.  

As is well discussed in the literature and broadly accepted, the scalar SU(3) singlets  have a reversed substructure compared to the pseudoscalars.   The lightest scalar SU(3)  singlet, $f_0(500)$, is made of about 76\% four-quarks, 21\% quark-antiquark and negligible glue (about 3\%); the second state around 1.5 GeV, $f_0(1370)$, is  made of about 79\% quark-antiquarks, 20\% four-quark and negligible glue.    The heaviest scalar SU(3) singlet is almost entirely  made of  glue and as we saw in Fig. \ref{F_m0m_m0p_after_chi_s} its mass range is such that it can be identified with either $f_0(1500)$ or $f_0(1710)$ (our simulations do not have the necessary resolution to pin down which of the two states is the case). 

\begin{figure}[!htb]
	\centering
	\includegraphics[width=5in]{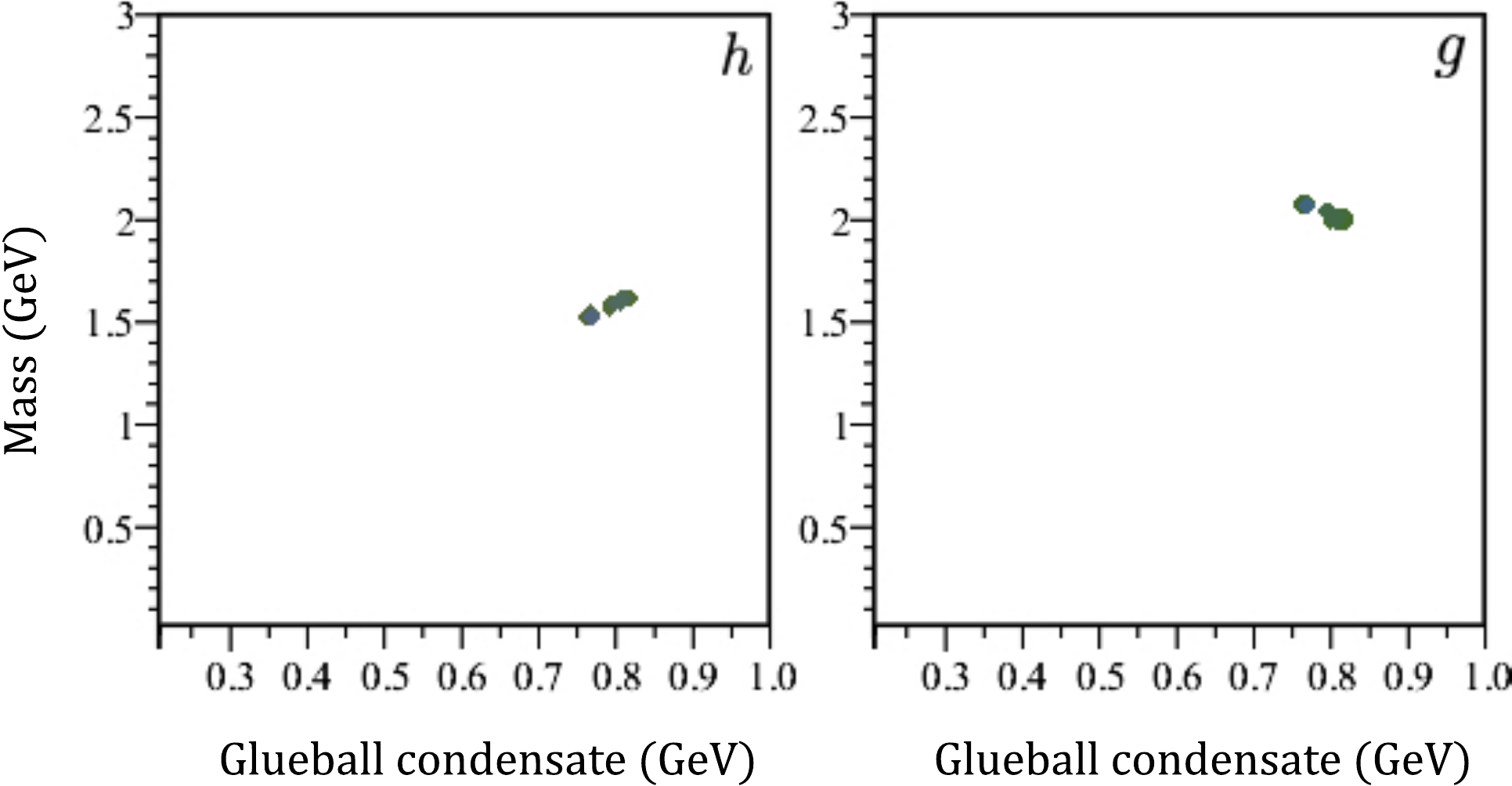}
	\caption{Numerical simulation results with $\chi_s(i) < 1.1 \chi_s^{\rm min}$ for the mass (in GeV) of the scalar glueball (left) and pseudoscalar glueball (right) versus the glueball condensate (in GeV).}
	\label{F_mh_mg_after_chi_s}
\end{figure}

Finally, we give our results for the mass of the pure scalar and pseudoscalr glueballs in Fig. \ref{F_mh_mg_after_chi_s}.  The scalar glueball mass is estimated around 1.6 GeV and the pseudoscalar glueball mass is about 2.0 GeV and  both of these results seem consistent with the lattice QCD estimates \cite{Morningstar}.

\clearpage

\section{Summary and conclusion}

In this work we presented a detailed numerical study of the mixing patterns of scalar and pseudoscalar mesons below and above 1 GeV using the generalized linear sigma model of Refs. \cite{Jora25,Jora26,19_FJL} in the SU(3) flavor symmetric limit.   The theoretical framework of the generalized linear sigma model contains two chiral nonets (a quark-antiquark nonet and a four-quark nonet) as well as two glueballs (a scalar glueball and a pseudoscalar glueball). The chiral symmetry together with  the trace and axial anomalies of QCD establish implicit relationships among the mass matrices and among the coupling constants in terms of the model parameters.  In the SU(3) limit, the model contains 10 meson states:  Two pseudoscalar octets, two scalar octets, three pseudoscalar singlets and three scalar singlets.  We started out by identifying the two pseudoscalar octets with $\pi(137)$ and $\pi(1300)$ and the two scalar octets with $a_0(980)$ and $a_0(1450)$.  
% -----------------------------
 Note that the identification of lighter scalar octet with $a_0(980)$ is consistent with the work of \cite{ChUA4}.  
%------------------------------ 
 We used the  experimental data on these octet masses,  $\pi(137)$ and $\pi(1300)$ decay constants,  a range of values for glueball condensate, together with limiting the singlet masses up to around 2 GeV as well as applying the minimum equations defining the vacuum of the model,  and numerically determined the boundaries of parameter space.   We then zoomed in on this parameter space and took a limit of the model in which the predicted  pseudoscalar singlet masses approach three of the experimental eta states.  The three etas that are most consistent with this model are:  $\eta(547)$, $\eta(1405)$ and $\eta(2225)$.  
%--------------------------
Our model predictions for eta states around 1.5 GeV is consistent with the results found in \cite{ChUA6,ChUA8} where it is shown that some of these states can be considered as being dynamically generated through interactions of lighter mesons.
%-----------------------------------
We then further added experimental inputs on decay widths and decay rations of scalar SU(3) singlets and found that these three  singlet states can be identified with $f_0(500)$, $f_0(1370)$ and either $f_0(1500)$ or $f_0(1710)$.  We found that in the limit where the experimental inputs on scalar and pseudoscalar singlets are taken into account the model has  predictions for quark and glue components of the physical states with relatively low uncertainties.   We found that the two pseudoscalar SU(3) octets $\pi(137)$ and $\pi(1300)$ are dominantly made of quark-antiquarks and four-quarks, respectively, whereas the predictions for scalar octet is reverse of this with $a_0(980)$ being a dominantly four-quark state and $a_0(1450)$ a quark-antiquark state.   Similarly, the model predicts the substructure of the singlet states with relatively low uncertainty.   The three pseudoscalar singlets (from lightes to heaviest) are dominantly quark-antiquarks, four-quarks and glue, whereas the  three scalar singlets (from lightes to heaviest) are dominantly four-quarks, quark-antiquarks, and glue.  The numerical simulations show that the scalar glueball condendate is around 0.8 GeV in close agreement with our previous work in \cite{19_FJL}.   The scalar and pseudoscalar glueball masses were determined to be respectively around 1.6 GeV and 2.0 GeV consistent with lattice QCD results \cite{Morningstar}.  All numerical estimates are summarized in appendix  \ref{A_num_res}.

\begin{figure}[!htb]
	\centering
	\includegraphics[width=3in]{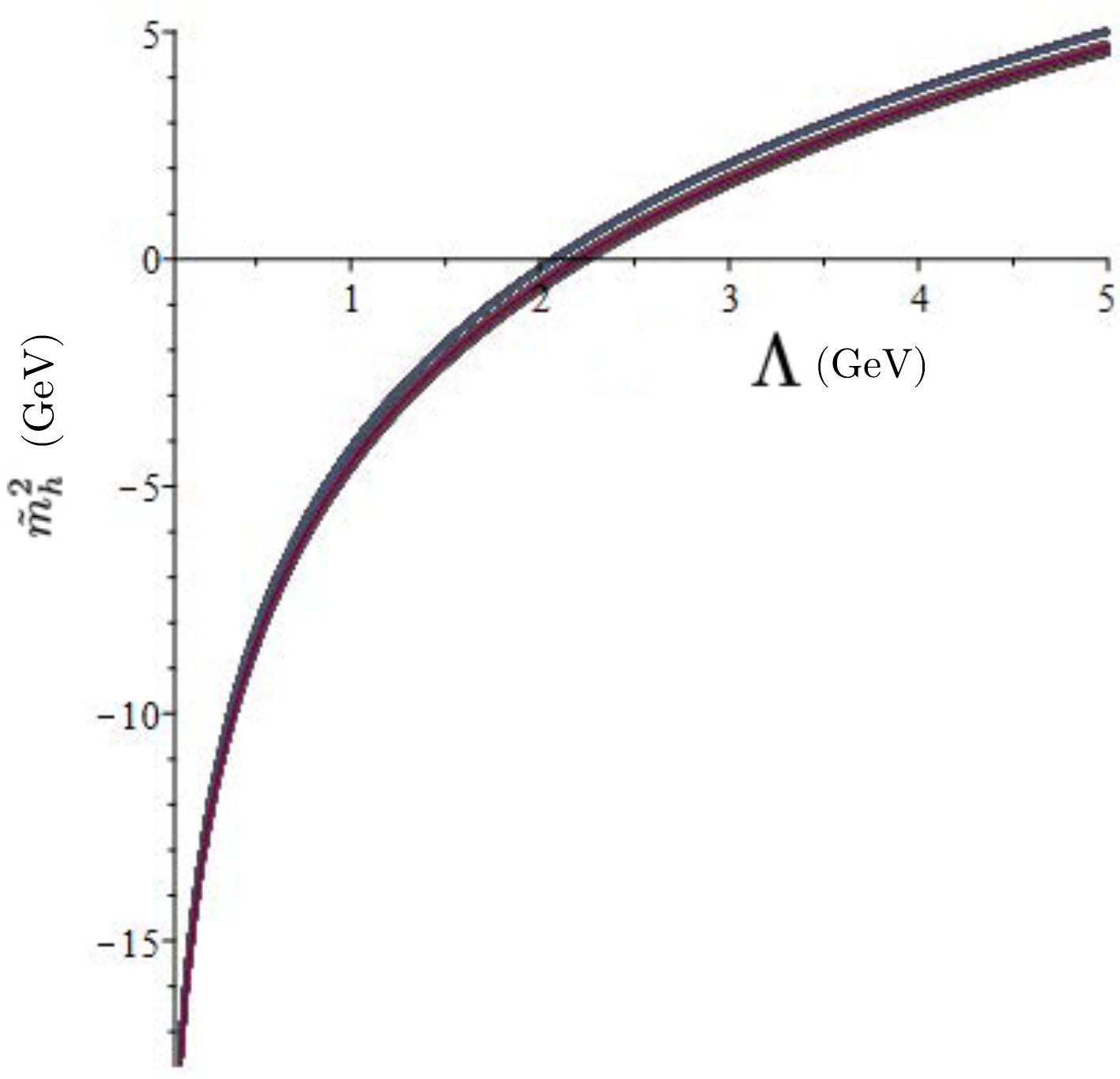}
	
	\caption{Numerical simulation results for the squared mass (in GeV$^2$) of the bare scalar glueball  the characteristic scale $\Lambda$ (in GeV).}
	\label{F_mhtilde_vs_h0}
\end{figure}

It is interesting to highlight the delicate role that the scalar glueball plays in this model.   First, we note that the scalar glueball mass [last equation in (\ref{Y20})] not only depends on pure glue ($u_5$ term), but also on the   quark condensates as well as the characteristic scale $\Lambda$ through logarithmic terms (which diverge in extreme limits of $\Lambda \rightarrow 0$ and $\Lambda \rightarrow  \infty$).   However, when the vacuum condition (\ref{E_Vmin_h}) is imposed, the scalar glueball mass becomes independent of $u_5$ and $\Lambda$ (and therefore free from any potential divergencies) and can be entirely expressed in terms of the condensates only.  We can solve (\ref{E_Vmin_h}) for $u_5$ and substitute in $\left( Y_0^2\right)_{33}$ in (\ref{Y20}) to show
\begin{equation}
m_h^2 =  
-{{36\,\beta\,u_4 \, \alpha^2}\over h_0
} 
- 12\, \alpha^2 u_1  
-12\,u_3 \, \beta^2  
+16\, h_0^2 \lambda_1 
\end{equation} 
which is independent of $\Lambda$ and shows that scalar glueball gets its physical mass both from interaction with quarks (the first three terms) as well as from the trace anomaly (the last term).
 The intricate role that the scalar glueball plays is to interact with quarks to stabilize the QCD vacuum which results in acquiring a  finite mass in the process.    
Therefore, the two parameters $u_5$ and $\Lambda$ are left undetermined in our analysis, although we can still express the {\it ``bare''} scalar glueball mass ${\widetilde m}_h^2 \ne m_h^2$   with  ${\widetilde m}_h^2 = 12\, u_5\, h_0^2$ in term of $\Lambda$ through the vacuum condition (\ref{E_Vmin_h}) as shown in Fig. \ref{F_mhtilde_vs_h0}.    The figure shows that the bare scalar glueball mass is sensitive to the characteristic scale and it is unphysical when $\Lambda$ is less than about 2 GeV, a manifestation of the complex gluon dynamics in the non-perturbative region.    In the case of pseudoscalar glueball there is no distinction between the physical and bare masses.  The pseudoscalar glueball receives its mass    $m_g^2 = {\widetilde m}_g^2 = 2\, u_6\, h_0^2$ from interaction with  scalar glueball.  As can be seen in Fig. \ref{F_mh_mg_after_chi_s} (the right figure) this mass is close to the phyiscal mass of the heaviest pseudoscalar 
SU(3) singlet state in this study.

%\begin{figure}[!htb]
%	\centering
%	\includegraphics[width=3.5in]{200914_mmp2g_mgtilde_vs_h0.pdf}
%	
%	\caption{Numerical simulation results for the mass (in GeV) of the pure pseudoscalar glueball mass versus the glueball condensate (in GeV).}
%	\label{F_mgtilde_vs_h0}
%\end{figure}

The analysis in the SU(3) flavor limit presented in this work, is a reasonable starting point for exploring the mixing of singlet quark composites with glueballs.    In this limit,  the model is more restricted,  and hence,  more tractable.   However,  the full analysis of this model should ultimately include SU(3) symmetry breaking effects, which, expectedly provide additional insights and  present new challenges.    This work paves the way for the full analysis that will include SU(3) breaking effects and will be analyzed in a future project.

\section*{Acknowledgments}

A.H.F. gratefully acknowledges  the support of College of Arts and Sciences of SUNY Poly.

\clearpage

%%%%%%%%%%%%%%%%%%%%%%%%%%%%

\appendix

\section{Partial decay widths of scalars in this model} \label{A_formulas}

The partial decay widths of the scalar mesons are computed in this model using the formulas:

\begin{eqnarray}
\Gamma[\psi_{8_i^+} \rightarrow \psi_{8^-_j}\psi_{0^-_k}] &=& \frac{q_{8^+_i8^-_j0^-_k}\gamma^2_{8^+_i8^-_j0^-_k}}{8\pi m^2_{8^+_i}}\nonumber \\
\Gamma[\psi_{0_i^+} \rightarrow \psi_{8^-_j}\psi_{8^-_k}] &=& \frac{3q_{0^+_i8^-_j8^-_k}\gamma^2_{0^+_i8^-_j8^-_k}}{8\pi m^2_{0^+_i}}\nonumber \\
\Gamma[\psi_{0_i^+} \rightarrow \psi_{0^-_j}\psi_{0^-_k}] &=&
 \frac{q_{0^+_i0^-_j0^-_k}\gamma^2_{0^+_i0^-_j0^-_k}}{4\pi m^2_{0^+_i}}
\end{eqnarray}
where the center of mass momenta are:
\begin{eqnarray}
q_{8^+_i8^-_j0^-_k} &=& 
\frac{1}{2m_{8^+_i}}
\sqrt{[m^2_{8^+_i}-(m_{8^-_j} + m_{0^-_k})^2][m^2_{8^+_i}-(m_{8^-_j} - m_{0^-_k})^2 ]} \nonumber \\ 
q_{0^+_i8^-_j8^-_k} &=& 
\frac{1}{2m_{0^+_i}}
\sqrt{[m^2_{0^+_i}-(m_{8^-_j} + m_{8^-_k})^2][m^2_{0^+_i}-(m_{8^-_j} - m_{8^-_k})^2 ]} \nonumber \\
q_{0^+_i0^-_j0^-_k} &=& 
\frac{1}{2m_{0^+_i}}
\sqrt{[m^2_{0^+_i}-(m_{0^-_j} + m_{0^-_k})^2][m^2_{0^+_i}-(m_{0^-_j} - m_{0^-_k})^2 ]} 
\end{eqnarray}
and the physical coupling constants are:
\begin{eqnarray}
\gamma_{8^+_i8^-_j0^-_k} &=& \left\langle \frac{\partial^3 V}{\partial ( \Psi_{8^+})_i \partial (\Psi_{8^-})_j \partial (\Psi_{0^-})_k} \right\rangle_0
= \sum\limits_{lmn}
\left\langle \frac{\partial^3 V}{\partial(B_{8^+})_l \partial(B_{8^-})_m \partial(B_{0^-})_n}\right\rangle_0 [K_{8^+}]_{li}[K_{8^-}]_{mj}[K_{0^-}]_{nk} \nonumber \\
%
% ---------------------------------------------------------------------------------
%
\gamma_{0^+_i8^-_j8^-_k} &=& \frac{1}{\sqrt{2}}\left\langle \frac{\partial^3 V}{\partial ( \Psi_{0^+})_i \partial (\Psi_{8^-})_j \partial (\Psi_{8^-})_k} \right\rangle_0
= \frac{1}{\sqrt{2}}\sum\limits_{lmn}\left\langle \frac{\partial^3 V}{\partial(B_{0^+})_l \partial(B_{8^-})_m \partial(B_{8^-})_n}\right\rangle_0 [K_{0^+}]_{li}[K_{8^-}]_{mj}[K_{8^-}]_{nk} \nonumber \\
%
% ---------------------------------------------------------------------------------
%
\gamma_{0^+_i0^-_j0^-_k} &=& \frac{1}{{2}}\left\langle \frac{\partial^3 V}{\partial ( \Psi_{0^+})_i \partial (\Psi_{0^-})_j \partial (\Psi_{0^-})_k} \right\rangle_0
= \frac{1}{{2}}\sum\limits_{lmn}\left\langle \frac{\partial^3 V}{\partial(B_{0^+})_l \partial(B_{0^-})_m \partial(0_{8^-})_n}\right\rangle_0 [K_{0^+}]_{li}[K_{0^-}]_{mj}[K_{0^-}]_{nk} \nonumber \\
\end{eqnarray}

The nonzero ``bare'' couplings are:

\begin{eqnarray}
\left\langle \frac{\partial^3 V}{\partial(B_{8^+})_1 \partial(B_{8^-})_1 \partial(B_{0^-})_1}\right\rangle_0
&=& 
{ {4   \left( 2\, u_2 \, \alpha^4 -  h_0^4 \, \lambda_2 \right) }
\over 
{ \sqrt{3}\, \alpha^3}
} 
\nonumber \\
%%%%%%%%%%%%%%%%%%%%%%%%%%%%%%%%%%%%%%%%%%%%%%%%%%%%%%%%%%%%%%%%%%%%%%%%%%%%%%%%%%%%%%%%%%%%%%%%%%%
\left\langle \frac{\partial^3 V}{\partial(B_{8^+})_1 \partial(B_{8^-})_1 \partial(B_{0^-})_2}\right\rangle_0
&=&  { { 4\, u_4 \, h_0} \over {\sqrt {3}}}
\nonumber \\
%%%%%%%%%%%%%%%%%%%%%%%%%%%%%%%%%%%%%%%%%%%%%%%%%%%%%%%%%%%%%%%%%%%%%%%%%%%%%%%%%%%%%%%%%%%%%%%%%%%
\left\langle \frac{\partial^3 V}{\partial(B_{8^+})_1 \partial(B_{8^-})_1 \partial(B_{0^-})_3}\right\rangle_0
&=&  -\,{ {\gamma_1\, h_0^3} \over {6\, \alpha^2} }
\nonumber \\	
%%%%%%%%%%%%%%%%%%%%%%%%%%%%%%%%%%%%%%%%%%%%%%%%%%%%%%%%%%%%%%%%%%%%%%%%%%%%%%%%%%%%%%%%%%%%%%%%%%%
% ----------------------------------------------------------------------------------------------------	
\left\langle \frac{\partial^3 V}{\partial(B_{8^+})_1 \partial(B_{8^-})_2 \partial(B_{0^-})_1}\right\rangle_0
&=&  {\frac {2\, \sqrt {3}\, h_0 \, \left( 6\, u_4  \,  \alpha^2
		\beta - h_0^3 \, \lambda_3 \right) }{9 \, \alpha^2\, \beta}}
\nonumber \\	
% ----------------------------------------------------------------------------------------------------	
%%%%%%%%%%%%%%%%%%%%%%%%%%%%%%%%%%%%%%%%%%%%%%%%%%%%%%%%%%%%%%%%%%%%%%%%%%%%%%%%%%%%%%%%%%%%%%%%%%%
\left\langle \frac{\partial^3 V}{\partial(B_{8^+})_1 \partial(B_{8^-})_2 \partial(B_{0^-})_2}\right\rangle_0
&=&  - \left\langle \frac{\partial^3 V}{\partial(B_{8^+})_2 \partial(B_{8^-})_1 \partial(B_{0^-})_2}\right\rangle_0 = 
{\frac {2\, \sqrt {3} \, h_0^4\, \lambda_3}{9\, \alpha\, \beta^2}}
\nonumber \\	
% ----------------------------------------------------------------------------------------------------	
%%%%%%%%%%%%%%%%%%%%%%%%%%%%%%%%%%%%%%%%%%%%%%%%%%%%%%%%%%%%%%%%%%%%%%%%%%%%%%%%%%%%%%%%%%%%%%%%%%%
\left\langle \frac{\partial^3 V}{\partial(B_{8^+})_1 \partial(B_{8^-})_2 \partial(B_{0^-})_3}\right\rangle_0
&=&  - \left\langle \frac{\partial^3 V}{\partial(B_{8^+})_2 \partial(B_{8^-})_1 \partial(B_{0^-})_3}\right\rangle_0 = {\frac {h_0^3\,  \left( \gamma_1-1 \right) }{18\, \alpha\,\beta} }
\nonumber \\	
% ----------------------------------------------------------------------------------------------------	
%%%%%%%%%%%%%%%%%%%%%%%%%%%%%%%%%%%%%%%%%%%%%%%%%%%%%%%%%%%%%%%%%%%%%%%%%%%%%%%%%%%%%%%%%%%%%%%%%%%
\left\langle \frac{\partial^3 V}{\partial(B_{8^+})_2 \partial(B_{8^-})_1 \partial(B_{0^-})_1}\right\rangle_0
&=&  {\frac {2\,  \sqrt {3}\, h_0\, \left( 6\,  u_4 \, \alpha^2\, \beta 	+ h_0^3 \, \lambda_3 \right) }
	{9\, \alpha^2\beta}}
\end{eqnarray}

\begin{eqnarray}
\left\langle \frac{\partial^3 V}{\partial(B_{0^+})_1 \partial(B_{8^-})_1 \partial(B_{8^-})_1}\right\rangle_0
&=& \,{\frac {4\,  \left( 2\, \alpha^4\, u_2 - h_0^4 \,\lambda_2 \right) }{\sqrt{3}\, \alpha^3}} 
\nonumber \\
% -------------------------------------------------------------------------------------------------	
\left\langle \frac{\partial^3 V}{\partial(B_{0^+})_1 \partial(B_{8^-})_1 \partial(B_{8^-})_2}\right\rangle_0
&=& 
\left\langle \frac{\partial^2 V}{\partial(B_{0^+})_1 \partial(B_{8^-})_2 \partial(B_{8^-})_1}\right\rangle_0
= {\frac {2\, \sqrt {3}\, h_0 \, \left( 6\, u_4   \, \beta  \,  \alpha^2 -  h_0^3\, \lambda_3 \right) }
	{9\, \beta\, \alpha^2}}
\nonumber \\
% -------------------------------------------------------------------------------------------------	
\left\langle \frac{\partial^3 V}{\partial(B_{0^+})_2 \partial(B_{8^-})_1 \partial(B_{8^-})_1}\right\rangle_0
&=&  { { 4\, u_4 \,  h_0  }\over \sqrt{3}}
\nonumber \\
% -------------------------------------------------------------------------------------------------	
\left\langle \frac{\partial^3 V}{\partial(B_{0^+})_2 \partial(B_{8^-})_1 \partial(B_{8^-})_2}\right\rangle_0
&=& \left\langle \frac{\partial^3 V}{\partial(B_{0^+})_2 \partial(B_{8^-})_2 \partial(B_{8^-})_1}\right\rangle_0 =
 -\,{\frac {2\, h_0^4 \, \lambda_3}{3\, \sqrt{3}\, \alpha\,\beta^2}}
\nonumber \\
% -------------------------------------------------------------------------------------------------	
\left\langle \frac{\partial^3 V}{\partial(B_{0^+})_3 \partial(B_{8^-})_1 \partial(B_{8^-})_1}\right\rangle_0
&=& {\frac {4\,  u_4  \, \beta \,  \alpha^2 +  4\,  \alpha^2 \, h_0  \, u_1 + 8\, h_0^3\, \lambda_2}  {\alpha^2}}
\nonumber \\
% -------------------------------------------------------------------------------------------------	
\left\langle \frac{\partial^3 V}{\partial(B_{0^+})_3 \partial(B_{8^-})_1 \partial(B_{8^-})_2}\right\rangle_0
&=& \left\langle \frac{\partial^3 V}{\partial(B_{0^+})_3 \partial(B_{8^-})_2 \partial(B_{8^-})_1}\right\rangle_0 = 
{\frac {4 \left(3\, u_4  \,  \beta\,  \alpha^2  +  2\,  h_0^3\, \lambda_3\right)}{3\, \alpha\,\beta} }
\nonumber \\
% -------------------------------------------------------------------------------------------------	
\left\langle \frac{\partial^3 V}{\partial(B_{0^+})_3 \partial(B_{8^-})_2 \partial(B_{8^-})_2}\right\rangle_0
&=&
4\, u_3  \,  h_0
\end{eqnarray}

\begin{eqnarray}
\left\langle \frac{\partial^3 V}{\partial(B_{0^+})_1 \partial(B_{0^-})_1 \partial(B_{0^-})_1}\right\rangle_0
&=& {\frac { 4\, \left( 2\,  u_2  \,   \alpha^4 - h_0^4
		\left( \lambda_2 + \lambda_3/3 \right)  \right) } {\sqrt{3}\, \alpha^3 }}
\nonumber \\
%---------------------------------------------------------------------------------------------------------
\left\langle \frac{\partial^3 V}{\partial(B_{0^+})_1 \partial(B_{0^-})_1 \partial(B_{0^-})_2}\right\rangle_0
&=& \left\langle \frac{\partial^3 V}{\partial(B_{0^+})_1 \partial(B_{0^-})_2 \partial(B_{0^-})_1}\right\rangle_0 = -\, {{8\, u_4  \, h_0 }\over  {\sqrt {3}}}
\nonumber \\
%---------------------------------------------------------------------------------------------------------
\left\langle \frac{\partial^3 V}{\partial(B_{0^+})_1 \partial(B_{0^-})_1 \partial(B_{0^-})_3}\right\rangle_0
&=&  \left\langle \frac{\partial^3 V}{\partial(B_{0^+})_1 \partial(B_{0^-})_3 \partial(B_{0^-})_1}\right\rangle_0 = -\,{\frac { h_0^3 \left( 2\,\gamma_1+1 \right) }{18\,  \alpha^2} }
\nonumber \\
%---------------------------------------------------------------------------------------------------------
\left\langle \frac{\partial^3 V}{\partial(B_{0^+})_2 \partial(B_{0^-})_1 \partial(B_{0^-})_1}\right\rangle_0
&=& -\, {8 \, u_4  \,   h_0  \, \over {\sqrt {3}} }
\nonumber \\
%---------------------------------------------------------------------------------------------------------
\left\langle \frac{\partial^3 V}{\partial(B_{0^+})_2 \partial(B_{0^-})_2 \partial(B_{0^-})_2}\right\rangle_0
&=&  -{\frac {4\, h_0^4\, \lambda_3}{3\, \sqrt{3}\,\beta^3} }
\nonumber \\
%---------------------------------------------------------------------------------------------------------
\left\langle \frac{\partial^3 V}{\partial(B_{0^+})_2 \partial(B_{0^-})_2 \partial(B_{0^-})_3}\right\rangle_0
&=&  \left\langle \frac{\partial^3 V}{\partial(B_{0^+})_2 \partial(B_{0^-})_3 \partial(B_{0^-})_2}\right\rangle_0 = {\frac { h_0^3 \left( 1- \gamma_1 \right) }  {18\, \beta^2}  }
\nonumber \\
%---------------------------------------------------------------------------------------------------------
\left\langle \frac{\partial^3 V}{\partial(B_{0^+})_3 \partial(B_{0^-})_1 \partial(B_{0^-})_1}\right\rangle_0
&=& {\frac { \left( 24\,\lambda_2 + 8\,\lambda_3 \right) h_0^3  +
		12\,  u_1 \,  h_0  \,  \alpha^2 -  24\,\beta\, u_4   \,   \alpha^2}{3\, \alpha^2} }
\nonumber \\
%---------------------------------------------------------------------------------------------------------
\left\langle \frac{\partial^3 V}{\partial(B_{0^+})_3 \partial(B_{0^-})_1 \partial(B_{0^-})_2}\right\rangle_0
&=&  \left\langle \frac{\partial^3 V}{\partial(B_{0^+})_3 \partial(B_{0^-})_2 \partial(B_{0^-})_1}\right\rangle_0 = -8\,\alpha\,  u_4
\nonumber \\
%---------------------------------------------------------------------------------------------------------
\left\langle \frac{\partial^3 V}{\partial(B_{0^+})_3 \partial(B_{0^-})_1 \partial(B_{0^-})_3}\right\rangle_0
&=&  \left\langle \frac{\partial^3 V}{\partial(B_{0^+})_3 \partial(B_{0^-})_3 \partial(B_{0^-})_1}\right\rangle_0 = {\frac {  h_0^2 \,  \left( 2\,\gamma_1 + 1 \right) }
	{2\, \sqrt{3}\, \alpha}}
\nonumber \\
%---------------------------------------------------------------------------------------------------------
%---------------------------------------------------------------------------------------------------------
\left\langle \frac{\partial^3 V}{\partial(B_{0^+})_3 \partial(B_{0^-})_2 \partial(B_{0^-})_2}\right\rangle_0
&=&  {\frac {8\,  \left(  h_0^2\, \lambda_3 + 3/2\, u_3  \,  \beta^2
		\right) h_0}{3\, \beta^2}}
\nonumber \\
%---------------------------------------------------------------------------------------------------------
%---------------------------------------------------------------------------------------------------------
\left\langle \frac{\partial^3 V}{\partial(B_{0^+})_3 \partial(B_{0^-})_2 \partial(B_{0^-})_3}\right\rangle_0
&=&  \left\langle \frac{\partial^3 V}{\partial(B_{0^+})_3 \partial(B_{0^-})_3 \partial(B_{0^-})_2}\right\rangle_0 =
 {\frac { h_0^2 \left( \gamma_1-1 \right) } {2\, \sqrt{3}\, \beta}
}\nonumber \\
%---------------------------------------------------------------------------------------------------------
\left\langle \frac{\partial^3 V}{\partial(B_{0^+})_3 \partial(B_{0^-})_3 \partial(B_{0^-})_3}\right\rangle_0
&=& 4\,u_6\, h_0
\end{eqnarray}

\clearpage

\section{Numerical values of the model parameters and predictions}\label{A_num_res}

For completeness,  in this appendix we give the numerical values of our simulations for various quantities of interest that we already discussed in previous sections mostly in terms of graphs.    Table \ref{T_par_values} gives the values of the model parameters for the two cases discussed in subsection \ref{ss_scalar_const} (we have given these numerical values in many digits to make it possible for the readers to reproduce our results).    Table \ref{T_m_phys_values} presents the physical masses in this model,  both at the minimum of $\chi_s$ for each of the two cases, as well as their average and standard deviation when considering 10\% variation around either minima.  Table  \ref{T_m_bare_values} gives the bare (unmixed) masses.    The components of physical states are given in Table \ref{T_comp_values} at the minimum of $\chi_s$ for both cases and their averaged values are given in Table   
\ref{T_comp_values_ave}.   The rotation matrices are given in Table \ref{T_rot_mats} and the decay widths and decay ratios are presented in Table \ref{T_decay_W_R}.

\begin{table}[ht]
\caption{Model parameters for the two cases where we identify the heaviest scalar SU(3) singlet with either  $f_0(1500)$ or $f_0(1710)$ assessed  by $\chi_s(1)$ and $\chi_s(2)$ defined in Eq. (\ref{chi_s}). The first two columns give the model parameters at the corresponding  minimum of each $\chi_s$ (given in many digits to allow sufficient accuracy for reproduction of the results given in this work). The last column gives the averages and standard deviations of parameters within 10\% of either minimum.
}
	\centering
	\begin{tabular}{M{60pt} || M{80pt} | M{80pt} |  r @{} M{12pt} @{} l N} %chktex 44
		Variable & Computed at $\chi_s^{\text{min}}(1)$ = 14.4 & Computed at $\chi_s^{\text{min}}(2)$ = 17.0 & \multicolumn{3}{c}{Mean $\pm$ $\sigma$} \\
		\toprule
		\toprule
		$\alpha \times 10^2$     & 5.656436159 &    5.812847151       & 5.765        &$\pm$           & 0.063    &\\ [6pt]
		$\beta \times 10^2$    & 3.302818815 & 3.019058250            & 3.107        &$\pm$           & 0.116     &\\ [6pt]
		$u_1 \times 10^{-1}$     & 1.622597054 & 1.336844501            & 1.456        &$\pm$           & 0.103     &\\ [6pt]
		$u_2 \times 10^{-2}$    & $-$7.990847622  & $-$5.378679133             & $-$6.440        &$\pm$           & 0.923     &\\ [6pt]
		$u_3$     & 1.094533000 & 1.160760311             & 1.146        &$\pm$           & 0.032     &\\ [6pt]
		$u_4$    & $-$0.9502544234 & $-$1.215013415             & $-$1.062       &$\pm$           & 0.102     &\\ [6pt]
		$u_6$   & 2.975765957 & 3.641199465             & 3.297        &$\pm$           & 0.247     &\\ [6pt]
		$\lambda_2 \times 10^2$     & $-$3.881733931 & $-$3.884961191             & $-$3.887        &$\pm$           & 0.005     &\\ [6pt]
		$\lambda_3 \times 10^3$    & 2.670453409 & 2.718862309             & 2.743        &$\pm$           & 0.080     &\\ [6pt]
		$\gamma_1 \times 10$   & 7.4906844 & 1.31692715             & 1.016        &$\pm$           & 0.209     &\\ [6pt]
		$h_0 \times 10$         & 8.149662587 & 7.643986219             & 7.880        &$\pm$           & 0.185     &\\ [6pt]
		$f_{8_-}' \times 10^4$         & $-$7.435438514 & $-$7.49587234             & $-$7.295        &$\pm$           & 0.212     &\\ [6pt]
%		\bottomrule
%		\bottomrule              
	\end{tabular}
\label{T_par_values}
\end{table}

\begin{table}[ht]
	\caption{Physical masses (in GeV) for the two cases where  we identify the heaviest scalar SU(3) singlet with either   $f_0(1500)$ or $f_0(1710)$
		assessed by $\chi_s(1)$ and $\chi_s(2)$ in Eq. (\ref{chi_s}). The first two columns give the masses at the corresponding minimum of each  $\chi_s$ and the last column gives the average and standard deviation of masses within 10\% of each minimum.   Some of the masses have been inputted as indicated. }
	\centering
	\begin{tabular}{M{60pt} || M{40pt} | M{40pt} |  r @{} M{12pt} @{} l N} %chktex 44
		Variable &Computed at  $\chi_s^{\text{min}}(1)$ & Computed at $\chi_s^{\text{min}}(2)$ & \multicolumn{3}{c}{Mean $\pm$ $\sigma$} \\
		\toprule
		\toprule
		$m_{8^-}$   (Input)  & 0.137 & 0.137             &       &     ---      &      &\\ [6pt]
		$m_{8^-}'$    & 0.140 & 0.131            & 1.357        &$\pm$           & 0.032     &\\ [6pt]
		\hline
		$m_{8^+}$   (Input)   & 0.980 & 0.980            &     &    ---       &      &\\ [6pt]
		$m_{8^+}'$  (Input)  & 1.474  & 1.474             &     & --- &      &\\ [6pt]
		\hline
		$m_{0^+}$     & 0.675 & 0.626             & 0.643        &$\pm$           & 0.020     &\\ [6pt]
		$m_{0^+}'$    & 1.152 & 1.213             & 1.178        &$\pm$           & 0.024     &\\ [6pt]
		$m_{0^+}''$   & 1.641 & 1.542             & 1.589        &$\pm$           & 0.036     &\\ [6pt]
		\hline
		$m_{0^-}$     & 0.550 & 0.548             & 0.548        &$\pm$           & 0.001     &\\ [6pt]
		$m_{0^-}'$    & 1.404 & 1.405             & 1.407        &$\pm$           & 0.003     &\\ [6pt]
		$m_{0^-}''$   & 2.220 & 2.220             & 2.219        &$\pm$           & 0.002     &\\ [6pt]
%		\bottomrule
%		\bottomrule              
	\end{tabular}
\label{T_m_phys_values}
\end{table}

\begin{table}[ht]
	\caption{Bare masses (in GeV) for the two cases where  we identify the heaviest scalar SU(3) singlet with either   $f_0(1500)$ or $f_0(1710)$
		assessed by $\chi_s(1)$ and $\chi_s(2)$ in Eq. (\ref{chi_s}). The first two columns give the masses at the corresponding minimum of each  $\chi_s$ and the last column gives the average and standard deviation of masses within 10\% of either minimum.  Note that for pseudoscalars the pure quark-antiquark pseudoscalar octet $\eta_8$  is lighter than the pure four-quark pseudoscalar octet $\eta_8'$, and  the pure quark-antiquark pseudoscalar singlet $\eta_0$  is lighter than the pure four-quark pseudoscalar octet $\eta_0'$. This is in contrast with the scalar masses where the pure quark-antiquark scalar octet $f_8$  is {\it heavier} than the pure four-quark scalar octet $f_8'$, and  the pure quark-antiquark scalar singlet $f_0$  is {\it heavier} than the pure four-quark scalar singlet $f_0'$.   This reversed mass spectrum for scalar mesons is understood based on the MIT bag model \cite{jaffe}. }
	\centering
	\begin{tabular}{M{60pt} || M{40pt} | M{40pt} |  r @{} M{12pt} @{} l N} %chktex 44
		Bare Masses & Computed at $\chi_s^{\text{min}}(1)$ & Computed at $\chi_s^{\text{min}}(2)$ & \multicolumn{3}{c}{Mean $\pm$ $\sigma$} \\
		\toprule
		\toprule
		$m_{\eta_8}$   & 0.722 &  0.624 &  0.662       &$\pm$    & 0.037      &\\ [6pt]
		$m_{\eta_8'}$  & 1.206 &  1.165 &  1.192       &$\pm$     &0.019     
		&\\ [6pt]
		\hline 
		$m_{f_8}$      & 1.296 &  1.333   &  1.308      &$\pm$    &  0.018   &\\ [6pt]
		$m_{f_8'}$     & 1.206 &  1.165   &  1.192      &$\pm$    & 0.019     &\\ [6pt]
		\hline 
		$m_{\eta_0}$  & 1.036 &  0.954    &   0.981     &$\pm$     &  0.034    &\\ [6pt]
		$m_{\eta_0'}$ & 1.474 &  1.427    &   1.467     &$\pm$     &  0.028    &\\ [6pt]
		$m_{g}$       & 1.988 &  2.063    &   2.020     &$\pm$     &  0.029    &\\ [6pt]
		\hline 
		$m_{f_0}$     & 1.062 & 1.121  &    1.089     &$\pm$      &  0.024    &\\ [6pt]
		$m_{f_0'}$    & 0.857 & 0.823  &    0.831     &$\pm$      &  0.020   &\\ [6pt]
		$m_h$         & 1.617 & 1.519  &    1.565     &$\pm$      &  0.036    &\\ [6pt]
%		\bottomrule
%		\bottomrule              
	\end{tabular}
\label{T_m_bare_values}
\end{table}

\begin{table}[ht]
	\caption{
		Simulation results for the quark-antiquark, four-quark, and glueball components of physical states
 for the two cases where  we identify the heaviest scalar SU(3) singlet with either   $f_0(1500)$ or $f_0(1710)$
		assessed by $\chi_s(1)$ and $\chi_s(2)$ in Eq. (\ref{chi_s}).   The two vertical divisions give the components  at the corresponding minimum of each  $\chi_s$.}	
	\centering
	\begin{tabular}{M{80pt} | *{2}{ | M{40pt} M{40pt} M{40pt} }N} %chktex 44
		\multirow{3}{*}{States} & \multicolumn{3}{c}{ Computed at $\chi_s^{\text{min}}(1)$} & \multicolumn{3}{c}{Computed at  $\chi_s^{\text{min}}(2)$}&\\ [6pt]
		& $q\overline{q}$ & $qq\overline{q}\overline{q}$ & glue & $q\overline{q}$ & $qq\overline{q}\overline{q}$ & glue &\\ [6pt]
		\toprule
		\toprule 
		
		$\psi_{8^+}^{(1)}$ &
		40.71  &     
		59.29  &      
		---    &      
		32.67  &
		67.33  &
		---  
		&\\ [6pt] % Row 1
		
		$\psi_{8^+}^{(2)}$ &
		59.29  &     
		40.71  &      
		---    &      
		67.33  &
		32.67  &
		---  
		&\\ [6pt] % Row 2
	\hline	
		%		\midrule
		
		$\psi_{8^-}^{(1)}$ &
		74.08  &     
		25.92  &      
		---    &      
		78.29  &
		21.71  &
		---  
		&\\ [6pt] % Row 1
		
		$\psi_{8^-}^{(2)}$ &
		25.92  &     
		74.08  &      
		---    &      
		21.71  &
		78.29  &
		---  
		&\\ [6pt] % Row 2
		
		%		\midrule
		\hline
		$\psi_{0^+}^{(1)}$ &
		23.24  &     
		73.46  &      
		3.30  &      
		20.03  &
		77.15  &
		2.82  
		&\\ [6pt] % Row 1
		
		$\psi_{0^+}^{(2)}$ &
		76.61  &     
		23.07  &      
		0.32  &      
		79.82  &
		18.55  &
		1.63  
		&\\ [6pt] % Row 2
		
		$\psi_{0^+}^{(3)}$ &
		0.15  &     
		3.47 &      
		96.38  &      
		0.16  &
		4.29  &
		95.55  
		&\\ [6pt] % Row 3
		
		%		\midrule
		\hline
		$\psi_{0^-}^{(1)}$ &
		60.94  &     
		21.95  &      
		17.11  &      
		68.91  &
		19.85  &
		11.24  
		&\\ [6pt] % Row 1
		
		$\psi_{0^-}^{(2)}$ &
		35.03  &     
		58.80  &      
		6.17  &      
		28.08  &
		66.82  &
		5.10  
		&\\ [6pt] % Row 2
		
		$\psi_{0^-}^{(3)}$ &
		4.03  &     
		19.25  &      
		76.72  &      
		3.01  &
		13.34  &
		83.65  
		&\\ [6pt] % Row 3
		
		%		\bottomrule
		%		\bottomrule
		
	\end{tabular}
	\label{T_comp_values}
\end{table}

\begin{table}[ht]
	\caption{Simulation results for the quark-antiquark, four-quark, and glueball components of physical states  averaged over the two cases where  we identify the heaviest scalar SU(3) singlet with either   $f_0(1500)$ or $f_0(1710)$.	  The averaging is done over all simulations that result in $\chi_s$ values that are within 10\% of either  $\chi_s$ minimum.}
	\centering
	\begin{tabular}{M{80pt} | *{3}{ | r @{} M{12pt} @{} l }N} %chktex 44
		\multirow{3}{*}{States} &\multicolumn{9}{c}{ $\chi_s(i) < 1.1\times\chi_s^{\text{min}}(i)$} &\\ [6pt]
		&  \multicolumn{3}{c}{$q\overline{q}$} & \multicolumn{3}{c}{$qq\overline{q}\overline{q}$} & \multicolumn{3}{c}{glue}  &\\ [6pt]
		\toprule
		\toprule 
		
		$\psi_{8^+}^{(1)}$ &
		38  & $\pm$ & 4 &
		62  &  $\pm$ & 4 &    
		\multicolumn{3}{c}{---}  
		&\\ [6pt] % Row 1
		
		$\psi_{8^+}^{(2)}$ &
		62  & $\pm$ & 4 &
		38  &  $\pm$ & 4 &
		\multicolumn{3}{c}{---}  
		&\\ [6pt] % Row 2
		
		%		\midrule
	\hline	
		$\psi_{8^-}^{(1)}$ &
		77  &  $\pm$ & 2 &      
		23  &  $\pm$ & 2 &      
		\multicolumn{3}{c}{---}  
		&\\ [6pt] % Row 1
		
		$\psi_{8^-}^{(2)}$ &
		23  &  $\pm$ & 2 &     
		77  &  $\pm$ & 2 &      
		\multicolumn{3}{c}{---}  
		&\\ [6pt] % Row 2
		
		%		\midrule
	\hline	
		$\psi_{0^+}^{(1)}$ &
		21  & $\pm$ & 2 &     
		76  & $\pm$ & 2 &      
		3  & $\pm$ & 0.2       
		&\\ [6pt] % Row 1
		
		$\psi_{0^+}^{(2)}$ &
		79  & $\pm$ & 2 &     
		20  & $\pm$ & 3 &      
		1  & $\pm$ & 1       
		&\\ [6pt] % Row 2
		
		$\psi_{0^+}^{(3)}$ &
		0.1  &  $\pm$ & 0.1  & 
		4  & $\pm$ & 0.3  &     
		96  & $\pm$ & 0.4     
		&\\ [6pt] % Row 3
		
		%	\midrule
	\hline	
		$\psi_{0^-}^{(1)}$ &
		66  & $\pm$ & 3  &    
		20  & $\pm$ & 2  &      
		14  & $\pm$ & 2        
		&\\ [6pt] % Row 1
		
		$\psi_{0^-}^{(2)}$ &
		30  & $\pm$ & 3  &     
		63  & $\pm$ & 3  &      
		7  & $\pm$ & 3        
		&\\ [6pt] % Row 2
		
		$\psi_{0^-}^{(3)}$ &
		3  & $\pm$ & 0.4  &     
		17  & $\pm$ & 3  &      
		80 & $\pm$ & 3  
		&\\ [6pt] % Row 3
		
		%		\bottomrule
		%		\bottomrule
		
	\end{tabular}
	\label{T_comp_values_ave}
\end{table}

\begin{table}[ht]
	\caption{Rotation matrices for the two cases where  we identify the heaviest scalar SU(3) singlet with either   $f_0(1500)$ or $f_0(1710)$
		assessed by $\chi_s(1)$ and $\chi_s(2)$ in Eq. (\ref{chi_s}).   The two vertical divisions give the components  at the corresponding minimum of each  $\chi_s$.}
	\centering
	\begin{tabular}{M{80pt} | *{2}{ | M{40pt} M{40pt} M{40pt} }N} %chktex 44
		Rotation Matrix & \multicolumn{3}{c}{ Computed at $\chi_s^{\text{min}}(1)$} & \multicolumn{3}{c}{ Computed at $\chi_s^{\text{min}}(2)$}&\\ [6pt]
		\toprule
		\toprule

		\multirow{3}{*}{$K_{8^+}^{-1}$}
		&	
		$-0.638$   &
		0.770    &
		---      &
		$-0.572$   &
		0.821    &
		---
		&\\ [6pt] % Row 1
		&	
		0.770    &
		0.638    &
		---      &
		0.821    &
		0.572    &
		---
		&\\ [6pt] % Row 2
		
		%		\midrule
		\hline
		\multirow{3}{*}{$K_{8^-}^{-1}$}
		&	
		$-0.861$    &
		$-0.509$    &
		---      &
		$-0.885$    &
		$-0.466$    &
		---
		&\\ [6pt] % Row 1
		&	
		0.509     &
		$-0.861$    &
		---      &
		0.466     &
		$-0.885$    &
		---
		&\\ [6pt] % Row 2
		
		%		\midrule
		\hline
		\multirow{4}{*}{$K_{0^+}^{-1}$}
		&	
		$-0.482$    &
		$-0.857$    &
		0.182     &
		$-0.447$    &
		$-0.878$    &
		0.168
		&\\ [6pt] % Row 1
		&	
		$-0.875$    &
		0.480     &
		$-0.056$    &
		$-0.893$    &
		0.431     &
		$-0.127$
		&\\ [6pt] % Row 2
		&	
		0.039     &
		0.186     &
		0.982     &
		$-0.040$    &
		0.207     &
		0.978
		&\\ [6pt] % Row 2
		
		%		\midrule
		\hline
		\multirow{4}{*}{$K_{0^-}^{-1}$}
		&	
		$-0.781$       &
		0.469       &
		0.414     &
		$-0.830$    &
		0.446     &
		0.335
		&\\ [6pt] % Row 1
		&	
		0.592     &
		0.767     &
		0.248     &
		0.530     &
		0.817     &
		0.226
		&\\ [6pt] % Row 2
		&	
		0.201     &
		$-0.439$    &
		0.876     &
		0.173     &
		$-0.365$    &
		0.915
		&\\ [6pt] % Row 2

		%		\bottomrule
		%		\bottomrule
		
	\end{tabular}
	\label{T_rot_mats}
\end{table}

\begin{table}[ht]
				\caption{Simulation results for the decay widths (in GeV) and decay ratios for the two cases where  we identify the heaviest scalar SU(3) singlet with either   $f_0(1500)$ or $f_0(1710)$
		assessed by $\chi_s(1)$ and $\chi_s(2)$ in Eq. (\ref{chi_s}).   The first two columns give the components  at the corresponding minimum of each  $\chi_s$, the last column gives the average and standard deviation around  either  minimum.}
	\centering
	\begin{tabular}{M{60pt} || M{40pt} M{40pt} |  r @{} M{12pt} @{} l N} %chktex 44
		Variable & Computed at $\chi_s^{\text{min}}(1)$ & Computed at $\chi_s^{\text{min}}(2)$ & \multicolumn{3}{c}{Mean $\pm$ $\sigma$} \\
		\toprule
		\toprule
		$\Gamma^{1}_{0\to88}$  & 1.574 & 0.946    & 1.318        &$\pm$           & 0.236     &\\ [6pt]
		$\Gamma^{2}_{0\to88}$  & 0.770 & 1.263    & 1.055        &$\pm$           & 0.190     &\\ [6pt]
		$\Gamma^{3}_{0\to88}$  & 0.030 & 0.0001    & 0.011        &$\pm$           & 0.010     &\\ [6pt]
		$\Gamma^{2}_{0\to00}$  & 0.071 & 0.137    & 0.117        &$\pm$           & 0.041     &\\ [6pt]
		$\Gamma^{3}_{0\to00}$  & 0.004 & 0.0002    & 0.002        &$\pm$           & 0.002     &\\ [6pt]
		$\Gamma^{1}_{8\to80}$  & 0.682 & 0.763    & 0.744        &$\pm$           & 0.032     &\\ [6pt]
		$\Gamma^{2}_{8\to80}$  & 0.461 & 0.830    & 0.655        &$\pm$           & 0.150     &\\ [6pt]
		$\Gamma^{2}_{r}$   & 10.870 & 9.204         & 9.614        &$\pm$           & 2.083     &\\ [6pt]
		$\Gamma^{3}_{r}$   & 6.993 & 0.4253        & 3.791        &$\pm$           & 2.653   &\\ [6pt]
%		$\chi_s(1)$      & 14.450    & ---              & 16.507        &$\pm$           & 1.476     &\\ [6pt]
%		$\chi_s(2)$      &---         &17.203          & 24.416        &$\pm$           & 4.876     &\\ [6pt]      
%		\bottomrule
%		\bottomrule              
	\end{tabular}
\label{T_decay_W_R}
\end{table}

\end{document}